\documentclass[%
reprint,
superscriptaddress,
nofootinbib,
 amsmath,amssymb,
 aps,
 pre,
]{revtex4-2}
\usepackage{booktabs}
\usepackage{xcolor}
\usepackage{graphicx}
\usepackage{dcolumn}
\usepackage{bm}
\usepackage{chngcntr}
\usepackage{verbatim}
\usepackage{amsmath}
\usepackage{mathtools}
\usepackage{amssymb}
\usepackage{amsfonts}
\usepackage{soul}
\usepackage{multirow}
\usepackage{array}
\usepackage{makecell}

\usepackage{overpic}
\usepackage{xcolor}
\usepackage{amsmath}
\usepackage{mathtools}
\usepackage{amssymb}
\usepackage{amsfonts}
\usepackage[colorlinks=true, allcolors=blue]{hyperref}

\begin{document}

\title{Dynamics of fluctuating populations in multi-state switching environments}

\author{Mauro Mobilia}
\email{M.Mobilia@leeds.ac.uk}
\affiliation{%
 Department of Applied Mathematics, School of Mathematics, University of Leeds, Leeds LS2 9JT, United Kingdom
}%
\homepage{https://eedfp.com}

\date{\today}

\begin{abstract}
Microbial populations generally evolve in fluctuating environments under time-varying conditions. These are often described by binary switching models, sometimes seen as coarse-grained feast--famine cycles, in which resource availability switches abruptly between abundant and scarce conditions. However, experimental studies suggest that feast--famine environments actually exhibit more complex temporal dynamics. 
Here, we study how two strains, one growing slightly slower than the other, compete for the same resources in fluctuating environments comprising a finite number of intermediate states, each having its own carrying capacity. Environmental switching between these states and their carrying capacities represents gradual changes in nutrient availability. This class of multi-state stochastic switching models can be interpreted as a coarse-grained description of feast--famine cycles and allows us to investigate strain competition under the gradual recovery and depletion of resources. 
By computational and analytical means, we characterise the population dynamics in these multi-state fluctuating environments. In particular, we study how the switching rates and distribution  of carrying capacities affect the population-size statistics, fixation probability, and mean fixation time. By comparing these results with their counterparts in binary environments, we clarify how the frequency and amplitude of environmental fluctuations influence population dynamics in coarse-grained feast--famine cycles.

\end{abstract}

\maketitle

\section{Introduction}
\label{sec:introduction}
Microbial populations live in volatile and fluctuating environments in which available resources, as well as temperature, pH, toxin concentration,
can fluctuate substantially over time~\cite{hibbingBacterialCompetitionSurviving2010,ISME2016,nguyen2021}. These abiotic variations, referred to as
environmental fluctuations (EF), can have an important influence on the evolution of biological populations~\cite{smits2017,cignarella2018,rodriguez-verdugoRateEnvironmentalFluctuations2019,abdul2021}. 
In small populations, demographic fluctuations (DF) are another important form of randomness and can result in  fixation - when one type takes over the population - or extinction~\cite{Ewens,Kimura,Review2018}. 
The  variations of population size and composition are often interdependent~\cite{Roughgarden,melbinger2010,cremer2011,cremer2012,melbinger2015,chuang2009,verdon2024habitat}, and this can lead to the coupling of EF and DF~\cite{Wienand2017,Wienand2018,west2020,Taitelbaum2020,Shibasaki2021,Taitelbaum2023,Hernandez2023,Asker2023,Hernandez2024,Asker2025,Hernandez2026}. The interplay of environmental variability and DF is particularly relevant when it gives rise to population bottlenecks, where new small microbial colonies are prone to fluctuations~\cite{Wahl02,patwaAdaptationRatesLytic2010,Brockhurst2007b}. It has recently been shown that environmental variability, bottlenecks and fluctuations are particularly relevant for the dynamics of  antimicrobial resistance~\cite{coatesAntibioticinducedPopulationFluctuations2018,mahrt2021bottleneck,nguyen2021,Hernandez2023,Hernandez2024,Hernandez2026,raatzPromotingExtinctionMinimizing2023,fruetSpatialStructureFacilitates2024a}. The influence 
 of different kinds of variability, such as heterogeneous rates and interactions, 
 on the eco-evolutionary dynamics of competing species 
 has also received much attention~\cite{He2011,Szolnoki2015,Esmaeili2018}.

Population dynamics in fluctuating environments have traditionally been studied both theoretically and experimentally using two-state, or binary, environmental switching~\cite{thattai2004,kussell2005,acar_stochastic_2008,abdul2021,Lambert2014,hufton2016,ashcroftFixationFinitePopulations2014,kalyuzhnyNeutralTheoryEnvironmental2015,meyer2020},
where individuals compete for resources under two alternating environmental conditions. For instance, in   Refs.~\cite{Wienand2017,Wienand2018,west2020,Taitelbaum2020,Shibasaki2021,Hernandez2023,Asker2023,Hernandez2024,Asker2025,Hernandez2026}, the time-varying environment is represented by a binary switching carrying capacity, whose fluctuations drive the population size. This in turn modulates demographic noise, leading to the coupling of EF and DF, which has a significant impact on the population size distribution and its fate, and is characteristic of cycles of abundant and scarce resources. 

Feast--famine cycles refer to alternating periods of nutrient abundance  (``feast'') and resource deprivation  (``famine'') experienced by many microbial populations~\cite{Srinivasan98,merritt2018,himeoka_dynamics_2020,niimi2026}. The temporal statistics of these nutrient fluctuations strongly influence long-term microbial adaptation by mediating the trade-off between  growth during resource-rich periods and survival during starvation, thereby shaping the evolution of different ecological strategies, including generalist and specialist nutrient-utilization strategies~\cite{Srinivasan98,merritt2018,niimi2026}. 
Competition between strains in an environment with a binary time-switching carrying capacity can be viewed as the simplest coarse-grained representation of a feast--famine cycle, where the two possible values of the carrying capacity correspond to the environmental states of feast and famine~\cite{Wienand2017,Wienand2018,west2020,Taitelbaum2020}. While this binary switching model has the advantage of being mathematically tractable, recent studies have highlighted the importance of multiple intermediate environmental conditions in population dynamics subject to feast--famine cycles. For instance, the coupled nutrient and population dynamics considered in Ref.~\cite{himeoka_dynamics_2020} naturally generate intermediate conditions between feast and famine, which are shown to play an important dynamical role. Related work has further investigated the evolutionary consequences of fluctuating nutrient availability for competing generalist and specialist nutrient-utilization strategies under stochastic resource supply~\cite{niimi2026}. More generally, experiments have demonstrated that the temporal statistics of EF, in particular their frequency and amplitude, strongly influence microbial population dynamics under feast--famine cycles~\cite{merritt2018}. Furthermore, the experimental comparison of eco-evolutionary dynamics in binary and ternary environments revealed important dynamical differences between abrupt environmental switching and gradual environmental deterioration through intermediate conditions~\cite{sanchez2013}. These studies therefore suggest that feast--famine environments possess a richer temporal and statistical structure than can, in general, be captured by binary switching models. This motivates us to investigate stochastic multi-state environmental models with intermediate environmental states characterized by distinct carrying capacities. 

\begin{figure*}
    \centering
    \setlength{\unitlength}{1cm}
    \begin{overpic}[width=0.8\textwidth]{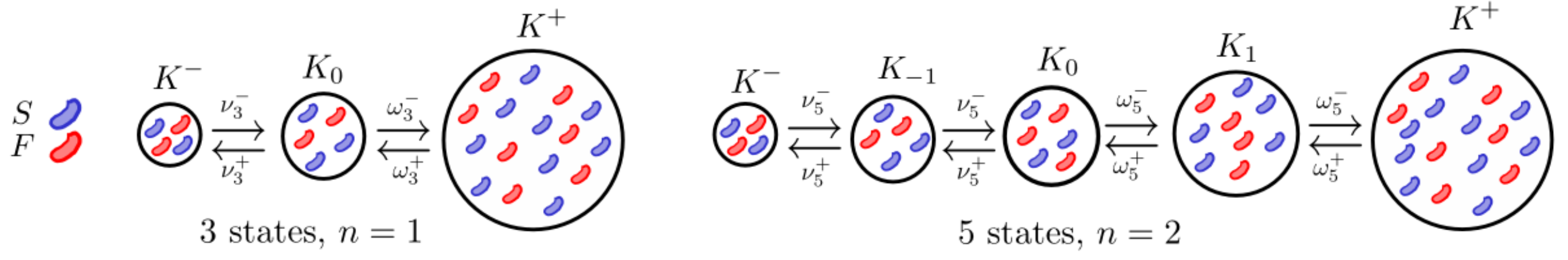}
    \put(1.25,15.2){(a)}
    \end{overpic}
    \begin{overpic}[width=0.8\textwidth]{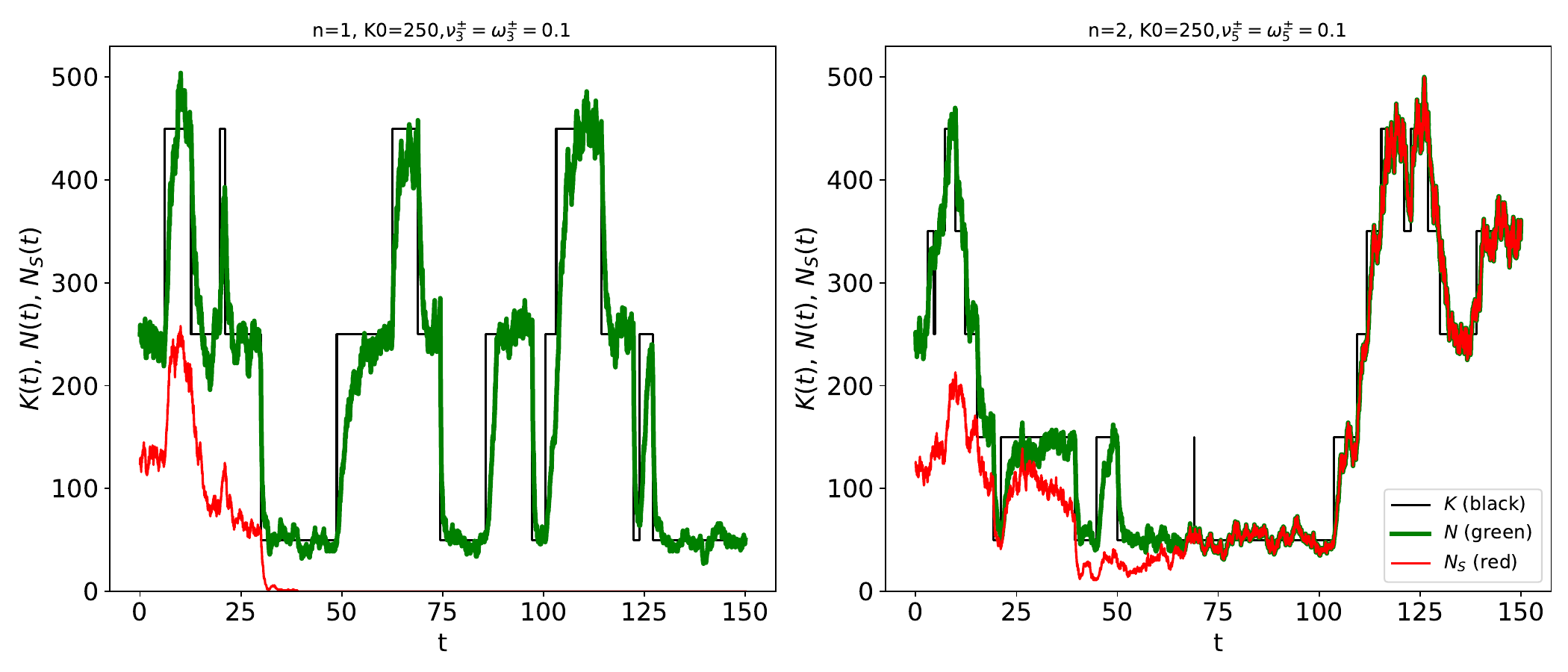}
     \put(1,42){(b)}
    \end{overpic}
    \caption{(a) Illustration of the the $3$-state ($n=1$, left) and  $5$-state ($n=2$, right) switching eco-evolutionary dynamics. (b) Typical sample paths of $K(t)$ (black), $N(t)$ (green), and $N_S(t)$ (red)  for the 3-state (left) and 5-state (right) switching models with $(K^-,K^+)=(50,450)$.
    The number of $F$ individuals is $N_F(t)=N(t)-N_S(t)$ (not shown).
    (b, left) $n=1$, $K_0=250$, $K(t)\in \{50,250,450\}$, $\nu^{\pm}_3=\omega^{\pm}_3=0.1$ (intermediate switching regime; see text); (b, right) $n=2$, $K_0=250$, $K(t)\in \{50, 150, 250, 350, 450\}$, $\nu^{\pm}_5=\omega^{\pm}_5=0.1$ (intermediate switching). In (b, left), 
    simultaneous $S$
    extinction and $F$ fixation (not shown)  occurs at $t\approx 35$, whereas in (b, right) there is $S$ fixation (with extinction of $F$, not shown) at $t\approx 70$. After the 
     fixation of a strain and  extinction of the other, the population consists of only the fixed type ($F/S$ on the left/right panel) and, 
    driven by the switches of $K(t)$,  
    $N(t)$ keeps fluctuating. 
    Other parameters in (b) are $(s,x_0)=(0.02,0.5)$.
    }
    \label{fig:Fig1-cartoon}
\end{figure*}

Specifically, here we  study the dynamics of two strains, one slightly slower than the other, competing for the same resources in multi-state fluctuating environments. The time-varying environment consists of multiple intermediate states, ordered from the harshest (``famine'') to the mildest (``feast'') conditions. Each environmental state is associated with its own carrying capacity, whose value increases progressively from harsh to mild states; see Fig.~\ref{fig:Fig1-cartoon}(a). 
We therefore consider a class of multi-state stochastic switching models that provide a coarse-grained representation of feast--famine cycles with intermediate environmental states. 
These stochastic individual-based models allow us to  describe, at population level,  the strain competition  when the recovery and depletion of resources is gradual~\cite{sanchez2013,merritt2018,himeoka_dynamics_2020,niimi2026}. This is an important feature of feast--famine cycle dynamics that is not captured by binary models. To characterise the multi-state switching dynamics, we ask: {\it How do the environmental switching rates and the distribution of carrying capacities influence the population size statistics, as well as the probability and mean time of fixation? How do these quantities differ from their binary-state counterparts?}\\ 
We address these questions by  using simulation and analytical means
to compute these quantities and compare them
 in multi-state and binary switching models. 
Since it has been experimentally shown  feast--famine cycle dynamics depend strongly on the frequency and amplitude of EF, we particularly study 
the influence of the switching rates (EF frequency) and distribution of  carrying capacities -- encoding the EF amplitude -- on the population size  distribution  and on its fixation properties. This sheds further light on how the frequency and amplitude of environmental fluctuations shape the complex dynamics of coarse-grained feast--famine cycles.

 In the next section, we introduce the class of multi-state feast--famine models that we study and outline our methods. In Sec.~\ref{Sec:Results}, we present our results, first for the population size distribution and average population size (Sec.~\ref{Sec:PSD}) and then discuss in detail the influence of the environment on the fixation probability of the slow strain (Sec.~\ref{Sec:Fixation}). 
 Sec.~\ref{Sec:Discussion} is dedicated to a
  discussion of  our findings and to our conclusions.  Additional technical details, as well as supplementary results,
 are given in a series of appendices.

\section{Models \& Methods}
\label{Sec:Model}

We consider a well-mixed population of time-fluctuating size $N(t)=N_{S}(t) + N_{F}(t)$ that at time $t$
consists of $N_{S}$  individuals of the slow growing strain
$S$, and $N_F$ of a faster growing 
type $F$. Each $F$ individual has a baseline fitness $f_F=1$, while all slow growers  
have fitness $f_S=1-s$. We assume that $S$ and $F$ compete for the same resources, with population growth limited by the carrying capacity denoted by $K$. The fraction of slow growers in the population is $x=N_S/N$, and the  average population  fitness is thus  
$\bar f=x f_S + (1-x)f_F=1-sx$. 
Therefore, the per capita growth rates of 
$F$ and $S$ are respectively $1/\bar f$ and $(1-s)/\bar
f$, with the growth of both strains limited by a logistic death rate $N/K$~\cite{Roughgarden,allenIntroductionStochasticProcesses2010b,cremer2011,Wienand2017,Wienand2018,Taitelbaum2020,Taitelbaum2023,Asker2025}. The population dynamics
evolve in continuous time. For analytical tractability, we often assume
  $0<s\ll 1$, giving  a small selective advantage to $F$ over $S$ and yielding 
 a time scale separation~\cite{Wienand2017,Wienand2018,west2020,Taitelbaum2020,Taitelbaum2023,Asker2025}; see below. However, many results of Sec.~\ref{Sec:Fixation} remain valid for  $0<s<1$.
 
In close relation to the Moran process~\cite{Moran,Ewens,Blythe07,traulsen2009stochastic,antal2006fixation} (see Sec.~\ref{Sec:Static-Moran}), a reference model in mathematical biology, 
the competition (selection) dynamics is  represented by a multivariate birth-death process~\cite{allenIntroductionStochasticProcesses2010b,Ewens} defined by the birth (or division) and death of an individual of type $\alpha\in\{S,F\}$ according to ~\cite{Wienand2017,Wienand2018,west2020,Taitelbaum2020,Shibasaki2021,Taitelbaum2023}
\begin{align}
    N_{\alpha}\stackrel{T^+_{\alpha}}{\longrightarrow} N_{\alpha}+1  \text{ (birth)} \quad\text{and} \quad N_{\alpha}\stackrel{T^-_{\alpha}}{\longrightarrow} N_{\alpha}-1 \text{ (death)},
\label{eq:BD}
\end{align}
occurring at the transition rates
\begin{equation}
\label{eq:Transition_rates}
        T^+_\alpha = \frac{f_\alpha}{\overline{f}} N_\alpha \quad \text{and}\quad
        T^-_\alpha = \frac{N}{K}N_\alpha,
\end{equation} %
where the carrying capacity $K$ is constant in a static environment and is a  randomly time-varying quantity in fluctuating environments; see below. 
(For notational simplicity we have dropped the time dependence from $\overline{f}$ and transition rates).
In this formulation,  selection operates
 on birth  events~\cite{Wienand2017,Wienand2018,Taitelbaum2020,Shibasaki2021,Taitelbaum2023,west2020,Taitelbaum2023,Asker2023,Hernandez2023,Hernandez2024,Asker2025,Hernandez2026,marrec2021,abbaraMutantFateSpatially2024,moawad2024}. This can be generalized to include selection on deaths; see e.g.~Refs.~\cite{melbinger2010,cremer2011,yagoobiCategorizingUpdateMechanisms2023}.

\subsection{Multi-state environmental dynamics}
\label{Sec:Model-EV}
The carrying capacity provides a coarse-grained description of environmental limitations on population growth~\cite{Roughgarden}, 
and 
is closely related to chemostat setups that are commonly used in laboratory-controlled experiments~\cite{Chemostat-book,abdul2021,Lambert2014,nguyen2021,Li2022,Shibasaki2021}.   A simple and biologically relevant way to represent  environmental variability and to capture 
the interdependence of demographic and EF 
therefore consists of letting the carrying capacity be a randomly time-varying quantity~\cite{Wienand2017,Wienand2018,west2020,Taitelbaum2020,Shibasaki2021,Taitelbaum2023,Hernandez2023,Asker2023,Hernandez2024,Asker2025,Hernandez2026}. Environmental variability is thus often described by binary-state models~\cite{thattai2004,kussell2005,acar_stochastic_2008,abdul2021}, with the carrying capacity typically taking two possible values describing sudden and radical changes of nutrient resources~\cite{Wienand2017,Wienand2018,west2020,Taitelbaum2020,Shibasaki2021,Hernandez2023,Asker2023,Hernandez2024,Asker2025,Hernandez2026}; see Appendix~\ref{appendix:background}. 
Here, we introduce a class of multi-state switching models where, in addition to extreme abundance (feast) and scarcity (famine), there are $2n-1$ intermediate states  ($n\geq 1$), each with its own carrying capacity; see Fig.~\ref{fig:Fig1-cartoon}(a). The switching between the ($2n+1$) environmental states and their carrying capacities can hence be viewed as describing a coarse-grained multi-state feast--famine cycle, with the switching rate and distribution of the carrying capacities encoding the frequency and amplitude of nutrient fluctuations~\cite{Srinivasan98,merritt2018,himeoka_dynamics_2020,niimi2026}.
Although it does not explicitly model nutrient dynamics, this formulation provides an individual-based description  
 of the gradual degradation and recovery of environmental conditions that is  characteristic of feast--famine dynamics~\cite{merritt2018,himeoka_dynamics_2020,niimi2026}.

In this context, environmental variability is encoded 
 by  the continuous-time multi-state coloured noise $\xi(t)=i\in\{-n,\dots,0,\dots,n\}$; see Appendix~\ref{appendix:EV} that is a $(2n+1)$-state generalisation of the classical binary telegraph (dichotomous) noise~\cite{Bena2006,Ridolfi11,HL06,acar_stochastic_2008,hufton2016,Wienand2017,Wienand2018,west2020,Shibasaki2021,Hernandez2023,Asker2023,Hernandez2024,Asker2025,Hernandez2026}. Each of the possible values    $\xi(t)=i$ corresponds to a specific environmental state with its own  
carrying capacity  $K(\xi(t))=K_i$. 
The states $i=-n,\dots,n$ are ordered from the  harshest (``famine'') to the 
 mildest (``feast''). The famine and feast states are labelled by $-n$ and $n$, respectively, and have carrying capacities $K_{-n}\equiv K^-$ and $K_{n}\equiv K^+$. (The notation $K^{\pm}$ is introduced to facilitate the comparison with the binary case.) We respectively refer to  $i\in\{1-n,\dots,-1\}$ and $i\in\{1,\dots,n-1\}$ as the intermediate harsh and mild states, while $i=0$ is the ``median state'' (or ``centre state''; see below) and its carrying capacity is $K_0$.
The carrying capacities are ordered and increase from harsh to mild states according to 
\begin{equation}
\label{eq:Korder}
K^- \equiv K_{-n}<\dots<K_{-1}<K_0<K_1<\dots<K_{n}\equiv K^+.
\end{equation}
when $K_0=(K^++K^-)/2$, $i=0$ is called the centre state.

The transitions between the environmental  states $i$ and $i+1$
occur with rates $\nu^{\pm}_{2n+1}$ for harsh states ($i=-n,\dots, -1$),  and with rates 
$\omega^{\pm}_{2n+1}$ for states 
$i=1,\dots, n-1$, according to
\begin{equation}
\label{eq:xi}
\begin{aligned}
\xi=i&\xrightleftharpoons[\nu^+_{2n+1}]{\nu^-_{2n+1}}\xi=i+1 \quad \text{ ($i=-n,\dots,-1$),} \\
\xi=i&\xrightleftharpoons[\omega^+_{2n+1}]{\omega^-_{2n+1}}\xi=i+1 \quad \text{ ($i=0,\dots,n-1$).}
\end{aligned}
\end{equation}
The carrying capacity $ K(t)\in\{K_i\}_{i=-n}^{n}$ is thus a random variable whose value switches at  rates  $\nu^{\pm}_{2n+1}$ and $\omega^{\pm}_{2n+1}$ according to 
\begin{equation}
 \label{eq:Kswitch}
\begin{aligned}
  K_i&\xrightleftharpoons[\nu^+_{2n+1}]{\nu^-_{2n+1}}K_{i+1} \text{ for $i=-n,\dots,-1$},\\
   K_i &\xrightleftharpoons[\omega^+_{2n+1}]{\omega^-_{2n+1}}K_{i+1} \text{ for $i=0,\dots,n-1$}.
 \end{aligned}
\end{equation}
When $\nu^{\pm}_{2n+1}\neq \omega^{\pm}_{2n+1}$, 
transitions among harsh states and among mild states occur at different rates. 
For the sake of simplicity, we introduce the parameter $\epsilon>-1$ and 
henceforth assume $\omega^{\pm}_{2n+1}=(1+\epsilon)\nu^{\pm}_{2n+1}$. Therefore,  the transitions between states $i> 0$ occur at a higher rate than those between states $i< 0$ when $\epsilon>0$, whereas it is the opposite when  $-1<\epsilon<0$.
The  only physical constraint on  the values of $K_i$ is given by
the ordering relation \eqref{eq:Korder}. While various choices are possible, for simplicity, we
consider $\{K_i\}_{i=-n}^{n}$ to be uniformly (linearly) distributed in $[K^-,K_0]$ and $[K_0,K^+]$, with
\begin{equation}
  \label{eq:Ki}
  K_i= \begin{cases}
         K^{-}+ (K_0-K^{-})\left(1+\frac{i}{n}\right) & \text{if }  i=-n,\dots, 0,\\
        K_0+(K^{+}-K_{0})\frac{i}{n} &\text{if } i=1,\dots, n.
        \end{cases}
 \end{equation}
When $n=1$ we thus have $K\in\{K^-,K_0,K^+\}$, while in the five-state case ($n=2$), $K\in\{K^-, (K_0+K^-)/2, K_0,  (K_0+K^+)/2,K^+\}$; 
see Appendix~\ref{appendix:EV} and Fig.~\ref{fig:Fig1-cartoon}(a). Hence, the spacing between consecutive carrying capacities in $\{K_i\}_{i=-n}^{0}$ is smaller than that in $\{K_i\}_{i=0}^{n}$ when $K_0<(K^++K^-)/2$, whereas the converse holds when $K_0>(K^++K^-)/2$. When $K_0=(K^++K^-)/2$,  $\{K_i\}_{i=-n}^{n}$ are uniformly spaced between $K^-$ and $K^+$, and $i=0$ corresponds to the ``centre state''.

To meaningfully compare the environmental variability  in multi-state and binary models, we require that the effective binary switching rates  $\widetilde{\nu}^{\pm}$ of the 
 transitions $K^-\xrightleftharpoons[\widetilde{\nu}^+]{\widetilde{\nu}^-}K^+$~\cite{Wienand2017,Wienand2018,west2020,Taitelbaum2020} between the  extreme carrying capacities are such that the mean times for these transitions be the same as under multi-state switching. We notice that
 in ($2n+1$)-state switching models, the transitions
$K^{\mp} \to K^{\pm}$ consists of $n$  moves, each of mean waiting time $1/\nu_{2n+1}^{\mp}$, and $n$ other transitions of expected waiting time $1/\omega_{2n+1}^{\mp}$ each. Matching the mean times of the transitions
$K^{\mp} \to K^{\pm}$ under binary and multi-state switching thus  
yields (see Appendices \ref{appendix:background} and \ref{appendix:EV}) 
\[
\widetilde{\nu}^{\pm}=\frac{\nu^{\pm}_{2n+1}\omega^{\pm}_{2n+1}}{n(\nu^{\pm}_{2n+1}+\omega^{\pm}_{2n+1})}=\frac{\nu^{\pm}_{2n+1}}{n}~\left(\frac{1+\epsilon}{2+\epsilon}\right).
\]
The effective rates $\widetilde{\nu}^{\pm}$ allow us to compare the environmental dynamics across binary and $(2n+1)$-state switching models. 
It is convenient  to write  $\nu_{2n+1}\equiv (\nu^-_{2n+1} + \nu^+_{2n+1})/2$,  $\widetilde{\nu}\equiv (\widetilde{\nu}^- + \widetilde{\nu}^+)/2$
and characterise the bias between forward ($K_i\to K_{i+1}$) and backward switching ($K_{i+1}\to K_i$) by $\delta=
(\nu^-_{2n+1} - \nu^+_{2n+1})/(2\nu_{2n+1})
=(\widetilde{\nu}^- - \widetilde{\nu}^+)/(2\widetilde{\nu})$
~\cite{Taitelbaum2020,Taitelbaum2023,Hernandez2023,Hernandez2024,Hernandez2026}, with $|\delta|< 1$. 
We thus have 
\begin{equation}
 \label{eq:nu}
\nu^{\pm}_{2n+1}=\nu_{2n+1}(1\mp \delta), \quad \omega^{\pm}_{2n+1}=\nu_{2n+1}(1+\epsilon)(1\mp \delta), 
\end{equation}
and  the effective binary rates $\widetilde{\nu}^{\pm} 
 =\widetilde{\nu}(1\mp\delta)$, where
\begin{equation}
 \label{eq:nutilde}
 \widetilde{\nu}=
 \frac{\nu_{2n+1}}{n} \left(\frac{1+\epsilon}{2+\epsilon}\right), 
\end{equation}
 with transitions from states $i\to i+1$ occurring at a higher rate than those from $i+1\to i$ when $\delta>0$  ($i\neq 0$). This means that $\delta>0$ corresponds to spending on average more time in states $i+1$
than $i$  ($i\neq 0$).

In this work,  environmental noise and the carrying capacity are always at stationarity,
with  their distribution denoted by ${\bm \pi}=(\pi_i)$ and characterised  
by  the parameters $n$ and $\rho\equiv (1+\delta)/(1-\delta)$. In Appendix~\ref{appendix:EV}, we show that 
\begin{equation}
\label{eq:pi}
\begin{aligned}
\pi_i=
\begin{cases}
\left(\frac{1-\rho}{1-\rho^{2n+1}}\right)\rho^{i+n}, \quad  &\text{ if } \rho\neq 1 \; (\delta\neq 0) \;\\
\frac{1}{2n+1}, \quad &\text{ if } \rho= 1 \; (\delta=0).\,
\end{cases}
\end{aligned}
\end{equation}
This distribution is a function of $n$ and $\delta$, 
but is independent of $\epsilon$ (see below). 
Since $\pi_i/\pi_{-i}=\rho^{2i}$, mild states ($i>0$) are more likely than harsh states ($i>0$) when $\rho>1$ ($\delta>0$) and conversely when $\rho<1$ ($\delta<0$).
The distribution $\pi_i=1/(2n+1)$ is  uniform when $\delta=0$ (symmetric forward/backward switching)~\cite{Wienand2017,Wienand2018,west2020,Shibasaki2021,Taitelbaum2023,Asker2025}. 
Here, $\xi$ and $K$ are always at stationarity, and their initial value in  each simulation run is  drawn from ${\bm \pi}$ according to Eq.~\eqref{eq:pi}; see Appendix~\ref{appendix:simulations}.
 
In this study,  environmental dynamics is thus represented by a 
 class of multi-state switching models, inspired by chemostat systems~\cite{Chemostat-book,abdul2021,Lambert2014,nguyen2021,Li2022,Shibasaki2021},  
 characterised by the parameters $\{n, K^{\pm},K_0,\nu_{2n+1},\delta,\epsilon\}$. 
These multi-state individual-based models, defined by 
Eqs.~\eqref{eq:BD}-\eqref{eq:Ki}, satisfy the  master equation \eqref{eq:ME}, and  are simulated using the Gillespie algorithm~\cite{Gillespie76}; see Appendix~\ref{appendix:simulations}.

\subsection{Mean-field dynamics in a static environment}
\label{Sec:Static-MF}
It is useful to consider
the dynamics in a static environment, where the carrying capacity is constant $K=\bar{K}\gg 1$. In this setting, after a transient, the population size is well approximated by the constant carrying capacity, $N\approx \bar{K}$~\cite{Wienand2017,Wienand2018,west2020,Taitelbaum2020,Shibasaki2021,Hernandez2023,Asker2023,Hernandez2024,Asker2025,Hernandez2026}. 
When $N$ and $\bar{K}$ are very large, demographic fluctuations
are negligible and a mean-field description of the dynamics is in order. Here, it is defined
by the following rate equations for $N$ and $x$~\cite{Wienand2017,Wienand2018,west2020,Taitelbaum2020}:
\begin{equation}
\label{eq:MF}
 \begin{aligned}
 \dot{N}&=\sum_{\alpha=\{S,F\}} \left(T_{\alpha}^+- T_{\alpha}^-\right)=N\left(1-\frac{N}{\bar{K}}\right),\\
  \dot{x}&=\frac{T_{S}^+- T_{S}^-}{n}-x\frac{\dot{N}}{N}=-\frac{sx(1-x)}{1+sx},
 \end{aligned}
\end{equation}
where the dot indicates the time derivative and we have used the expressions of $T_{\alpha}^{\pm}$ given by Eqs.~\eqref{eq:Transition_rates}, with $f_{S}=1-s$ and $f_F=1$  and $\overline{f}=1-sx$~\cite{Wienand2017,Wienand2018,west2020,Taitelbaum2020}.
The logistic rate equation for $N$ leads to the population size reaching the carrying capacity, $N\to \bar{K}$, on a timescale $t\sim 1$. Since $\dot{x}<0$ for $0<x<1$, the fraction of $S$ cells $x$
decreases on a slower timescale $t\sim \mathcal{O}(1/s)$, and eventually vanishes ($x\to 0$). We notice that, when $s\ll 1$, there is a timescale separation, with $N$ and $x$ acting respectively as the fast and slow variables. Further, Eqs.~(\ref{eq:MF}) show that the dynamics of $N$ and $x$ decouple at mean-field level within a static environment.  As discussed below, in this class of switching models, eco-evolutionary coupling  arises from fluctuations at the individual-based level.

\subsection{Finite population in a static environment: Moran model}
\label{Sec:Static-Moran}
It is also relevant to consider a finite population of constant size $N=\bar{K}$  in a static environment, and study its fixation properties. This is achieved  in terms of the Moran model~\cite{Moran,Ewens}, which is a reference evolutionary process~\cite{antal2006fixation,blythe2007stochastic,
traulsen2009stochastic}.

In a static environment where the carrying capacity is finite and  constant, with $K=\bar{K}\gg 1$, the population is unlikely to go extinct in an observable time~\cite{assaf2017,Assaf2010,AM11,Wienand2017,Wienand2018,Taitelbaum2020}, and its size fluctuates about the carrying capacity, with $N \approx \bar{K}$. In this setting, it is reasonable to assume  $N = \bar{K}$ (Moran approximation); see Eqs. \eqref{eq:MF}. The dynamics of the population make-up 
 is thus described by tracking the number $N_S$ of $S$ cells present in a population 
 of size $N=\bar{K}$ evolving according to the Moran process~\cite{Moran,Ewens}. (The 
 number of $F$ is $N_F=N-N_S=\bar{K}-N_S$.)
Within this Moran approximation, the population makeup $(N_S, N_F)= (N_S, \bar{K}-N_S)$ changes according to ~\cite{Moran,Ewens,antal2006fixation,
Blythe07,traulsen2009stochastic}
\begin{equation}
\label{Eq:Moran}
\begin{aligned}
    (N_S, N_F) &\stackrel{{\widetilde T}_S^{+}}{\longrightarrow} (N_S+1, N_F-1), 
    \nonumber\\ (N_s, N_F)&\stackrel{{\widetilde T}_S^{-}}{\longrightarrow}   (N_S-1, N_F+1),
    \end{aligned}
\end{equation}
where the Moran transition rates ${\widetilde T}_{S}^{\pm}$ are defined in terms of those of Eq.~\eqref{eq:Transition_rates} with $K=\bar{K}$~\cite{Wienand2017,Wienand2018,west2020,Taitelbaum2020,Hernandez2023,Asker2023,Hernandez2024,Asker2025,Hernandez2026}:
\begin{equation}
 \label{eq:MoRates}
 \begin{aligned}
 {\widetilde T}_S^{+}(x, \bar{K})&=\frac{T^+_{S} T^-_{F}}{\bar{K}} =\frac{(1-s)}{1-sx}~\bar{K} x\left(1-x\right),\\
 {\widetilde T}_S^{-}(x, \bar{K})&=\frac{T^-_{S} T^+_{F}}{\bar{K}} =\frac{1}{1-sx}~\bar{K} x\left(1-x\right),
 \end{aligned}
\end{equation}
with $x=N_S/N=N_S/\bar{K}$.
These Moran transition rates correspond to the increase and decrease of the number of $S$ by a unit.

Since the population size is constant, every birth of an $S$ is accompanied by the death of an $F$ , and vice versa. 
This Moran process is characterised by the absorbing states $(N_S,N_F)=(\bar{K},0)$ ($S$ fixation) and $(N_S,N_F)=(0,\bar{K})$ ($F$ fixation).
The  probability and mean time of fixation of this process can be computed analytically~\cite{Ewens,antal2006fixation,Blythe07,traulsen2009stochastic}.
When the initial fraction of $S$ individuals is $x_0=N_S(0)/N(0)=N_S(0)/\bar{K}$, 
the fixation probability of $S$ in a finite population of size $\bar{K}$ is~\cite{antal2006fixation,traulsen2009stochastic,Taitelbaum2020,Asker2023,Asker2025}
\begin{equation}
 \label{eq:phi_M}
 \phi_{M}(x_0,\bar{K})=\lim_{t\to \infty}\mathbb{P}\left(N_S(t)=\bar{K}\right)=\frac{(1-s)^{-\bar{K} x_0}-1}{(1-s)^{-\bar{K}}-1},
 \end{equation}
where $\mathbb{P}\left(\cdot\right)$ denotes the probability that the condition $\left(\cdot\right)$ is satisfied. 
The unconditional Moran mean fixation time, denoted by  $\tau_{M}(x_0,\bar{K})$, 
can also be computed exactly~\cite{antal2006fixation,traulsen2009stochastic,Assaf2010,Taitelbaum2020,Asker2023,Asker2025} and its expression is given of Eq.~\eqref{eq:MFT_M} in Appendix~\ref{appendix:MFT}.

In what follows, $x_0$ will be treated as a parameter, and we simplify the notation by generally writing respectively $\phi_{M}(\bar{K})$ and $\tau_{M}(\bar{K})$ for $\phi_{M}(x_0,\bar{K})$ and $\tau_{M}(x_0,\bar{K})$, with the value of $x_0$ clear from the context.

\section{Results: Population size distribution and fixation  in fluctuating environments}
\label{Sec:Results}
In the mean-field setting, the slow strain $S$ unavoidably goes extinct by Gause's competitive exclusion principle~\cite{gause1932,hardin1960}; see Eq.~\eqref{eq:MF}. When $K$ is constant and finite, $S$
  has a fixation probability that vanishes exponentially in a static environment; see Eq.~\eqref{eq:phi_M}. This picture radically changes  in fluctuating environments where the 
 carrying capacity varies endlessly by switching back and forth between the values $\{K_i\}_{i=-n}^n$~\cite{Wienand2017,Wienand2018,west2020,Taitelbaum2020,Shibasaki2021,Taitelbaum2023,Hernandez2023,Asker2023,Hernandez2024,Asker2025,Hernandez2026}. 
 
Here, to understand how the environmental variability statistics influence the dynamics of the multi-state switching models, we study the impact of
$\{K_i\}_{i=-n}^n$ and  switching rates $\nu^{\pm}_{2n+1}$ on the  population size distribution (PSD), $S$ fixation probability and unconditional mean fixation time. 
According to Eq.~\eqref{eq:Ki}, the distribution of 
$\{K_i\}_{i=-n}^n$, and therefore the amplitude of EF, is controlled by  $K_0$,  $K^{\pm}$ and $n$. For the comparison with the effective binary environments, where the carrying capacity switches at effective rates  $\widetilde{\nu}^{\pm}$ between the extreme values $K^{\pm}$ according to 
$K^-\xrightleftharpoons[\widetilde{\nu}^+]{\widetilde{\nu}^-}K^+$, 
the relation between $\widetilde{\nu}^{\pm}=\widetilde{\nu}(1\mp \delta)$ and the multi-state switching rate  $\nu_{2n+1}$ 
 is set by Eq.~\eqref{eq:nutilde}.

Our simulation data are obtained by sampling $10^5$ realizations and considering $50$-$100$ values of $\nu_{2n+1}$ in the ranges $[10^{-2},10^2]$, $[10^{-3},10]$, and $[10^{-4},10]$. 
In  Figs.~\ref{fig:Fig2-hist}-\ref{fig:Fig8_FigMFT_n2}, $(K^-,K^+)\in \{(50,450),(25,475)\}$,  $\delta\in\{0, \pm 0.1, \pm 0.2, \pm 0.25\}$, $\epsilon\in\{-0.8, -0.5, 0, 0.5, 5\}$ and $(s,x_0)\in \{(0.02,0.5),(0.05,0.6), (0.1,0.7)\}$. 
Simulation methods and further  details are given in Appendix \ref{appendix:simulations}.

 \subsection{Population size distribution in fluctuating environments}
 \label{Sec:PSD}
 This class of models is characterised by a long-lived, or quasi-stationary, 
population size distribution (PSD) followed by the eventual extinction of the entire population
after a very long time (practically unobservable when
$K\gg 1$~\cite{Wienand2017,Wienand2018,Taitelbaum2020,assaf2017}). Here, we focus on 
timescale $t\gtrsim 1/s$ on which one strain is likely to
have taken over the population (fixation) while the other 
has gone extinct, and the population has settled 
into its long-lived PSD~\cite{Wienand2017,Wienand2018,west2020,Taitelbaum2020,Shibasaki2021,Taitelbaum2023,Hernandez2023,Asker2023,Hernandez2024,Asker2025,Hernandez2026}; see Fig.~\ref{fig:Fig1-cartoon}(b) and  Appendix~\ref{appendix:MFT}.
\begin{figure*}
\centering
\setlength{\unitlength}{1cm}
\begin{overpic}[width=0.24\textwidth]{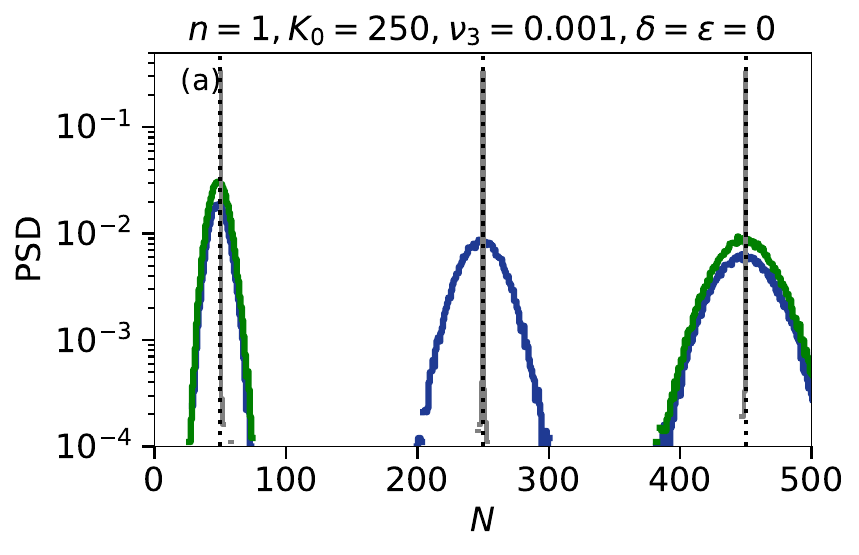}
\end{overpic}
\hfill
\begin{overpic}[width=0.24\textwidth]{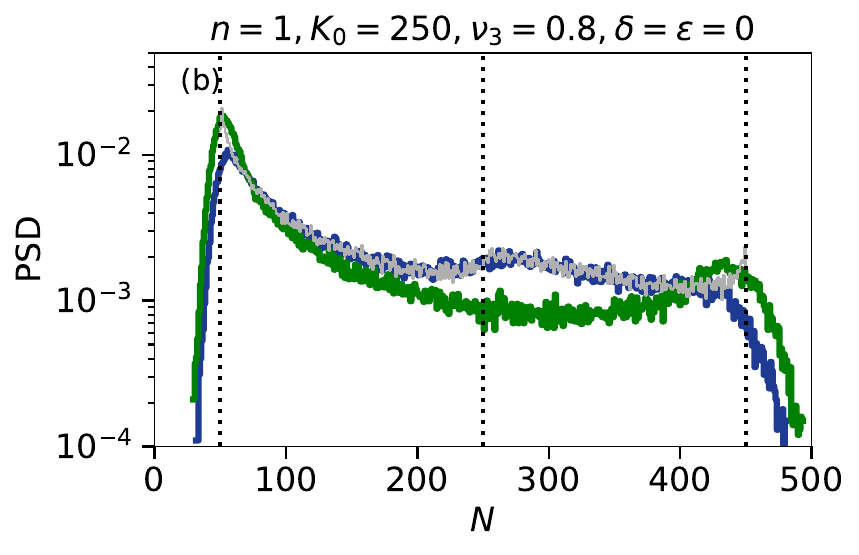}
\end{overpic}
\hfill
\begin{overpic}[width=0.24\textwidth]{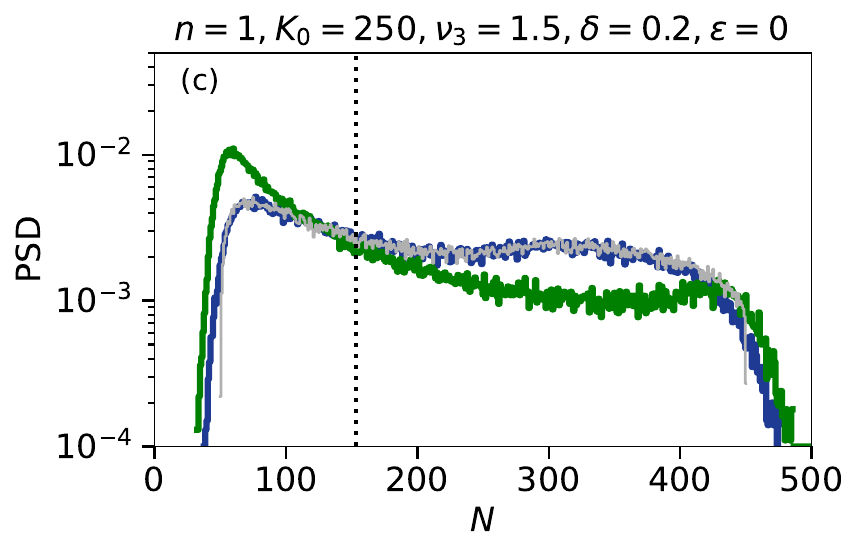}
\end{overpic}
\hfill
\begin{overpic}[width=0.24\textwidth]{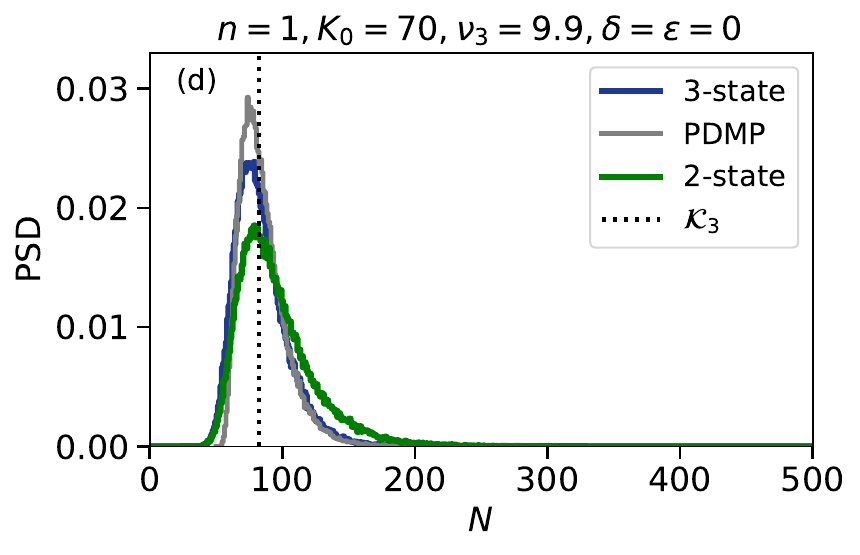}
\end{overpic}
\vspace{0.5em}
\begin{overpic}[width=0.24\textwidth]{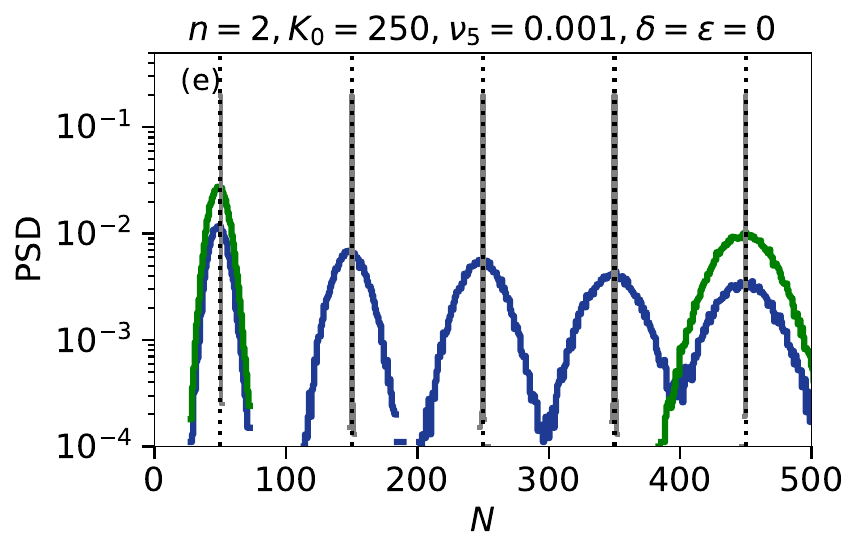}
\end{overpic}
\hfill
\begin{overpic}[width=0.24\textwidth]{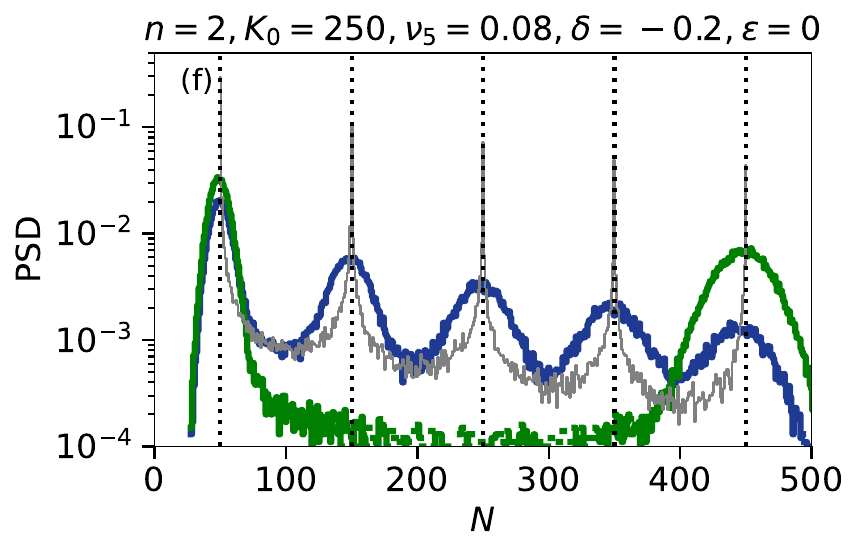}
\end{overpic}
\hfill
\begin{overpic}[width=0.24\textwidth]{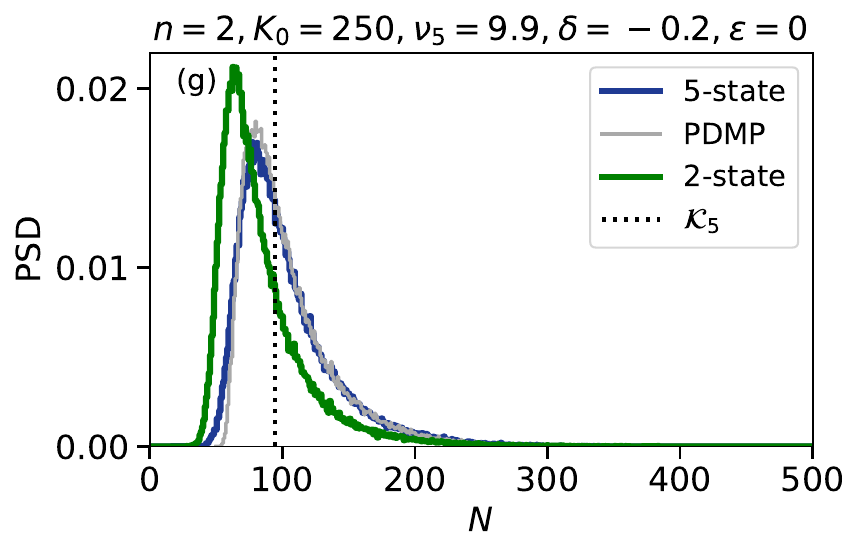}
\end{overpic}
\hfill
\begin{overpic}[width=0.24\textwidth]{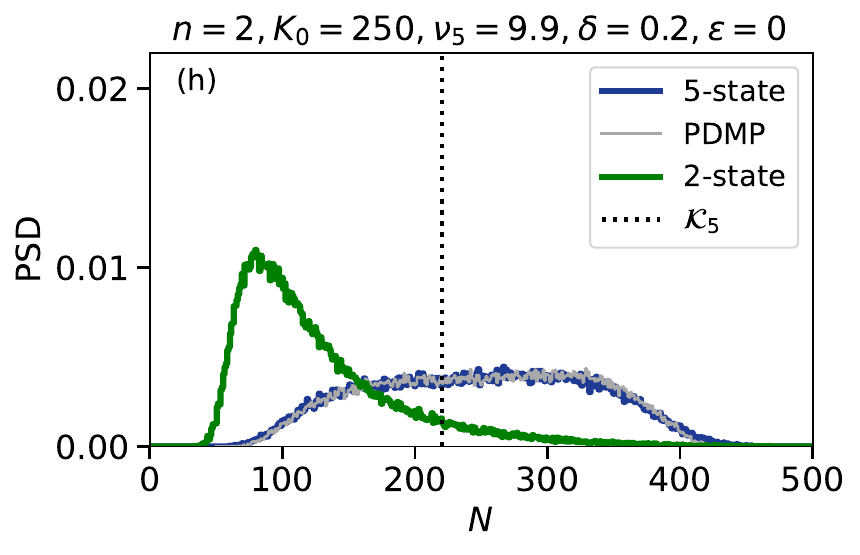}
\end{overpic}
\caption{Histograms of the quasi-stationary  PSD,  $p_{\nu_{2n+1}}(N)$
(dark blue), for different switching rates for 3-state (a)-(d) and 5-state (b)-(h) multi-state switching models. In grey are shown the histograms 
of the $N$-PDMP (see Eq,~\eqref{eq:PDMP-nstate} and Appendix~\ref{appendix:simulations}). 
These are compared with those of the PSD  $p_{\widetilde{\nu}}(N)=p_{(\nu_{2n+1}/n)(1+\epsilon)/(2+\epsilon)}(N)$ of the effective two-state model (see text and Eq.~\eqref{eq:nutilde}) shown in green. 
Panels (a)-(c), (e) and (f) are in semi-log scale (values $<10^{-4}$ are not shown), while  (d), (g) and (h) are in linear scale.  
The vertical dotted lines in (a),(b),(d),(e) are eyeguides showing
$N=K_i$. In panels (c) (d),(g),(h) the  dotted lines show ${\cal K}_3\approx 153.23$ (c) and  ${\cal K}_3\approx 82.17$ (d), while in (g) and (h) they respectively indicate ${\cal K}_5\approx 94.70$  and  ${\cal K}_5\approx 220.18$.
In all panels, $(s,x_0,K^-,K^+)=(0.02,0.5,50,450)$.
 For the green histograms in (d),(g),(h), ${\cal K}_2=90$ (d), ${\cal K}_2\approx 77.59$ (g), and ${\cal K}_2\approx 107.14$ (h); see Eq.~\eqref{eq:curlyK2}.
Other parameters are: 
$(\nu_3,K_0,\delta,\epsilon)=\{(10^{-3},250,0,0)\text{ in (a)}, (0.8,250,0,0)\text{ in (b)}, (1.5,250,0.2,0)\text{ in (c)}, 
 (9.9,70,0,0)\text{ in (d)}\}$; and 
 $(\nu_5,K_0,\delta,\epsilon)=\{(10^{-3},250,0,0)\text{ in (e)},
 (0.08,250,-0.2,0)\text{ in (f)}, (9.9,250,-0.2,0)\text{ in (g)}, (9.9,250,0.2,0)\text{ in (h)}\}$. 
 In (a) and (c) the dotted lines coincide with $N$-PDMP results (grey lines).
 All histograms have been obtained after $t>2000$, as outlined in Appendix~\ref{appendix:simulations}.  See also Figs.~\ref{fig:Fig9}(a),(b).
}
\label{fig:Fig2-hist}
\end{figure*}

 In fluctuating environments, the endlessly switching carrying capacity drives the population size, with $N$ varying according to a birth-death logistic process $N\stackrel{T^{\pm}}{\longrightarrow} N\pm 1$, with transition rates $T^{+}=T_S^++T_F^+=N$ and $T^{-}=T_S^-+T_F^-=N^2/K$. This  process has an absorbing boundary at $N=0$ leading to the eventual collapse of the population after a time that grows dramatically with the system size~\cite{Wienand2017,Wienand2018,Taitelbaum2020,assaf2017}. Here, have  $K\geq 25$ 
 which is sufficiently large to ensure that the population will not go extinct in our simulations. Here, we first focus the (marginal)
 PSD, denoted by $p_{\nu_{2n+1}}(N)$, with $\delta$ and $\epsilon$ treated as parameters; see Appendices~\ref{appendix:ME}. In Fig.~\ref{fig:Fig2-hist}, this quantity is compared with the  (marginal) PSD of the effective binary switching model, denoted by $p_{\widetilde{\nu}}(N)=p_{(\nu_{2n+1}/n)(1+\epsilon)/(2+\epsilon)}(N)$ 
 ~\cite{Wienand2017,Wienand2018,Taitelbaum2020,assaf2017}; see Eq.~\eqref{eq:nutilde} and Appendix~\ref{appendix:background}. 
Since $N$ relaxes on a timescale $t\sim 1$; see Eq.~\eqref{eq:MF}, we distinguish three main regimes~\cite{Wienand2017,Wienand2018,west2020,Taitelbaum2020}:\\
 (i) When  \(\nu_{2n+1}\ll 1\), environmental switching is much slower than the ecological timescale (``slow switching''). \(N\) is thus approximately constant and close to one value of $\{K_i\}_{i=-n}^n$
 randomly drawn from $\pi_i$ at time $t=0$, i.e. $N\approx K_i$ with probability $\pi_i$. The PSD is therefore multimodal with 
 $p_{\nu_{2n+1}}(N)$ characterised by $2n+1$ peaks about each $K_i$,
 their intensity being set by $\pi_i$;  
 see Fig.~\ref{fig:Fig2-hist}(a),(e),(f). 
 \\
 (ii) When \(\nu_{2n+1}\gg 1\), the environment varies on a much faster timescale than the ecological dynamics (``fast switching''). In this regime, \(N\) is not able to track \(K(t)\) and environmental fluctuations self-average, with the population experiencing the effective carrying capacity $\mathcal{K}_{2n+1}$ whose inverse is obtained by averaging $1/K$ over ${\bm \pi}$
 ~\cite{Wienand2017,Wienand2018,west2020,Taitelbaum2020,Shibasaki2021,Taitelbaum2023,Hernandez2023,Asker2023,Hernandez2024,Asker2025,Hernandez2026} (see also Appendix~\ref{appendix:background}), yielding
 \begin{equation}
  \label{eq:curlyK}
  \mathcal{K}_{2n+1}=\left(\sum_{i=-n}^{n} \frac{\pi_i}{K_i}\right)^{-1}, 
 \end{equation}
where  $K_i$ and $\pi_i$  are given by Eqs.~\eqref{eq:Ki} and \eqref{eq:pi}.
The binary counterpart of  $\mathcal{K}_{2n+1}$, denoted by  $\mathcal{K}_{2}$, is given by Eq.~\eqref{eq:curlyK2}. 
In this fast-switching regime, $p_{\nu_{2n+1}}$ is unimodal and approximately centred at $\mathcal{K}_{2n+1}$; see Fig.~\ref{fig:Fig2-hist}(c),(g). 
\\
(iii) When \(\nu_{2n+1}\sim 1\), the time scales of environmental  and ecological dynamics are similar (``intermediate switching'') and the PSD has a complex shape; see Fig.~\ref{fig:Fig2-hist}(b),(c). 
When $\nu_{2n+1} \lesssim 1$, $N$  tracks the time-switching $K(t)$; see Fig.~\ref{fig:Fig1-cartoon}(b), yielding  ``population bottlenecks''  
when  transitions $K_{i+1}\to K_i$, with $K_{i+1}\gg 1$, occur~\cite{Hernandez2023,Asker2025,Hernandez2026}. Bottlenecks are of great relevance for bacterial dynamics as they lead to
new colonies consisting of a small number of individuals prone to fluctuations~\cite{Wahl02,patwaAdaptationRatesLytic2010,Brockhurst2007b,Hernandez2023,Asker2025,Hernandez2026}. 
When $\nu_{2n+1}\gtrsim 1$,  the peaks of $p_{\nu_{2n+1}}$ gradually merge; see Fig.~\ref{fig:Fig2-hist}(c), and as $\nu_{2n+1}$ increases the PSD morphs from being multimodal to unimodal. 

The blue histograms of Fig.~\ref{fig:Fig2-hist} illustrate how  environmental and demographic noise jointly
affect the PSD in multi-state feast--famine cycles. Clearly, due to random fluctuations, the population size can be above $K^+$ and below $K^-$, and thus 
$p_{\nu_{2n+1}}$ is not bounded to $[K^-,K^+]$.
Under slow switching, $\nu_{2n+1}\ll 1$, the intensity of the  $2n+1$ peaks of $p_{\nu_{2n+1}}$, set by $\pi_i$, is higher for  low  $K_i$ when $\delta<0$ (Fig.\ref{fig:Fig2-hist}(f)) and for high   $K_i$ when $\delta>0$. %
The PSD is right-tailed and the peaks around low $K_i$ are always sharper and narrower than those about higher $K_i$; see Fig.\ref{fig:Fig2-hist}~(a),(b),(c),(e),(f). This stems from the   fast decay and slower growth of $N$ that is typical of  logistic dynamics~\cite{Wienand2017}.
In the intermediate switching regime, $\nu_{2n+1}\sim 1$,
the PSD peaks gradually merge, while  $p_{\nu_{2n+1}}$ remains right-tailed and has an approximately flat profile between $N\approx K^{\pm}$; see Fig.\ref{fig:Fig2-hist}(b),(c). In the fast switching regime, $\nu_{2n+1}\gg 1$, when $\delta\leq 0$, $p_{\nu_{2n+1}}$ is a right-tailed distribution  sharply centred about ${\cal K}_{2n+1}$ given by Eq.~\eqref{eq:curlyK}; see Fig.\ref{fig:Fig2-hist}(d),(g).
When $\delta>0$ and $n>1$, the  mild  environmental states are 
likely to be populated, resulting in a broader PSD and the 
convergence towards $N\approx{\cal K}_{2n+1}$ requires a very high switching rate; see Fig.\ref{fig:Fig2-hist}(h).
Although $\epsilon$ modifies the environmental switching dynamics (Appendix~\ref{appendix:EV}), it leaves the stationary distribution ${\bm \pi}$ unchanged; see Eq.~\eqref{eq:pi}. Stochastic simulations show no noticeable dependence of the PSD on $\epsilon$ (for $-0.8\leq \epsilon\leq 5$), as illustrated by the comparison of Figs.~\ref{fig:Fig2-hist}(g),(h) and ~\ref{fig:Fig9}(a),(b), suggesting that $p_{\nu_{2n+1}}$ and related quantities are chiefly controlled by the stationary environmental statistics.

The comparison of the multi-state and two-state switching models PSD in Fig.~\ref{fig:Fig2-hist}
shows that the most striking differences arise in the slow switching regime, where, due to the intermediate environmental states, 
$p_{\nu_{2n+1}}$ has $2n-1$  peaks at $N\approx K_{1-n},\dots, K_{n-1}$ that are absent from the binary PSD; see Fig.~\ref{fig:Fig2-hist}(a),(e),(f). Moreover, when $\nu_{2n+1}\sim 1$, the probability that $N$ takes a value between $K^-$ and $K^+$ is higher in multi-state than binary environments; see Fig.~\ref{fig:Fig2-hist}(b),(c). 
Under fast switching,  the multi-state and binary-state PSD are both unimodal, but are
respectively centred at ${\cal K}_{2n+1}$
and ${\cal K}_{2}$,  given by Eqs.~\eqref{eq:curlyK} and \eqref{eq:curlyK2}, generally with ${\cal K}_{2n+1}>{\cal K}_{2}$ when $\delta>0$  (see Figs.~\ref{fig:Fig2-hist}(h) and \ref{fig:Fig9}(b); and see below).

Many features of the quasi-stationary PSD are captured by the piecewise deterministic Markov process~\cite{davisPiecewiseDeterministicMarkovProcesses1984,hufton2016,Wienand2017,Wienand2018} for the population size  ($N$-PDMP). This process is obtained by letting $N$ evolve deterministically between two environmental switches, 
according to the logistic equation with carrying capacity $K=K_i$ in the environmental state $\xi(t)=i\in\{-n,\dots,n\}$
until an environmental switch occurs~\cite{Wienand2017,Wienand2018,west2020,Taitelbaum2020,Taitelbaum2023,Hernandez2023,Asker2023,Hernandez2024,Asker2025,Hernandez2026}; see Appendix~\ref{appendix:background}. Hence, 
by generalizing the binary $N$-PDMP process of Eq.~\eqref{eq:PDMP-2state},
the $(2n+1)$-state $N$-PDMP is defined by
\begin{equation}
\label{eq:PDMP-nstate}
\dot{N}=
\begin{cases}
N\left(1-\frac{N}{K^-}\right) & \text{if } \xi=-n, \\
\qquad \quad \vdots & \qquad\vdots \\
N\left(1-\frac{N}{K_0}\right) & \text{if } \xi=0, \\
\qquad \quad \vdots & \qquad\vdots \\
N\left(1-\frac{N}{K^+}\right) & \text{if } \xi=n,
\end{cases}
\end{equation}
with $\dot{N}=N\left(1-\frac{N}{K_i}\right)$ when $\xi(t)=i\in\{-n,\dots,n\}.$ 
The marginal stationary probability density
of the binary $N$-PDMP \eqref{eq:PDMP-2state} can be computed explicitly  and is given by Eq.~\eqref{eq:NPDMP_marg}.
For ($2n+1$)-state switching models, 
the  $N$-PDMP marginal stationary probability density, denoted by $p^{\text {PDMP}}_{\nu_{2n+1}}$, cannot be obtained analytically but can 
 be efficiently computed from simulations of Eq.~\eqref{eq:PDMP-nstate}; see Appendix~\ref{appendix:simulations} and  Fig.~\ref{fig:Fig2-hist}.

\begin{figure*}
\centering
\setlength{\unitlength}{1cm}
\begin{overpic}[width=0.24\textwidth]{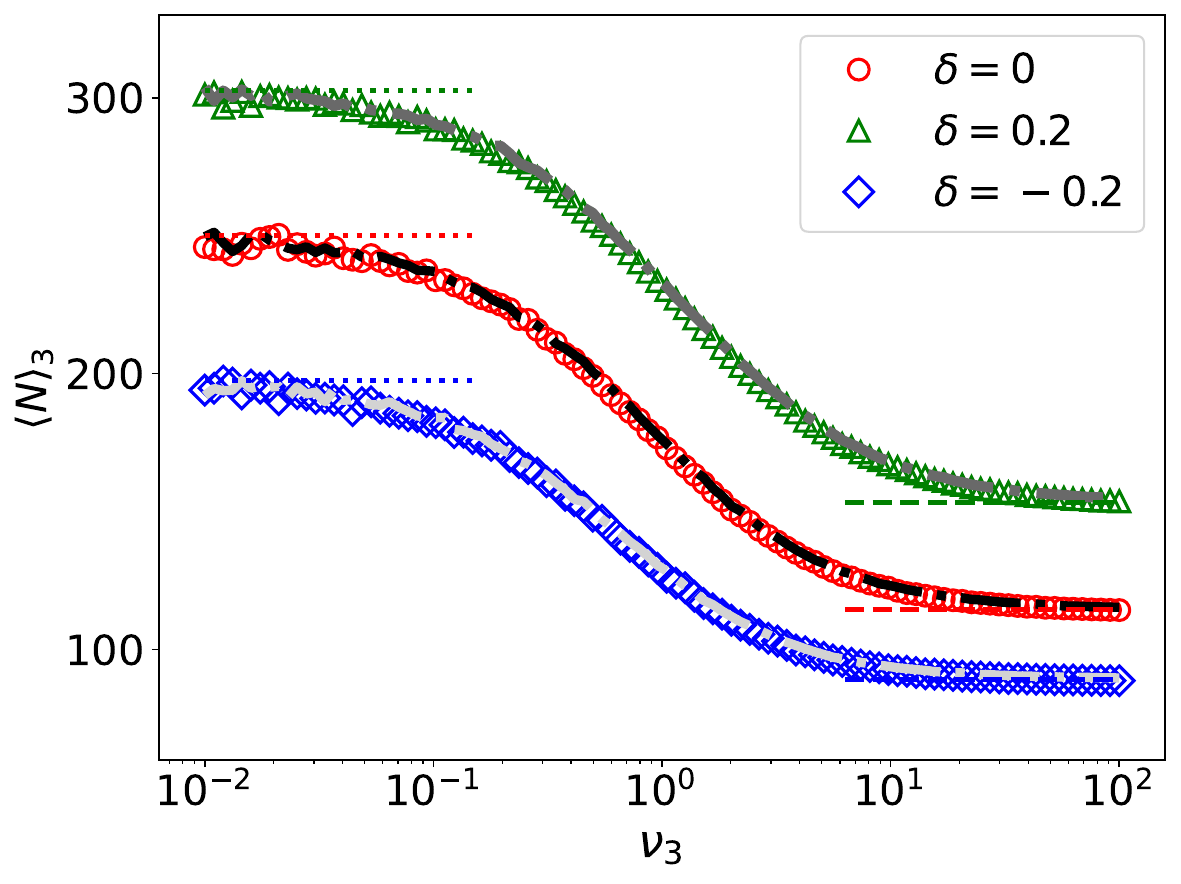}
    \put(2,77){(a)}
\end{overpic}
\hfill
\begin{overpic}[width=0.24\textwidth]{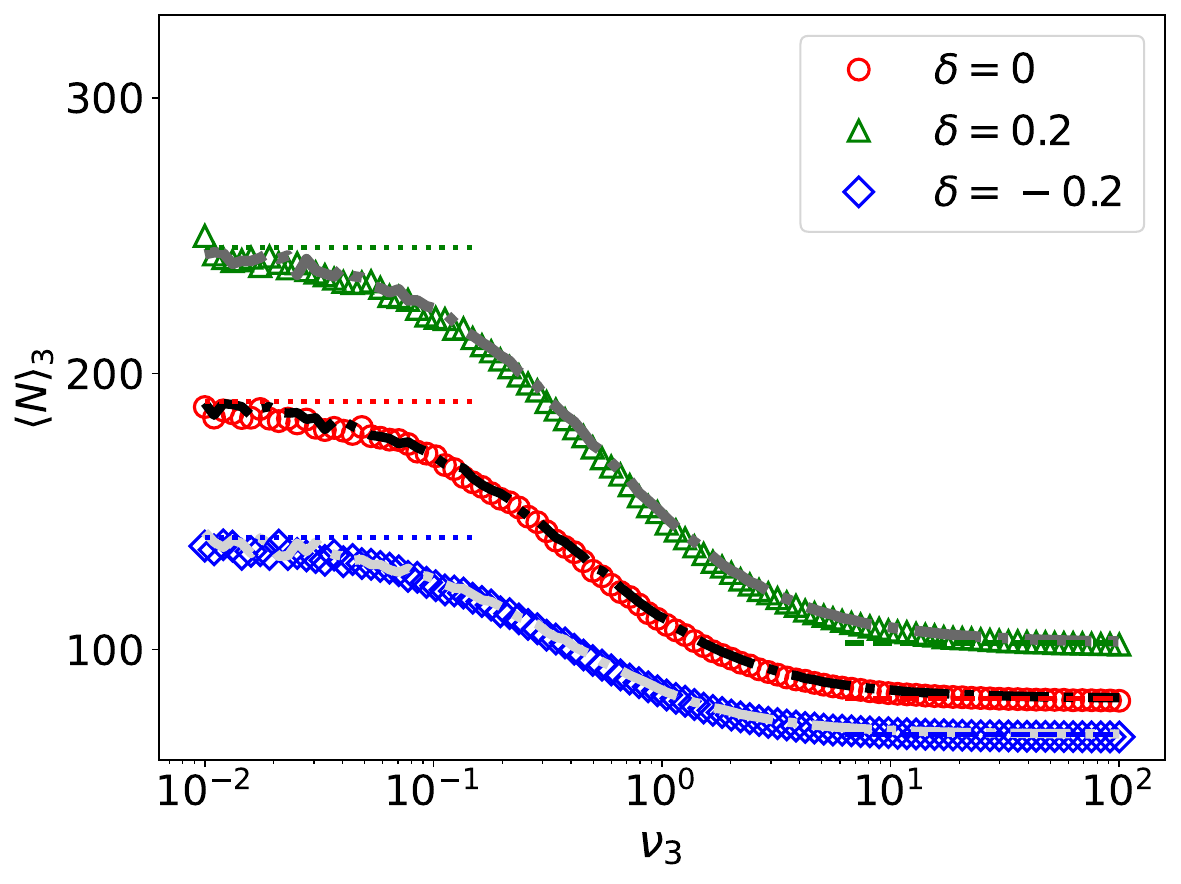}
    \put(2,77){(b)}
\end{overpic}
%
%
\begin{overpic}[width=0.24\textwidth]{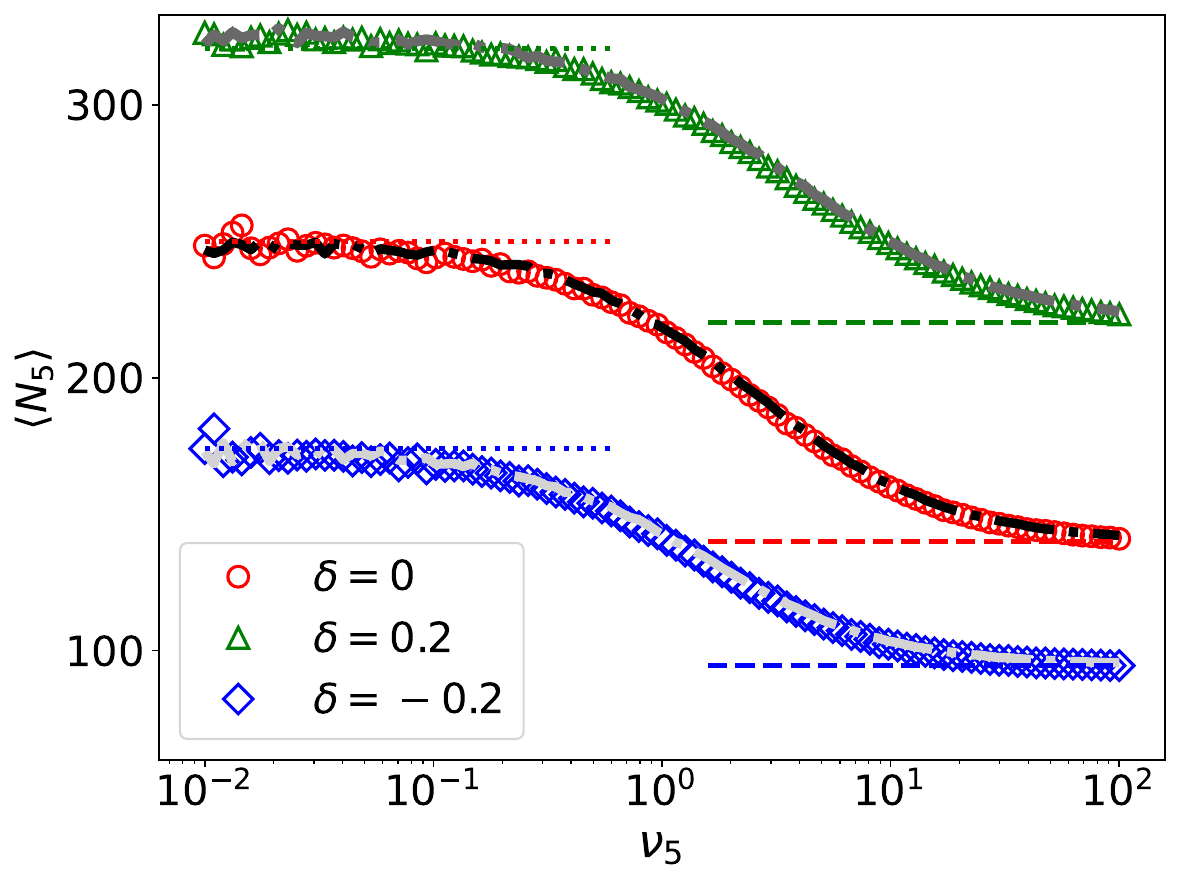}
    \put(2,77){(c)}
\end{overpic}
\hfill
\begin{overpic}[width=0.24\textwidth]{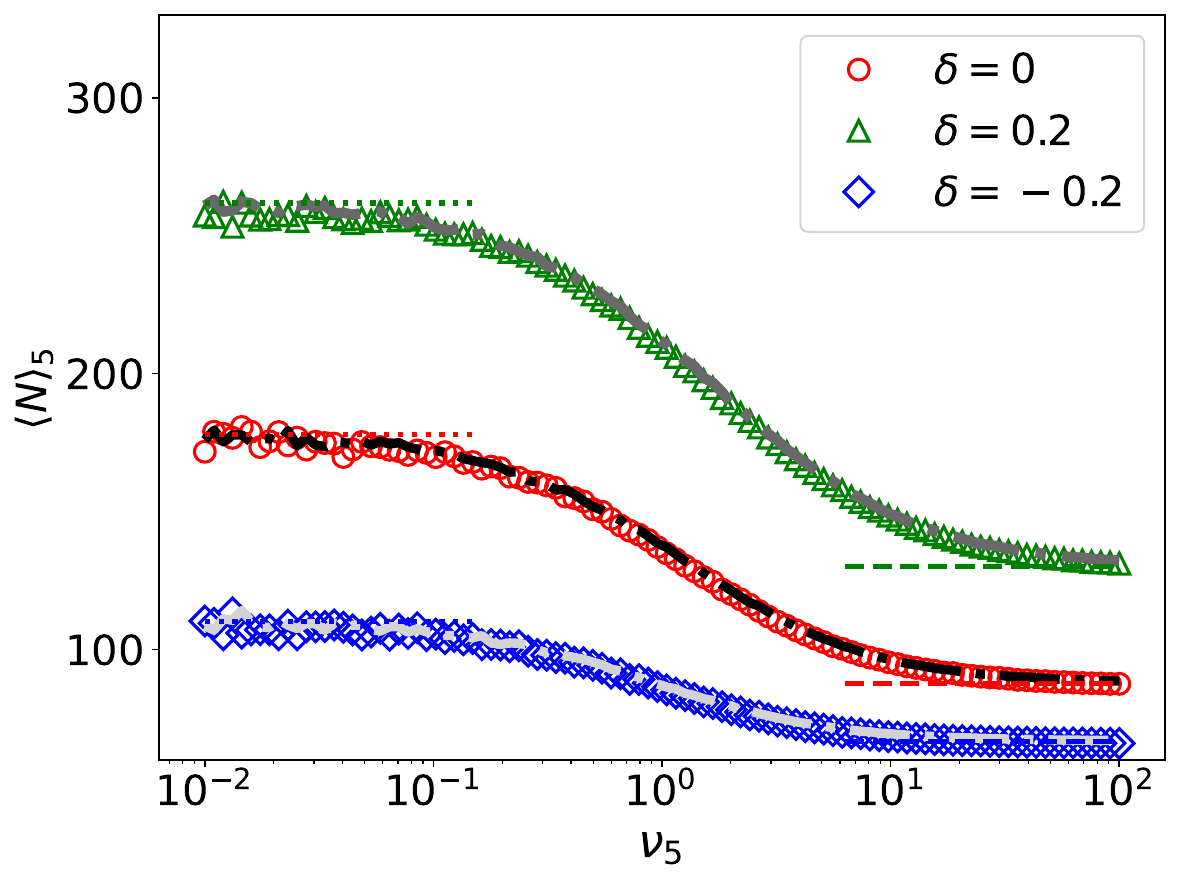}
    \put(2,77){(d)}
\end{overpic}
\caption{Average population size 
$\langle N\rangle_{2n+1}$ (markers) and $\langle N\rangle^{\text{PDMP}}_{2n+1}$
(dash-dotted curves) versus $\nu_{2n+1}$ for different values of $n, K_0$ and $\delta$; see text and \eqref{eq:avN}: 
(a)  $(n,K_0)=(1,250)$; (b)  $(n,K_0)=(1,70)$; (c)  $(n,K_0)=(2,250)$; (d)  $(n,K_0)=(2,70)$. In all panels, we have $\delta=0$ (red circles, black curves), $\delta=0.2$ (green triangles, dark gray curves), $\delta=-0.2$ (blue diamonds, grey curves). The dotted and dashed horizontal lines are eyeguides
respectively showing $\langle K\rangle_{2n+1}=\sum_{i=-n}^n K_i \pi_i$ and ${\cal K}_{2n+1}$ (given by Eq.~\eqref{eq:curlyK}) for each combination of $n,K_0,\delta$. 
Other parameters are $(K^-,K^+,\epsilon)=(50,450,0)$ in all panels.
In all the examples, 
$\langle N\rangle_{2n+1}$ and $\langle N\rangle^{\text{PDMP}}_{2n+1}$ have been computed after $t> 5000$, as outlined in Appendix~\ref{appendix:simulations}.
}
\label{fig:Fig3-meanN}
\end{figure*}

 Since the $N$-PDMP  ignores demographic fluctuations, it accounts  for  randomness only stemming from the random switching of $K(t)$  and the support of $p^{\text {PDMP}}_{\nu_{2n+1}}$ is $N\in [K^-,K^+]$. Figure ~\ref{fig:Fig2-hist}
illustrates that $p^{\text {PDMP}}_{\nu_{2n+1}}$ correctly reproduces all the peaks of $p_{\nu_{2n+1}}$; see Fig.~\ref{fig:Fig2-hist}(a),(b), while it cannot capture the width of the distribution under slow switching.
Under intermediate  switching, systematic deviations between $p^{\text {PDMP}}_{\nu_{2n+1}}$ and the PSD mainly arise 
  around $N\approx K^{\pm}$ and are caused by demographic fluctuations;  see Fig.~\ref{fig:Fig2-hist}(b). 
When $\nu_{2n+1}\gtrsim 1$, we find that  $p_{\nu_{2n+1}}\approx p^{\text {PDMP}}_{\nu_{2n+1}}$ indicating that most of the fluctuations arise from environmental switching and are captured by the $N$-PDMP;  see Figs.~\ref{fig:Fig2-hist}(c), (d),(g),(h) and Fig.~\ref{fig:Fig9}. 
 The PDMP stationary density also provides an accurate 
approximation of the  average long-term population size $\langle N\rangle_{2n+1}\equiv \sum_{N=0}^{\infty} Np_{\nu_{2n+1}}(N)$, that 
can be computed numerically as 
\begin{equation}
 \label{eq:avN}
 \langle N\rangle_{2n+1}\approx \langle N\rangle^{\text {PDMP}}_{2n+1}\equiv \int_{K^-}^{K^+} N~ p^{\text {PDMP}}_{\nu_{2n+1}}(N)~dN.
\end{equation}
In Figure~\ref{fig:Fig3-meanN}, we find an excellent agreement between
$\langle N\rangle_{2n+1}$ computed from 
stochastic simulations data and 
the  numerical computation of $\langle N\rangle^{\text {PDMP}}_{2n+1}$ (outlined in Appendix~\ref{appendix:simulations}). As in the binary switching case~\cite{Wienand2017,Wienand2018,west2020,Taitelbaum2020,Asker2023,Asker2025}, the $N$-PDMP approximation captures that  $\langle N\rangle_{2n+1}\approx \langle N\rangle^{\text {PDMP}}_{2n+1}$
is a decreasing function of $\nu_{2n+1}$ (at fixed  $\delta,\epsilon$), with   $\langle N\rangle_{2n+1}\approx \langle K\rangle_{2n+1}$ when $\nu_{2n+1}\ll 1$ and $\langle N\rangle_{2n+1}\approx \mathcal{K}_{2n+1}$ when $\nu_{2n+1}\gg 1$. At fixed $\nu_{2n+1}$, $\langle N\rangle_{2n+1}$  increases with $\delta$. In Fig.~\ref{fig:Fig3-meanN}, we notice that  the convergence $\langle N\rangle_{2n+1}\approx \mathcal{K}_{2n+1}$ is slower for $n=2$ than for $n=1$
when $(\delta,\epsilon, K_0)$ are kept fixed; see Fig.~\ref{fig:Fig3-meanN}(a),(c) and Fig.~\ref{fig:Fig3-meanN}(b),(d). 
This stems from the number of the number of  intermediate states increasing with $n$ and the effective rate $\widetilde{\nu}\propto \nu_{2n+1}/n$; see Eq.~\eqref{eq:nutilde}.

\subsection{Fixation in fluctuating environments}
\label{Sec:Fixation}
In this section, we study how environmental statistics, encoded in $\nu^{\pm}_{2n+1}, n, $ and $\{K_i\}_{i=-n}^{n}$, influence the $S$ strain fixation probability, denoted by $\phi_{2n+1}=\mathbb{P}\left(N_S(t)=N(t)>0|~t<\infty\right)$. This is the probability that $S$ takes over the entire population in a finite time, avoiding the risk of overall extinction. 
In Appendix \ref{appendix:MFT}, we investigate the effect of the environmental statistics on the unconditional mean fixation time (uMFT) for either of the two strains  to take over the population.

\subsubsection{Fixation under slow and fast environmental switching}
\label{Sec:slow-fast}

When $\nu_{2n+1}\ll s$ (slow switching), environmental switching is much slower than selection dynamics. It is therefore is likely that no switches occur on the evolutionary timescale  $t\sim 1/s$ (see Eq.~\eqref{eq:MF} and Appendix \ref{appendix:MFT}).
In this regime, the population is thus subject to the carrying capacity $K_i$ with a probability $\pi_i$, and an approximate expression of $\phi_{2n+1}$ is obtained from 
the Moran expression \eqref{eq:phi_M} by writing 
 $\phi_{2n+1}(\nu_{2n+1}\ll s)\approx \phi_{2n+1}^0$, where
\begin{equation}
 \label{eq:phi0}
 \phi_{2n+1}^0\equiv \sum_{i=-n}^{n} \phi_M(K_i)~\pi_i,
\end{equation}
$K_i$ and $\pi_i$ being given by Eqs.~\eqref{eq:Ki} and \eqref{eq:pi}. This corresponds to a ``quenched approximation''~\cite{Hernandez2024,Mobilia2023,Meyer2024}, assuming that fixation occurs while $N \approx K_i$ with a probability $\pi_i$. 
A similar reasoning for the binary switching model yields $\phi_{2}(\widetilde{\nu}\ll s)\approx \phi_{2}^0\equiv \frac{1-\delta}{2}\phi_M(K^-) + \frac{1+\delta}{2}\phi_M(K^+)$~\cite{Wienand2017,Wienand2018,Taitelbaum2020,Taitelbaum2023,Hernandez2023,Asker2023,Hernandez2024,Asker2025,Hernandez2026}; see  Appendix
~\ref{appendix:background}.  

When $\nu_{2n+1}\gg s$ (fast switching), there are  many environmental switches on the timescale $t\sim 1/s$ yielding the self-average of
environmental fluctuations. The carrying capacity
hence takes the  effective value ${\cal K}_{2n+1}$ given by Eq.~\eqref{eq:curlyK} and $N\approx {\cal K}_{2n+1}$. In the  fast switching regime, 
we can thus use an  ``annealed approximation''~\cite{Hernandez2024,Mobilia2023,Meyer2024} for
the $S$ fixation probability  by  writing
in  terms of the Moran result \eqref{eq:phi_M} 
$\phi_{2n+1}(\nu_{2n+1}\gg s)\approx \phi_{2n+1}^{\infty}$, with
\begin{equation} 
  \label{eq:phiinf}
\begin{aligned}
 \phi_{2n+1}^{\infty}\equiv \phi_M({\cal K}_{2n+1}). 
\end{aligned}
\end{equation}
Similarly, in the binary environment we have 
$\phi_{2}(\widetilde{\nu}\gg s)\approx \phi_{2}^{\infty}\equiv \phi_M({\cal K}_2)$~\cite{Wienand2017,Wienand2018,Taitelbaum2020,Taitelbaum2023,Hernandez2023,Asker2023,Hernandez2024,Asker2025,Hernandez2026}, where  ${\cal K}_2$ is given by Eq.~\eqref{eq:curlyK2}.

\begin{figure*}
\centering
    \includegraphics[width=\textwidth]{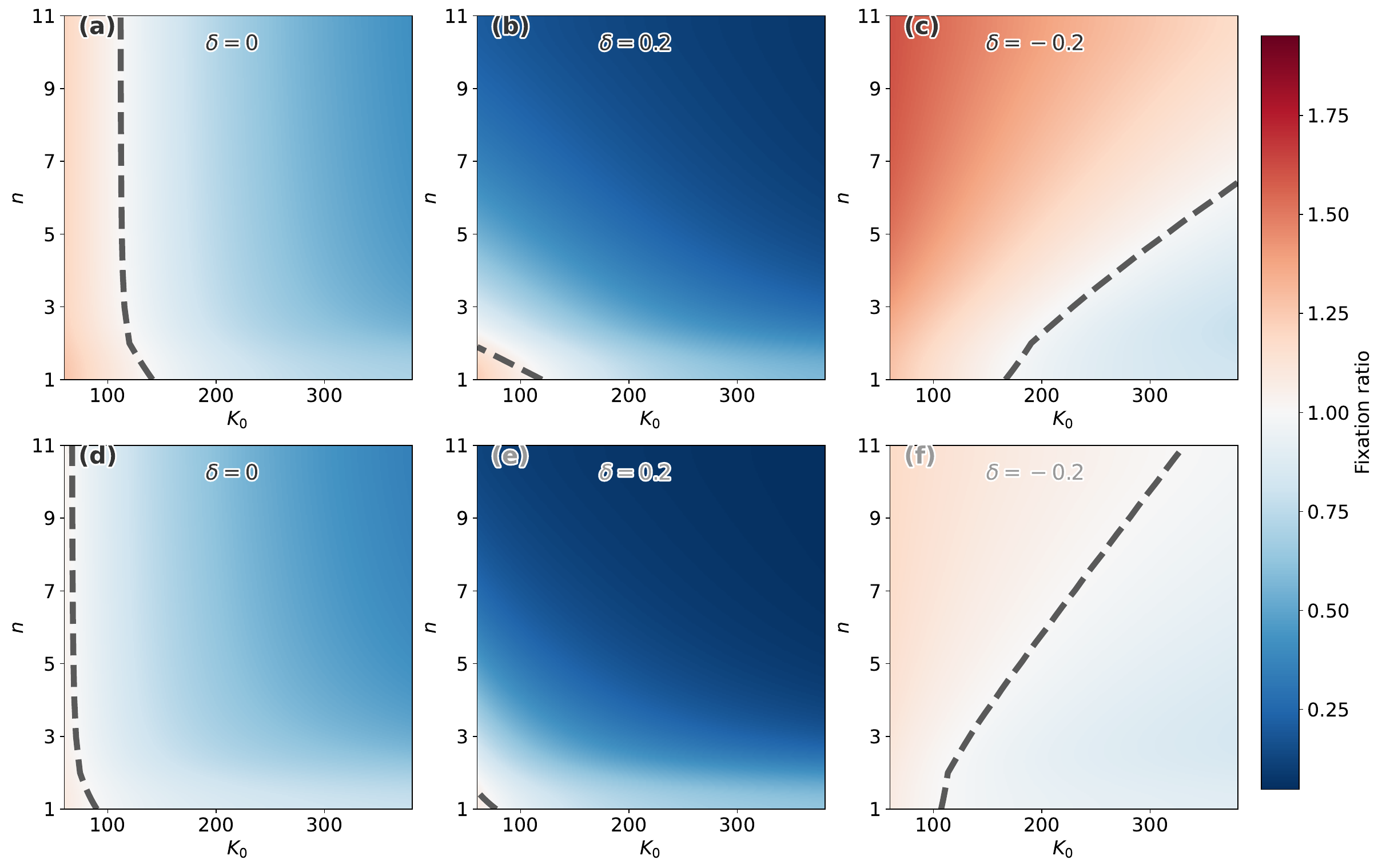}
    \caption{Comparison of the $S$  fixation probability in the regime of slow and fast switching in  multi-state ($n\geq 1$) and  binary switching models: 
    Heatmaps of the ratios $\phi_{2n+1}^{0}/\phi_{2}^{0}$ (slow switching) in (a)-(c) and $\phi_{2n+1}^{\infty}/\phi_{2}^{\infty}$ (fast switching) in (d)-(f) as colour-coded in the rightmost vertical bar; see text. Parameters are $(K^-,K^+,s,x_0,\epsilon)=(50,450,0.02,0.5,0)$ and $\delta=0$ in (a,d),  $\delta=0.2$ in (b,e), and $\delta=-0.2$ in (c,f).
    For reference, ${\cal K}_2=90$ for $\delta=0$ , ${\cal K}_2\approx 107.14$ for $\delta=0.2$, and ${\cal K}_2\approx 77.59$  for $\delta=-0.2$. In all panels, $\phi_{2n+1}^{0,\infty}/\phi_{2}^{0}<1$ blueish areas and $\phi_{2n+1}^{0,\infty}/\phi_{2}^{0}>1$ in yellow/orange regions.
     The black dashed lines (surrounded by a whitish cloud) indicate the contours  along which the ratios equal to one.}
    \label{fig:Fig4-heatmaps}
\end{figure*}

The heatmaps of Fig.~\ref{fig:Fig4-heatmaps}(a)-(c) 
and Fig.~\ref{fig:Fig4-heatmaps}(d)-(f)
respectively show the ratios 
$\phi_{2n+1}^{0}/\phi_{2}^{0}$ 
and $\phi_{2n+1}^{\infty}/\phi_{2}^{\infty}=\phi_M({\cal K}_{2n+1})/\phi_M({\cal K}_2)$ 
for different values of $\delta$ as a functions of $n$ and $K_0$ (for $s,x_0$ and $K^{\pm}$ kept fixed). The regions where $\phi_{2n+1}^{0,\infty}/\phi_{2}^{0,\infty}>1$ (orange/pink) are 
separated from those where $\phi_{2n+1}^{0,\infty}/\phi_{2}^{0,\infty}<1$ (blueish) by dashed lines along which  $\phi_{2n+1}^{0,\infty}=\phi_{2}^{0,\infty}$. Due to the strong dependence of $\pi_i$ on $\delta$; see Eq.~\eqref{eq:pi},
the shape and area of these regions vary greatly with $\delta$.
When $\delta\leq 0$ and  $K_0$ is sufficiently low, the $S$ fixation probability under slow and fast switching can be higher in multi-state than in binary environments for all $n\geq 1$; see Fig.~\ref{fig:Fig4-heatmaps}(a),(c),(d),(f). 
When $\delta<0$, environmental bias is towards states having a low carrying capacity, resulting in predominantly orange/pink-dominated
 heatmaps where $\phi_{2n+1}^{0,\infty}>\phi_{2}^{0,\infty}$
 with  $\phi_{2n+1}^{0}\lesssim \phi_{2}^{0}$ for sufficiently high values of $K_0$ (pale blue below  the dashed line); see Fig~\ref{fig:Fig4-heatmaps}(c,f).
  When $\delta>0$, environmental switching is biased towards  states with a high carrying capacity, yielding the predominantly blue-dominated heatmap of Fig.~\ref{fig:Fig4-heatmaps}(b,d), corresponding to $\phi_{2n+1}^{0,\infty}<\phi_{2}^{0,\infty}$ for most o$n, K_0$ pairs. (For example, $\phi_{2n+1}^{0}>\phi_{2}^{0}$  only for $K_0\lesssim 100$ when $n\geq 2$). The results of Fig.~\ref{fig:Fig4-heatmaps} are in agreement with the simulation data of Figs.~\ref{fig:Fig5_n1} and \ref{fig:Fig6_n2}, and illustrate
  that the distribution of carrying capacities $\{K_i\}_{i=-n}^{n}$ (here set by $n$ and $K_0, K^{\pm}$)
  can lead to complex fixation scenarios, significantly richer than those arising under binary switching.

In binary switching environments, the Moran approximations $\phi_{2}^{0,\infty}$ were found to work better under weak selection, $s\ll 1$, because fixation then occurs on a slower timescale~\cite{Wienand2018}. The evolutionary dynamics of $x(t)$ therefore samples many demographic fluctuations of $N(t)$, so that deviations from an effectively constant population size approximately average out. For stronger selection, $s={\cal O}(1)$, fixation events occur more rapidly and the coupling between fluctuations of $N(t)$ and $x(t)$ leads to larger deviations from the Moran predictions. As a result, the slow/fast-switching approximations $\phi_{2}\approx\phi_{2}^{0}$ and $\phi_{2}\approx\phi_{2}^{\infty}$
are more accurate when $s\ll1$ than for $s={\cal O}(1)$~\cite{Wienand2018}. In contrast, for multi-state environments, switches occur through intermediate carrying capacities. This gradual variation reduces the influence the selection strength, yielding $\phi_{2n+1}$ to be well approximated by $\phi_{2n+1}^{0}$ and $\phi_{2n+1}^{\infty}$ for weak and moderate selection strengths in the suitable switching regimes; see  Figs.~\ref{fig:Fig5_n1}(d) and \ref{fig:Fig6_n2}(d).

\subsubsection{Fixation under intermediate environmental switching}
\label{Sec:intermediate}
Under intermediate environmental switching, where $\nu_{2n+1}\sim s$ ($|\delta|<1$),  evolutionary and environmental dynamics take place on similar time scales, and $N$ has a nontrivial PSD; see  \ref{fig:Fig2-hist}(b),(c). 
It is generally difficult to obtain analytical results for the fixation in the intermediate switching regime. Here, in addition to stochastic simulations, we devise an efficient $N$-PDMP approximation, generalizing the binary-state approach of Refs.~\cite{Wienand2017,Wienand2018,west2020} (see Appendix~\ref{appendix:background}). The method relies on a timescale separation, and a suitable rescaling of the switching rate. When $s\ll 1$, $N$ settles in its (quasi-)stationary PSD on a much shorter timescale than
$t\sim 1/s$ (evolutionary timescale); see Eq.~\eqref{eq:MF}. $N$ is thus 
 considered  at quasi-stationarity, with  the $N$-PDMP providing a suitable approximation of its dynamics. 
In Appendix~\ref{appendix:EV}, we show that
the average number of switches on the  timescale $1/s$ is $\nu_{2n+1}'/s$, where
\begin{align} \label{eq:nuprime}
\hspace{-4mm}
\nu_{2n+1}' &= \nu_{2n+1} \left(\frac{1-\rho}{1-\rho^{2n+1}}\right) \Bigg[ 1+\delta +\frac{2\rho}{1-\rho}(1-\rho^{n-1})(2+\epsilon) \nonumber\\ &\qquad +\rho^n \bigl[2+\epsilon(1+\delta)\bigr] +\rho^{2n}(1+\epsilon)(1-\delta) \Bigg]. \end{align}
 When $\delta=\epsilon=0$, we have 
$ \nu_{2n+1}'
 =4n\nu_{2n+1}/(2n+1)$. 
Following Refs.~\cite{Wienand2017,Wienand2018,west2020,Taitelbaum2020,Asker2025}, we
can use the $N$-PDMP density  to approximate the $S$ fixation probability when $\nu_{2n+1}\sim s$
and  $1/\langle N \rangle \ll s\ll 1$. For this we average the Moran fixation probability $\phi_M(N)$, obtained from Eq.~\eqref{eq:phi_M}, over $p^{\text{PDMP}}_{\nu_{2n+1}'/s}(N)$, where  the  switching rate has been rescaled, $\nu_{2n+1}\to \nu_{2n+1}'/s$ using Eq.~\eqref{eq:nuprime}, to properly account for the expected number of switches occurring on the fixation timescale~\cite{Wienand2017,Wienand2018,west2020,Taitelbaum2020,Taitelbaum2023,Asker2025}.:
\begin{equation}
  \label{eq:phiPDMP}
 \phi_{2n+1}^{\text{PDMP}}\equiv \int_{K^-}^{K^+} \phi_M(N)~p^{\text{PDMP}}_{\nu_{2n+1}'/s}(N)~dN.
\end{equation}
In Appendix \ref{appendix:MFT}, this approach is used for the uMFT. A similar approximation for $\phi_{2}$ and uMFT in binary environments has been studied in Refs.~\cite{Wienand2017,Wienand2018,Taitelbaum2020}; see Appendix~\ref{appendix:background}.
In practice,   $\phi_{2n+1}^{\text{PDMP}}$ is efficiently computed numerically as outlined in Appendix~\ref{appendix:simulations}. The comparison with extensive simulations of the individual-based models; see  Figs.~\ref{fig:Fig5_n1} and \ref{fig:Fig6_n2}, shows that Eq.~\eqref{eq:phiPDMP} generally gives a very good approximation of the $S$ fixation probability, with  $\phi_{2n+1}\approx \phi_{2n+1}^{\text{PDMP}}$ over a broad range of values of the switching rate, $\nu_{2n+1}\sim s$, as well as $\nu_{2n+1}\ll s$ and $\nu_{2n+1}\gg s$ (slow/fast switching regimes). The same  agreement  is found in binary-switching environments~\cite{Wienand2017,Wienand2018,west2020,Asker2025}.

Below, for the sake of concreteness, we  discuss in some detail the cases of  ternary and five-fold fluctuating environments ($n=1$ and $n=2$).

\subsubsection{Fixation  in the three-state switching model}
\label{Sec:n1}

\begin{figure*}
\centering
\setlength{\unitlength}{1cm}
\begin{overpic}[width=0.32\textwidth]{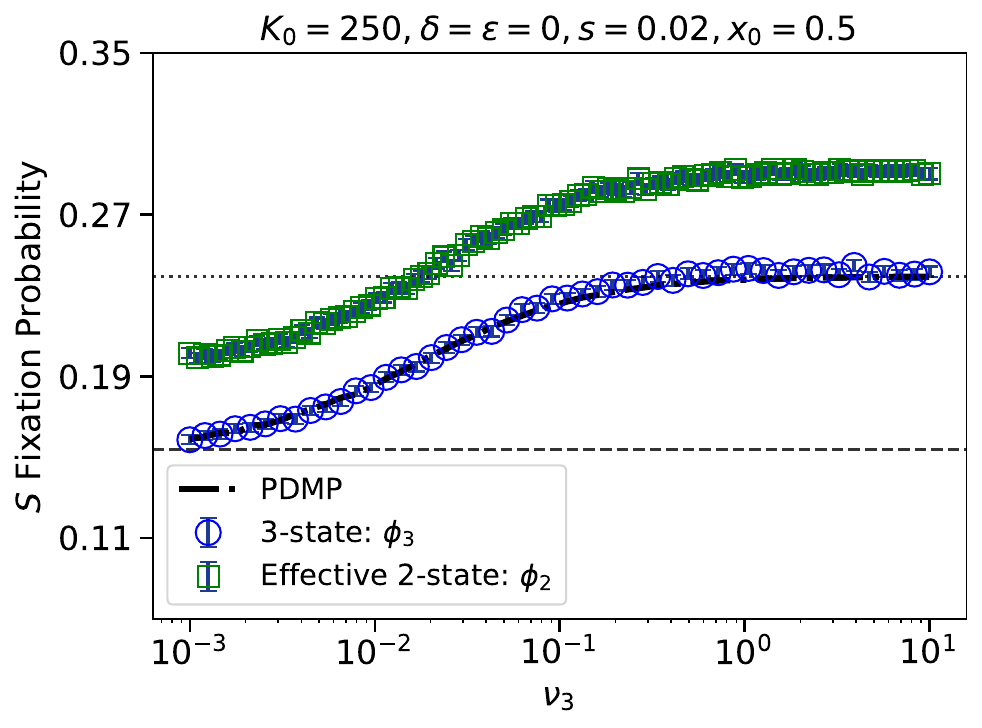
}
    \put(12,75){(a)}
\end{overpic}
\hfill
\begin{overpic}[width=0.32\textwidth]{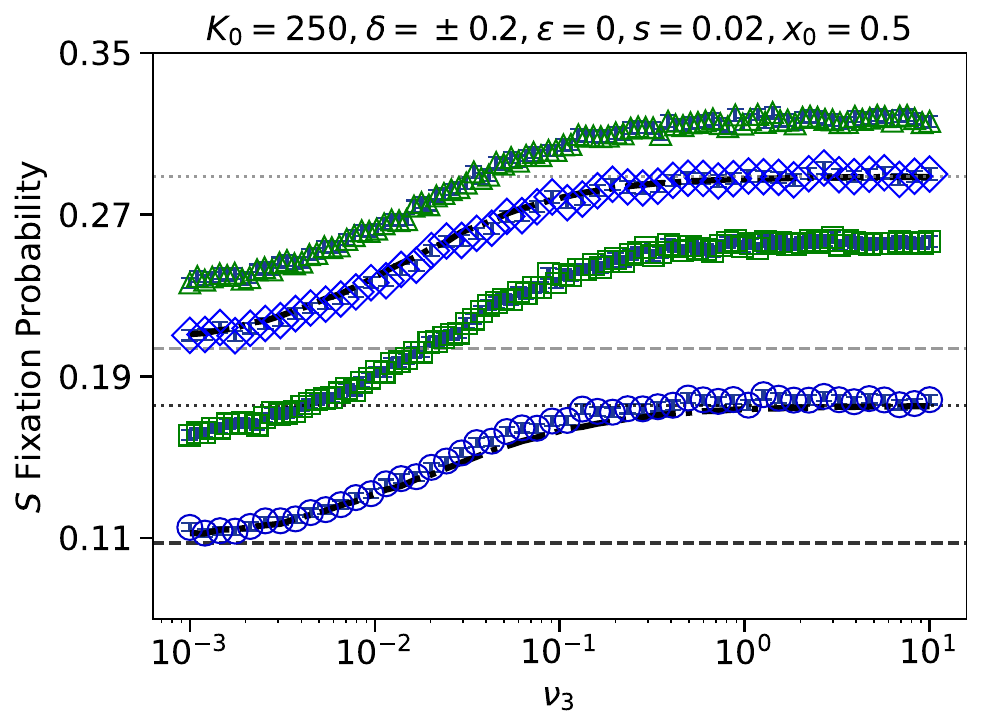
}
    \put(12,75){(b)}
\end{overpic}
\hfill
\begin{overpic}[width=0.32\textwidth]{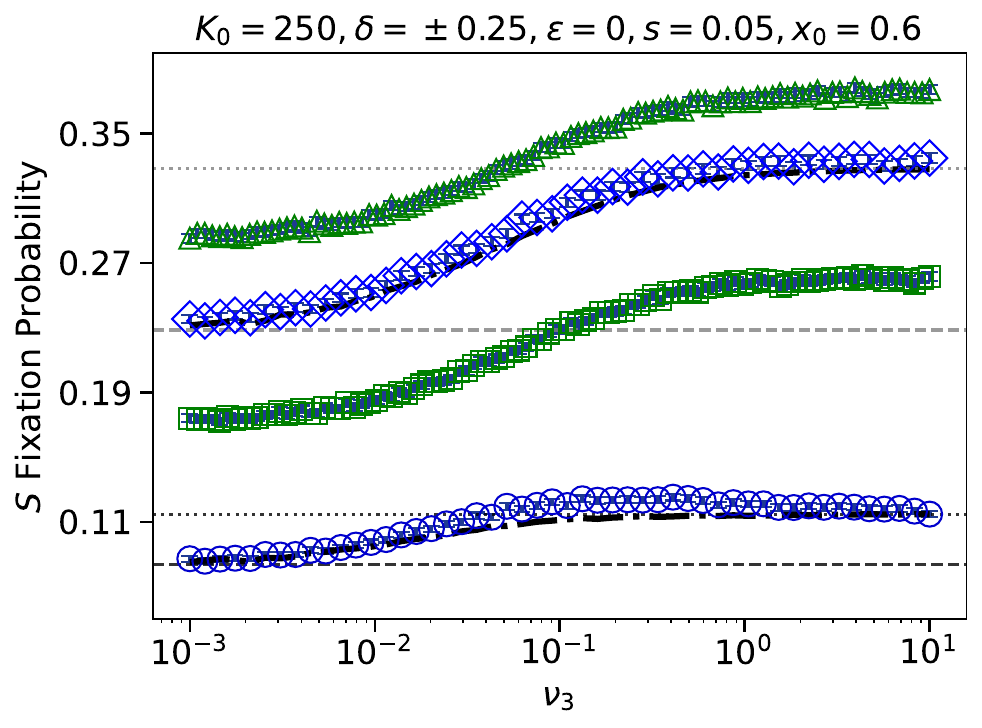
}
    \put(12,75){(c)}
\end{overpic}
\hfill
\vspace{1.5em}

\begin{overpic}[width=0.32\textwidth]{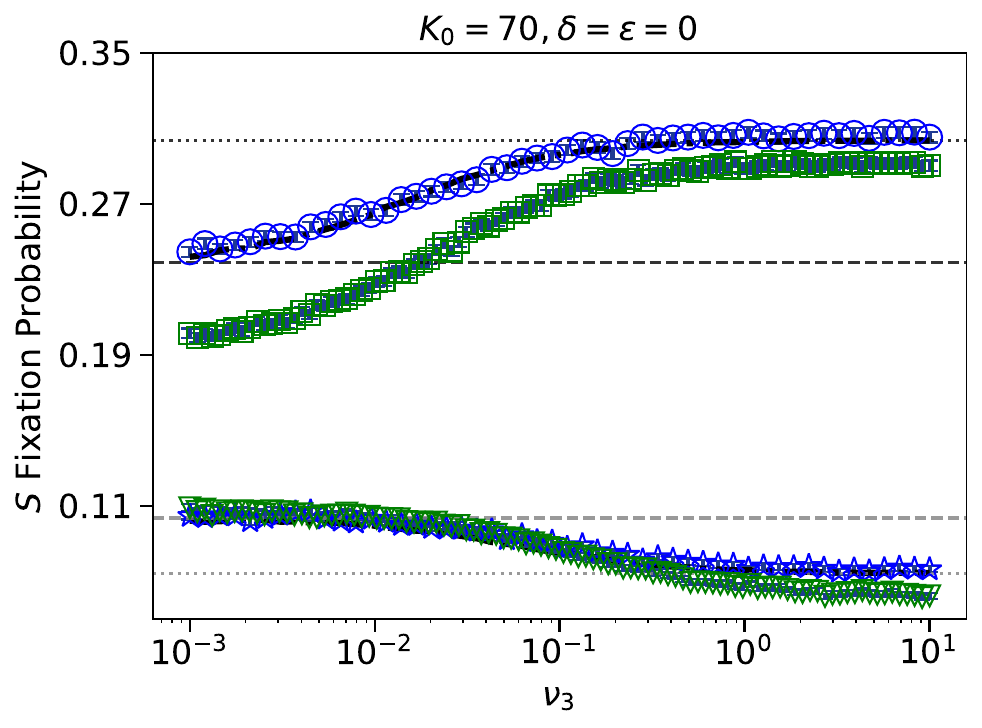
}
    \put(12,75){(d)}
\end{overpic}
\hfill
\begin{overpic}[width=0.32\textwidth]{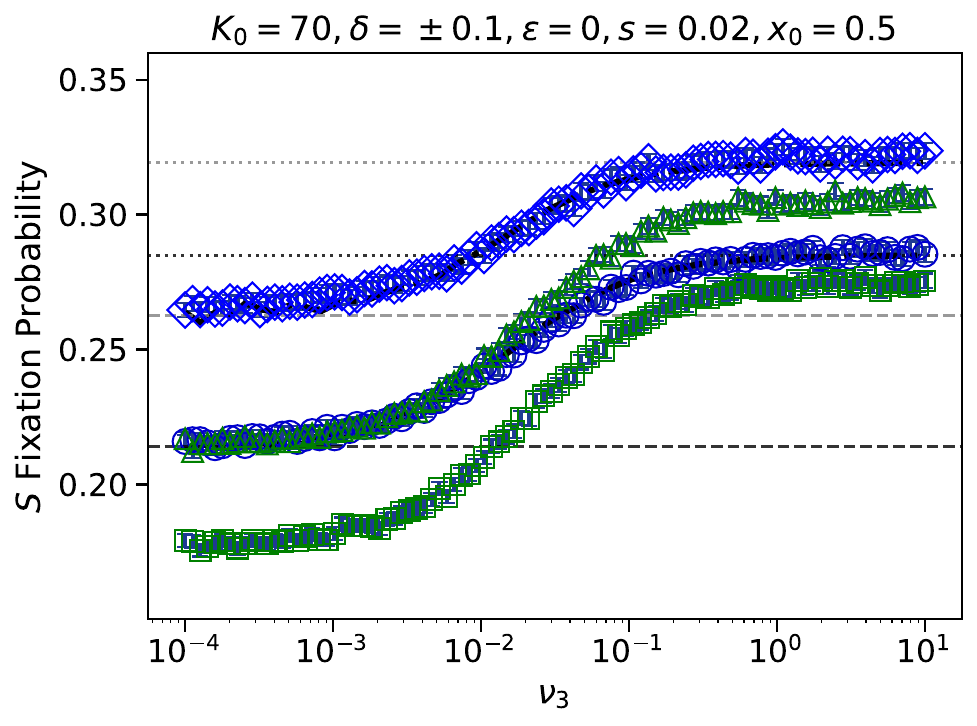
}
    \put(12,75){(e)}
\end{overpic}
\hfill
\begin{overpic}[width=0.32\textwidth]{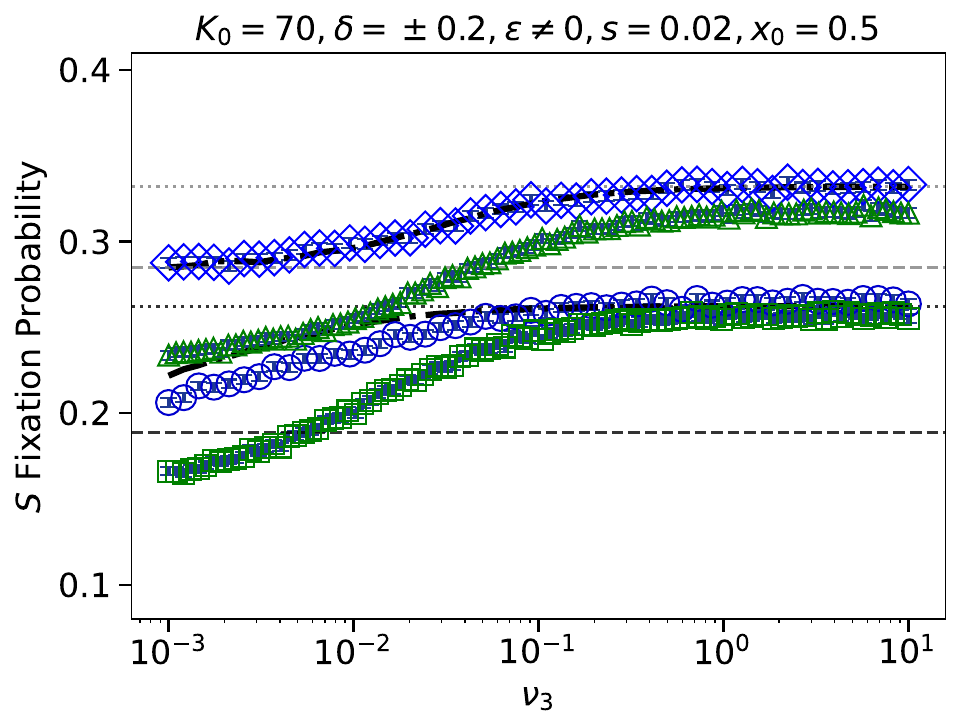
}
    \put(12,75){(f)}
\end{overpic}
\caption{$S$ fixation probability  in the three-state  model as a function of the switching rate $\nu_3$.  The centre state carrying capacity in  (a)-(c) is  $K_0=250$, while  $K_0=70$ in (d)-(f). Here, $n=1$, and  
$(K^-,K^+)=(50,450)$ in all panels except (c).
Symbols are from full stochastic simulations and black dashed-dotted curves (in all panels), often almost indistinguishable from markers, are from the $N$-PDMP-based approximation; see Eq.~\eqref{eq:phiPDMP}.  In blue, is shown $\phi_{3}$ vs. $\nu_3$  for different parameter sets:  $(\delta,\epsilon, s, x_0)=(0,0,0.02,0.5)$ (circles) in (a);
 $(\delta,\epsilon, s, x_0)=(0.2,0,0.02,0.5)$ (circles) and $(\delta,\epsilon, s, x_0)=(-0.2,0,0.02,0.5)$ (diamonds) in (b); $(\delta,\epsilon, s, x_0, K^-,K^+)=(0.25,0,0.05,0.6, 25, 475)$  (circles) and $(\delta,\epsilon, s, x_0, K^-,K^+)=(-0.25,0,0.05,0.6, 25, 475)$  (diamonds) 
 in (c);  
 $(\delta,\epsilon, s, x_0)=(0,0,0.02,0.5)$ (circles) and $(\delta,\epsilon, s, x_0)=(0,0,0.1,0.7)$ (crosses) in (d); $(\delta,\epsilon, s, x_0)=(0.1,0,0.02,0.5)$ (circles) and $(\delta,\epsilon, s, x_0)=(-0.1,0,0.02,0.5)$ (diamonds) 
 in (e); $(\delta,\epsilon, s, x_0)=(0.2,5,0.02,0.5)$ (circles) and $(\delta,\epsilon, s, x_0)=(-0.2,-0.5,0.02,0.5)$ (diamonds) in (f). 
Green markers  show simulation data  for the fixation probability $\phi_2$ of the effective two-state model; see text and Eq.~\eqref{eq:nutilde}. In (b,c,e,f), results for $\phi_2$ are shown as triangles when $\delta=-0.2$ and as squares when  $\delta=0.2$. In (d), simulations data of $\phi_2$  are shown as squares for $(s, x_0)=(0.02,0.5)$ and as downside triangles for $(s, x_0)=(0.1,0.7)$. 
For the $N$-PDMP-based approximation of
$\phi_{2}$, see Refs.~\cite{Wienand2017,Wienand2018,Taitelbaum2020}.
Horizontal dashed and dotted lines are eyeguides showing
 $\phi_3^{0}$ (dashed) and $\phi_3^{\infty}$ (dotted)  from Eqs.~\eqref{eq:phi0} and \eqref{eq:phiinf} for each parameter set.
Error bars are included in each case but are
typically too small to see (Appendix \ref{appendix:simulations}). 
}
\label{fig:Fig5_n1}
\end{figure*}


\begin{figure*}
\centering

\setlength{\unitlength}{1cm}

\begin{overpic}[width=0.32\textwidth]{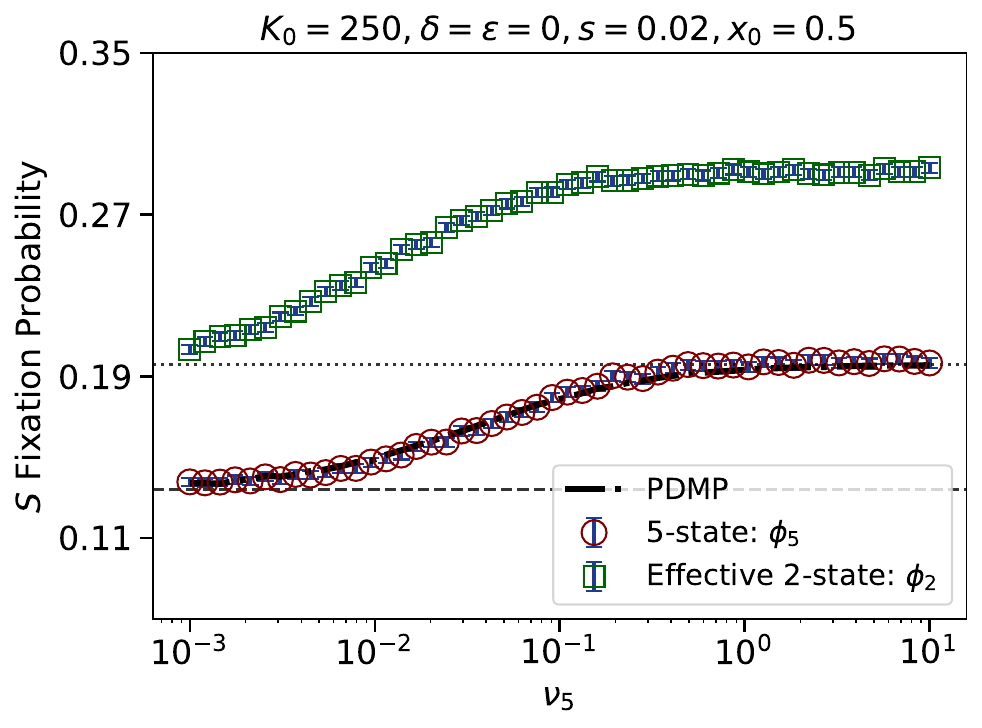
}
    \put(12,75){(a)}
\end{overpic}
\hfill
\begin{overpic}[width=0.32\textwidth]{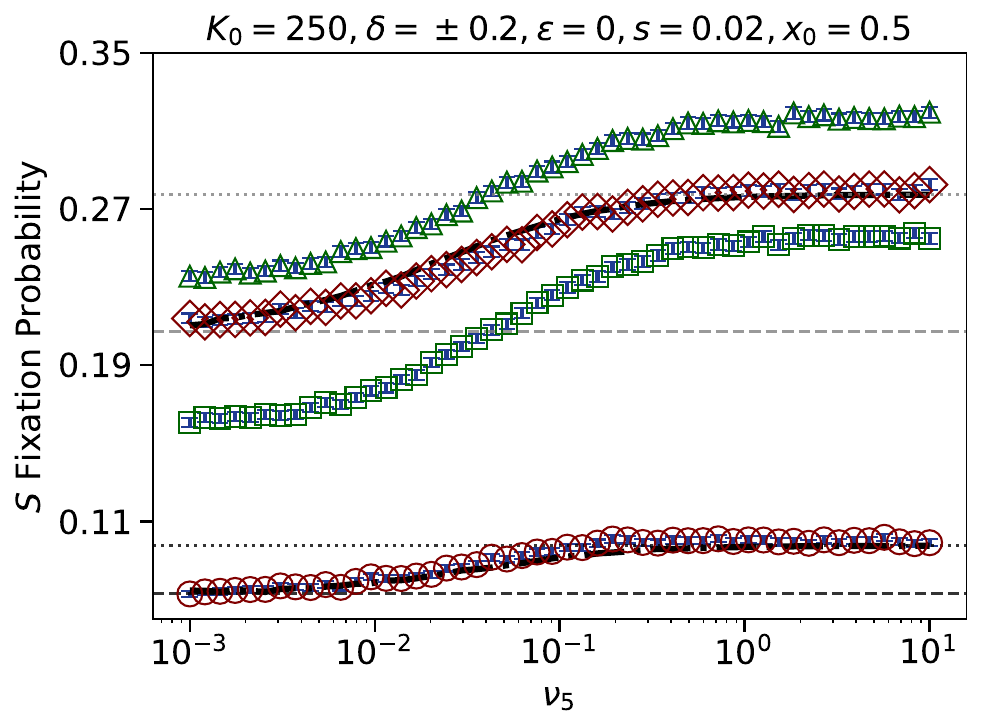
}
    \put(12,75){(b)}
\end{overpic}
\hfill
\begin{overpic}[width=0.32\textwidth]{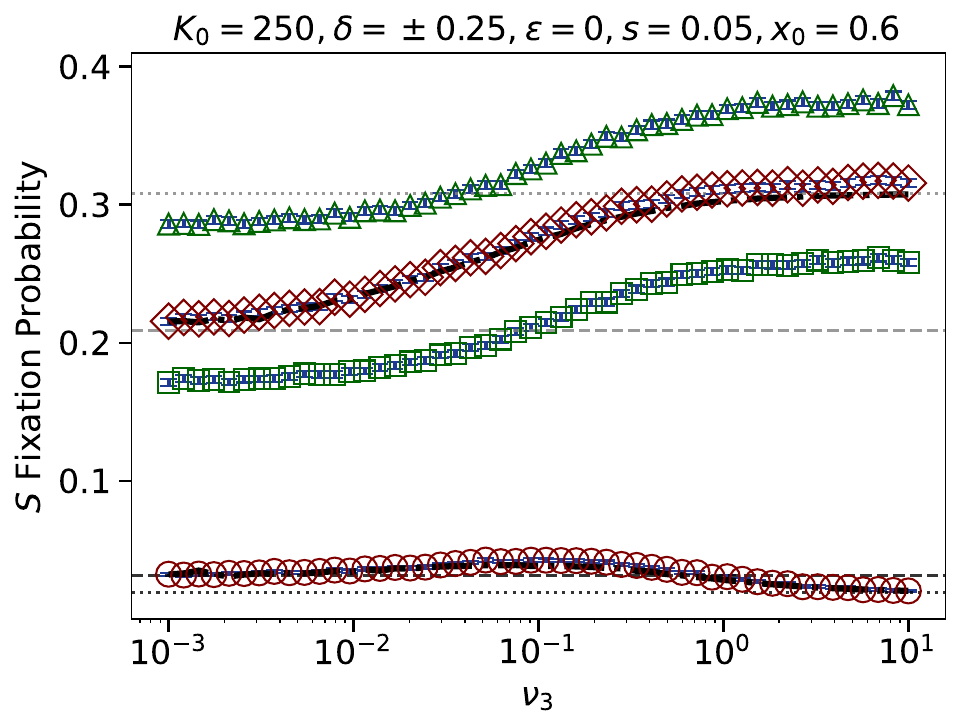
}
    \put(12,75){(c)}
\end{overpic}
\hfill
\vspace{1.5em}

\begin{overpic}[width=0.32\textwidth]{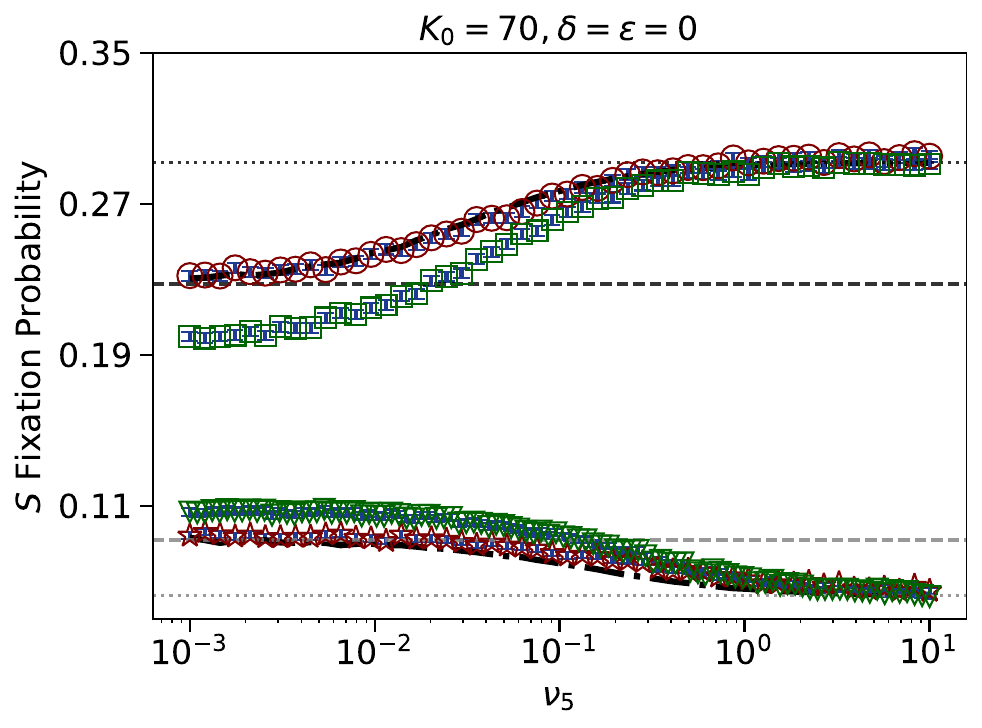
}
    \put(12,75){(d)}
\end{overpic}
\hfill
\begin{overpic}[width=0.32\textwidth]{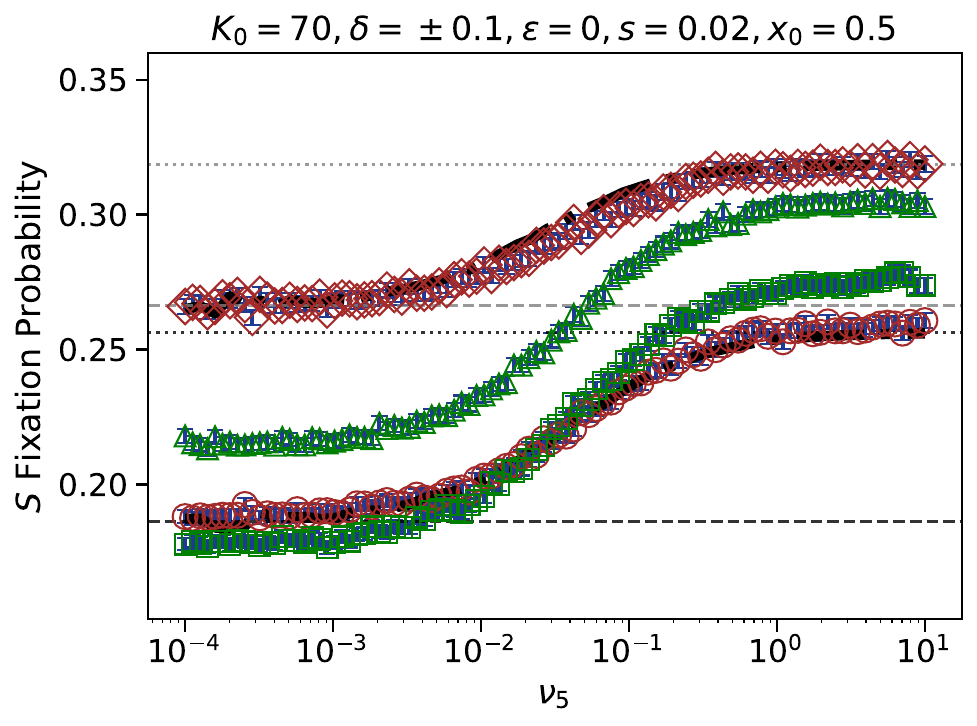
}
    \put(12,75){(e)}
\end{overpic}
\hfill
\begin{overpic}[width=0.32\textwidth]{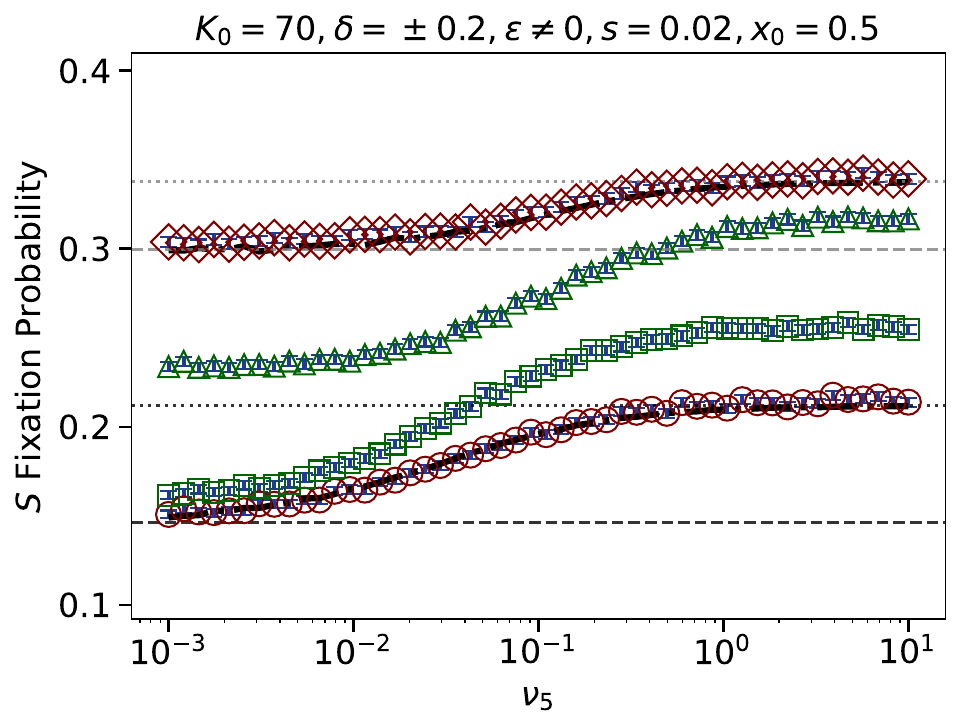
}
    \put(12,75){(f)}
\end{overpic}
\caption{$S$ fixation probability  in the five-state  model as a function of the switching rate $\nu_5$.  The centre state carrying capacity in  (a)-(c) is  $K_0=250$, while  $K_0=70$ in  (d)-(f). Here, $n=2$, and  
$(K^-,K^+)=(50,450)$ in all panels except (c).
Symbols are from full stochastic simulations and black dashed-dotted curves (in all panels),  almost indistinguishable from markers, are from the $N$-PDMP-based approximation; see Eq.~\eqref{eq:phiPDMP}.  In brown, is shown $\phi_{5}$ vs. $\nu_5$  for different parameter sets: 
 $(\delta,\epsilon, s, x_0)=(0,0,0.02,0.5)$ (circles) in (a);
 $(\delta,\epsilon, s, x_0)=(0.2,0,0.02,0.5)$ (circles) and $(\delta,\epsilon, s, x_0)=(-0.2,0,0.02,0.5)$ (diamonds) in (b); $(\delta,\epsilon, s, x_0, K^-,K^+)=(0.25,0,0.05,0.6, 25, 475)$  (circles) and $(\delta,\epsilon, s, x_0, K^-,K^+)=(-0.25,0,0.05,0.6, 25, 475)$  (diamonds) 
 in (c);  
 $(\delta,\epsilon, s, x_0)=(0,0,0.02,0.5)$ (circles) and $(\delta,\epsilon, s, x_0)=(0,0,0.1,0.7)$ (brown crosses) in (d); $(\delta,\epsilon, s, x_0)=(0.1,0,0.02,0.5)$ (circles) and $(\delta,\epsilon, s, x_0)=(-0.1,0,0.02,0.5)$ (diamonds) 
 in (e); $(\delta,\epsilon, s, x_0)=(0.2,5,0.02,0.5)$ (circles) and $(\delta,\epsilon, s, x_0)=(-0.2,-0.5,0.02,0.5)$ (diamonds) in (f). 
 As in Fig.~\ref{fig:Fig5_n1}, 
 green markers  show simulation data  for the fixation probability $\phi_2$ of the effective two-state model; see text and \eqref{eq:nutilde}. In (b,c,e,f), results for $\phi_2$ are shown as triangles when $\delta=-0.2$ and as squares when  $\delta=0.2$. 
 In (d), simulations data of $\phi_2$  are shown as squares for $(s, x_0)=(0.02,0.5)$ and as downside triangles for $(s, x_0)=(0.1,0.7)$. For the $N$-PDMP-based approximation of
$\phi_{2}$, see Refs.~\cite{Wienand2017,Wienand2018,Taitelbaum2020}.
 In (c), 
 $\phi_{5}$ varies weakly but non-monotonically with $\nu_5$. 
Horizontal dashed and dotted lines are eyeguides showing
 $\phi_5^{0}$ (dashed) and $\phi_5^{\infty}$ (dotted)  from Eqs.~\eqref{eq:phi0} and \eqref{eq:phiinf} for each parameter set.
Error bars are included in each case but are
 too small to see (Appendix \ref{appendix:simulations}). 
\label{fig:Fig6_n2}
}
\end{figure*}
When $n=1$, the  ternary environmental cycle consists of
the feast and famine states of carrying capacity $K^{\pm}$,
and the median state  whose carrying capacity is $K_0$.
Thus, in the three-state switching model $K(t)\in \{K^-, K_0, K^+\}$; see Fig.~\ref{fig:Fig1-cartoon}(a,left).

At the start of each simulation, the value of $K$ is randomly allocated  according to the distribution given by Eq.~\eqref{eq:pi} which, for  $i\in\{-1,0,1\}$,  explicitly reads
\begin{equation*}
 \label{eq:piiK3}
\pi_i=
\begin{cases}
 \left(\frac{1-\rho}{1-\rho^3}\right)~\rho^{i+1},  & \text{if $\rho\neq 1 \; (\delta\neq 0)$}\\
 \frac{1}{3} & \text{if $\rho= 1 \; (\delta= 0)$.}
\end{cases}
 \end{equation*}
In the regime of slow switching where $\nu_{2n+1}\ll s$, the $S$ fixation probability is approximately given by Eq.~\eqref{eq:phi0}, with $\phi_3(\nu_{2n+1}\ll s)\approx\phi_3^0$, where
 \begin{equation*}
 \hspace{-5mm}
 \label{eq:phi03}
\phi_3^0\equiv
\begin{cases}
 \left(\frac{1-\rho}{1-\rho^3}\right)\left[\phi_M(K^-) + \rho\phi_M(K_0)+\rho^2\phi_M(K^+)\right],  & (\rho\neq 1)\\
[\phi_M(K^-) + \phi_M(K_0)+\phi_M(K^+)]/3,  & (\rho= 1),
\end{cases}
 \end{equation*}
and $\phi_M(K_i)$ is given by Eq.~\eqref{eq:phi_M}. In binary environments, this approximation yields $\phi_2^0\equiv  \frac{1-\delta}{2}\phi_M(K^-) + \frac{1+\delta}{2}\phi_M(K^+)$~\cite{Wienand2017,Wienand2018,west2020,Taitelbaum2020,Taitelbaum2023} (Appendix~\ref{appendix:background}). From the comparison of these results, 
we find  that the $S$ fixation probability is higher in ternary environments ($\phi_3^{0}>\phi_2^{0}$)  when $\phi_M(K_0)>\phi_2^{0}$, and $\phi_3^{0}\leq \phi_2^{0}$ otherwise. The conditions for $\phi_3^{0}>\phi_2^{0}$ in the examples of Fig.~\ref{fig:Fig4-heatmaps}(a)-(c) 
are therefore  $K_0\lesssim 141.36$ ($\delta=0$), $K_0\lesssim 119.61$ ($\delta=0.2$), and $K_0\lesssim 166.50$ ($\delta=-0.2$). The horizontal dashed lines in Fig.~\ref{fig:Fig5_n1} show that  $\phi_3(\nu_3\ll s)\approx \phi_3^0$, and hence 
$\phi_3^0$ is a good approximation in the slow switching regime, with $\phi_3(\nu_{3}\ll s)\approx \phi_3^{0}<\phi_2(\nu_{3}\ll s)$ when $K_0=250$ (Fig.~\ref{fig:Fig5_n1}(a)-(c)), and $\phi_3(\nu_{3}\ll s)\approx \phi_3^{0}>\phi_2(\nu_{3}\ll s)$ when $K_0=70$ in Fig.~\ref{fig:Fig5_n1}(e),(f) and Fig.~\ref{fig:Fig5_n1}(d) when $s=0.02$.

In the fast switching regime, the population is subject to the effective carrying capacity given by Eq.~\eqref{eq:curlyK}, that here reads
\begin{equation*}
 \label{eq:curlyK3}
{\cal K}_3=
\begin{cases}
 \left(\frac{1-\rho^3}{1-\rho}\right)~\frac{K^-K_0K^+}{K_0K^+ + \rho K^- K^+ +\rho^2 K_0 K^-} & \text{if $\rho\neq 1 \; (\delta\neq 0)$}\\
 \frac{3K^-K_0K^+}{K_0K^+ + K^- K^+ + K_0 K^-} & \text{if $\rho= 1 \; (\delta= 0)$.}
\end{cases}
 \end{equation*}
Comparing this expression with its binary-state counterpart, ${\cal K}_2$, given by Eq.~\eqref{eq:curlyK2}, we find that  ${\cal K}_2>{\cal K}_3$ when $K_0<\left[\frac{(1+\delta)}{2K^-}+\frac{(1-\delta)}{2K^+}\right]^{-1}$.  Since $\phi_M(K)$ is a  decreasing function of $K$,  the $S$ fixation probability under fast switching is higher in 3-state than 2-state environments ($\phi_3^{\infty}>\phi_2^{\infty}$)
when $K_0<\left[\frac{(1+\delta)}{2K^-}+\frac{(1-\delta)}{2K^+}\right]^{-1}$, whereas it is the opposite ($\phi_3^{\infty}\leq \phi_2^{\infty}$) otherwise.
This is captured by the heatmaps of Fig.~\ref{fig:Fig4-heatmaps}(d)-(f) where for $n=1$ we find  $\phi_3^{\infty}>\phi_2^{\infty}$  when  $K_0<90$ in (d), if $K_0\lesssim 77.59$ in (e), and when $K_0\lesssim107.14$ in (f). 
The dotted lines in Fig.~\ref{fig:Fig5_n1}(a)-(c) illustrate  that, for given $(K^{\pm},s,x_0,\delta,\epsilon)$, $\phi_3(\nu_{3}\gg s)\approx \phi_3^{\infty}<\phi_2(\nu_{3}\gg s)$ when $K_0=250$;  while those in Fig.~\ref{fig:Fig5_n1}(d)-(f) show that $\phi_3(\nu_3\gg s)\approx \phi_3^{\infty}>\phi_2(\nu_3\gg s)$ when $K_0=70$.

In the intermediate switching regime ($\nu_3\sim s$), simulation results of Fig.~\ref{fig:Fig5_n1} show that    the $N$-PDMP-based approximation given by Eq.~\eqref{eq:phiPDMP}
 generally faithfully captures the properties of  $\phi_{3}$ in all regimes, with 
$\phi_{3}\approx \phi_{3}^{\text{PDMP}}$. In Fig.~\ref{fig:Fig5_n1}(a)-(e), 
dashed lines showing $\phi_{3}^{\text{PDMP}}$ are almost indistinguishable from $\phi_{3}$ for all switching rate values.

 Since the environmental switching is biased towards higher population size when $\delta>0$ (Fig.~\ref{fig:Fig2-hist}) and $\phi_M(N)$ is a decreasing function of $N$, $\phi_{3}$ decreases with $\delta$ at fixed value of $\nu_3$ (Fig.~\ref{fig:Fig2-hist}(b),(f)) and  is higher for lower values of $K_0$; compare 
 Fig.~\ref{fig:Fig5_n1}(a),(d) and Fig.~\ref{fig:Fig5_n1}(b),(f). This feature holds for arbitrary values of $n$; see below.  
 We have verified that for $-0.8\leq \epsilon\leq 5$ the parameter $\epsilon$ has no noticeable effect on $\phi_{3}$; e.g., compare 
 Fig.~\ref{fig:Fig5_n1}(b) and Fig.~\ref{fig:Fig10}(a); see Appendix~\ref{appendix:add-figures}. For high value of $\epsilon$, the $N$-PDMP-based approximation
overestimates the average number of switches prior to fixation, leading to $\phi_{3}^{\text{PDMP}}\gtrsim \phi_{3}$ when $\nu_3 \lesssim s$ in Fig.~\ref{fig:Fig5_n1}(f) and Fig.~\ref{fig:Fig10}(a) when $\epsilon=5$.

These results show that, depending on the value of $K_0$, ternary  switching  can either enhance or decrease the $S$ fixation probability with respect to the binary case: When $K_0$ is sufficiently low (e.g. $K_0=70$ in Fig.~\ref{fig:Fig5_n1}(d)-(f), the population size is small over significant periods, during which the selection pressure is balanced by demographic fluctuations. This leads to the $\phi_{3}$ to be larger than $\phi_{2}$ and closer to the fixation probability of the faster strain.


\subsubsection{Fixation in the five-state  switching model}
\label{Sec:n2}
When $n=2$, the  five-fold environmental cycle consists of
the feast, famine and median states  of respective carrying capacities $K_{\pm 2}= K^{\pm}$ and $K_0$,
 and the intermediate mild/harsh states of carrying capacity $K_{\pm 1}=(K_0+K^{\pm})/2$.
Hence, in the five-state switching model $K(t)=K_i\in \{K^-, K_{-1}, K_0, K_1, K^+\}$; see Fig.~\ref{fig:Fig1-cartoon}(a,right), with its value  randomly initiated from the distribution given by Eq.~\eqref{eq:pi}.

In the regime of slow switching where $\nu_{2n+1}\ll s$, the $S$ fixation probability is  approximated by $\phi_5^0$, from Eq.~\eqref{eq:phi0}. For given values of  $s,x_0,\delta, K^{\pm}$, by solving 
$\phi_5^0>\phi_2^0=\frac{1-\delta}{2}\phi_M(K^-) + \frac{1+\delta}{2}\phi_M(K^+)$, we find  the conditions on $K_0$ for which the $S$ fixation probability under slow switching is enhanced by five-state environments with respect to the binary case. For the  examples of Fig.~\ref{fig:Fig4-heatmaps}(a)-(c) with $n=2$, we find $\phi_5^{0}>\phi_2^{0}$  if $K_0\lesssim 119.96$ ($\delta=0$), $K_0\lesssim 52.74$ ($\delta=0.2$), and $K_0\lesssim 190.22$ ($\delta=-0.2$). 

 In the case $n=2$, the effective carrying capacity under fast switching, given by Eq.~\eqref{eq:curlyK}, explicitly reads
$
 \label{eq:curlyK5}
{\cal K}_5=\left(\frac{1-\rho^5}{1-\rho}\right)~\left[\frac{1}{K^-}+\frac{2\rho}{K_{0}+K^-}+\frac{\rho^2}{K_{0}}+\frac{2\rho^3}{K_0+K^+}+\frac{\rho^4}{K^+}\right]^{-1}$
when $\rho\neq 1$ ($\delta\neq 0$), which becomes
${\cal K}_5=5~\left[\frac{1}{K^-}+\frac{2}{K_{0}+K^-}+\frac{1}{K_{0}}+\frac{2}{K_0+K^+}+\frac{1}{K^+}\right]^{-1}$ when switching is symmetric ($\rho= 1$, $\delta=0$). Proceeding as in the ternary case, the comparison of ${\cal K}_5$ and ${\cal K}_2$ leads to the conditions for ${\cal K}_2>{\cal K}_5$ and thus for $\phi_5^{\infty}>\phi_2^{\infty}$. In Fig.~\ref{fig:Fig4-heatmaps}(d)-(f) we thus find that 
 $\phi_5^{\infty}>\phi_2^{\infty}$  when $K_0<74.40$ ($\delta=0$), if $K_0\lesssim 113.34$ ($\delta=-0.2$),  while 
 $\phi_5^{\infty}< \phi_2^{\infty}$  when $\delta=0.2$. 

The dotted and dashed lines in Fig.~\ref{fig:Fig6_n2} show that  $\phi_5(\nu_5\ll s)\approx \phi_5^0$ and $\phi_5(\nu_5\gg s)\approx \phi_5^{\infty}$, with 
$\phi_5^0$  and $\phi_5^{\infty}$ thus providing good approximations in the slow and fast switching regimes when $K_0=250$ and $K_0=70$.

When $\nu_5\sim s$ (intermediate switching), Figure~\ref{fig:Fig6_n2} shows that  the $N$-PDMP-based approximation, given by Eq.~\eqref{eq:phiPDMP}, is in good agreement with simulation results, 
yielding $\phi_{5}\approx \phi_{5}^{\text{PDMP}}$ for all parameter sets. Again, we notice that Eq.~\eqref{eq:phiPDMP} is a good approximation across the slow, intermediate and fast switching regimes, with dashed-dotted lines in Fig.~\ref{fig:Fig6_n2} almost indistinguishable from the data of $\phi_{5}$. In particular, $\phi_{5}^{\text{PDMP}}$ captures the non-trivial $\nu_5$-dependence of $\phi_{5}$, and even reproduces its weak peak in the intermediate regime (for $\nu_5\sim 0.1$) of Fig.~\ref{fig:Fig6_n2}(c). 
 As in ternary environments, $\phi_{5}$ at given $\nu_5$ decreases with $\delta$; see Fig.~\ref{fig:Fig6_n2}(b), and  is higher for lower values of $K_0$; compare 
 Fig.~\ref{fig:Fig6_n2}(a),(d) for $(s,x_0)=(0.02,0.5)$ and Fig.~\ref{fig:Fig6_n2}(b),(f).

 The comparison of Figs.~\ref{fig:Fig5_n1} and \ref{fig:Fig6_n2}, supported by the heatmaps of Figs.~\ref{fig:Fig4-heatmaps}, indicates that the $S$ fixation probability 
 decreases when the number of environmental states  increases, with the  environmental bias $\delta>0$ strongly reducing $\phi_{2n+1}$ as $n$ becomes bigger. This effect can be countered by reducing the value of $K_0$. Generally,  $\phi_{2n+1}$ and $\phi_{2}$ exhibit the same qualitative dependence on the switching rate, with $n$ affecting mostly the amplitude of $\phi_{2n+1}$, and $\phi_{2n+1}<\phi_{2}$ or $\phi_{2n+1}>\phi_{2}$ for all values of $\nu_{2n+1}$. However, for some parameters 
 $\phi_{2n+1}<\phi_{2}$ in a certain switching regime and $\phi_{2n+1}>\phi_{2}$ in another; see Fig.~\ref{fig:Fig6_n2}(e) where, for $\delta=0.1$, $\phi_{5}>\phi_{2}$ in the slow/intermediate regime and $\phi_{5}<\phi_{2}$ under intermediate/fast switching. 
 There are also parameter sets for which one of $\phi_{2n+1}$ or $\phi_{2}$ varies non-monotonically with $\nu_{2n+1}$ while the other is an increasing or decreasing function of the switching rate 
  (Figs.~\ref{fig:Fig5_n1}(c) and \ref{fig:Fig6_n2}(c)), or where $\phi_{2n+1}<\phi_{2}$ and $\phi_{2n+1}>\phi_{2}$ on two separate regimes of $\nu_{2n+1}$ (Fig.~\ref{fig:Fig6_n2}(e)).

 These results show that multi-state switching environments lead to various fixation scenarios, with the  $S$ fixation probability that can be enhanced or hindered with respect to the binary case, depending on the values of $K_0$ and $\nu_{2n+1}$, encoding the amplitude and frequency of  environmental fluctuations.

\section{Discussion \& conclusion}
\label{Sec:Discussion}
Microbial communities are commonly subject to changing environmental conditions that are often  described  in terms of two-state switching models. Strain competition subject to a
 binary time-switching carrying capacity can be seen as the simplest coarse-grained description of feast--famine cycles, characterised by alternating periods of nutrient abundance (feast) and scarcity (famine). Recent experimental and theoretical studies, however, show the importance of {\it multiple intermediate environmental states} on feast-famine cycle dynamics~\cite{merritt2018,himeoka_dynamics_2020,niimi2026}.
 Experiments have indeed demonstrated that the frequency and amplitude  of environmental fluctuations  have a strong impact on microbial feast--famine cycles, with gradual changes of nutrient conditions 
 influencing  nutrient-utilization strategies.  Eco-evolutionary dynamics in binary and ternary environments revealed important differences between abrupt  and gradual environmental deterioration through intermediate conditions~\cite{sanchez2013}.

This has motivated us to study a class of multi-state switching models for
the dynamics of two strains, one ($S$) slightly slower than the other, competing for the same resources in  multi-state fluctuating environments. Here, the time-varying environment consists of multiple ordered \((2n+1)\) environmental states, each associated with its own carrying capacity. To represent gradually changing conditions, there are switches between  nearest-neighbour  states and their carrying capacities: The  environment evolves as a  Markov chain over multiple carrying capacity states, which provides a simple description of gradual environmental variability. 
The individual-based switching models considered here are
 not intended as a mechanistic description of feast--famine experiments, but can be interpreted as a coarse-grained representation of fluctuating nutrient environments. In this picture, the multi-state carrying capacities are  associated with different nutrient conditions, the switching rates determine the frequency of environmental fluctuations, and the  distribution of carrying capacities quantifies their amplitude. This class of  models therefore captures, at the population level, the influence of the statistical properties of environmental variability on eco-evolutionary dynamics, while coarse-graining over microscopic processes such as nutrient uptake.  
 From this perspective, the \((2n+1)\)-state switching models provide an individual-based framework for investigating how the frequency and amplitude of environmental fluctuations jointly affect stochastic feast--famine cycle dynamics and their evolutionary outcomes.

By computational and analytical means, we have characterised the long-time dynamics and fate of these multi-state switching models. In particular, 
we have studied the influence of the switching rates and distribution of carrying capacities 
on the long-time population-size statistics and fixation properties, and compared these results with those obtained in binary environments. Due to the presence of intermediate states, the population size distribution (PSD) is generally broader and the environmental bias (parameter $\delta$) has a stronger effect than  under binary switching. In general, owed to the underlying logistic dynamics, the PSD is right-tailed.
 Under slow switching, the PSD is characterised by $(2n+1)$ peaks, whereas fast switching yields the same qualitative behaviour as in two-state environments. 
 Many features of the PSD are aptly captured by a description of the population size by a piecewise-deterministic Markov process ($N$-PDMP). 
The $S$ fixation probability in the multi-state switching models
can be either hindered or enhanced with respect to the binary case, depending on the switching rate, number of states and their carrying capacities. 
 Since, the slow strain is more likely to fixate the population under harsh than mild conditions, increasing the number of environmental states and the values of the carrying capacities  tends to
 reduce the $S$ fixation probability.  However, there are  environmental conditions under which 
 the fixation of 
 $S$ is more likely in the presence of intermediate environmental states than in the binary case; see Figs.~\ref{fig:Fig5_n1}(d)-(f) and \ref{fig:Fig6_n2}(d)-(f). The unconditional mean fixation time  scales
  with the evolutionary timescale ($1/s$), with
 the prefactor depending on 
 the frequency and amplitude of environmental variability.
The  probability and unconditional mean time of fixation can be obtained analytically in the limits of slow and fast switching. We have also devised an  $N$-PDMP-based approximation for the fixation properties under weak selection,  an efficient computational method
 that provides useful scaling information.

In the feast--famine cycles considered in Refs.~\cite{merritt2018,himeoka_dynamics_2020,niimi2026}, environmental variability is encoded in nutrient dynamics. In the class of  switching models studied here, multi-state
carrying capacities provide a coarse-grained description  of varying nutrient levels. This individual-based framework has allowed us to 
analyse how the frequency and amplitude of environmental fluctuations shape population dynamics, highlighting that  environmental complexity  leads to richer eco-evolutionary scenarios. In the future, this approach could be generalised in different directions. 
In addition to straightforward extensions (e.g., $2n$ environmental states, other distributions of the carrying capacities), we can mention the limit of a large number of environmental states that will help shed further light on the notoriously difficult problem of eco-evolutionary dynamics under continuous environmental noise; see, e.g., Refs.~\cite{assafCooperationDilemmaFinite2013,kalyuzhnyNeutralTheoryEnvironmental2015,melbinger-vergassola2015,nguyen2021,Taitelbaum2023}.
This framework can also be employed to investigate the coarse-grained effect of 
varying nutrient levels on the spread of cooperative antimicrobial resistance in well-mixed settings~\cite{Yurtsev13,Hernandez2023}, or in spatially structured metapopulations inspired by chemostat systems or batch cultures~\cite{Abbara2023,moawad2024,Li2022,verdon2024habitat,Hernandez2026}. More generally, the present work shows that  multi-state stochastic switching environments provide a simple yet versatile framework for investigating how the statistical properties of environmental fluctuations shape eco-evolutionary dynamics.

\vspace{2mm}
{\noindent{\bf Data availability statement:}
 Simulation data and codes  that support the findings of this study are openly available at the following URL/DOI: 10.5518/1899~\cite{codes-data-MM}.}

\section*{Acknowledgments}
Partial support from the U.K. Engineering and Physical Sciences Research Council (EPSRC) under the Grant No. EP/V014439/1 is gratefully acknowledged.

\appendix
\section{Further details on the class of switching models}
\label{appendix:model-details}
\subsection{Background: Competition in binary-state switching models}
\label{appendix:background}
As a background for this study, we review the 
main features of the competition dynamics in binary-state switching models; see Refs.~\cite{Wienand2017,Wienand2018,west2020,Taitelbaum2020} for further details.

In two-state switching models, with cyclically alternating mild and harsh conditions, the time-varying carrying capacity $K(t)\in \{K^-,K^+\}$ takes only two possible values: $K^+$ and $K^-$ representing respectively the carrying capacity of the ``feast'' and ``famine'' states ($K^+>K^-$)~\cite{Wienand2017,Wienand2018,west2020,Taitelbaum2020,Shibasaki2021,Taitelbaum2023,Hernandez2023,Asker2023,Hernandez2024,Asker2025,Hernandez2026}. It is thus convenient to write the
 carrying capacity of two-state switching models as
\begin{equation}
    \label{eq:KD(t)}
    K(t)=\frac{1}{2}\left[K^++K^- +\xi_{\text{D}}(t)(K^+-K^-),\right]
\end{equation}
where $\xi_{\text{D}}(t)\in\{-1,1\}$ is the coloured dichotomous  Markov noise (DMN)~\cite{Bena2006,HL06,Ridolfi11},  also called telegraph process,  encoding the  binary environmental variability and 
driving the switching of $K$. The DMN 
switches between $\pm 1$ according to $\xi_{\text{D}} \to -\xi_{\text{D}}$ at rate $\widetilde{\nu}^{\pm}$ when $\xi_{\text{D}}=\pm 1$~\cite{Bena2006,HL06,Ridolfi11}. It is useful to write $\widetilde{\nu}^{\pm}$ in terms of the mean switching rate $\widetilde{\nu}\equiv (\widetilde{\nu}^{-}+\widetilde{\nu}^{+})/2$ and \(\delta\equiv (\widetilde{\nu}^{-}-\widetilde{\nu}^{+})/(2\widetilde{\nu})\) (environmental switching bias), with $|\delta|< 1$. When the DMN is at stationarity, all instantaneous correlations become time-independent while its autocovariance is a function of the time difference. This means that at stationarity $\xi_{\text{D}}=\pm 1$ with a probability $(1\pm \delta)/2$, and 
the 
stationary 
average of the DMN is $\langle \xi_{\text{D}} (t)\rangle=\delta$, while its autocovariance reads $\langle \xi_{\text{D}}(t)\xi_{\text{D}}(t')\rangle-\langle \xi_{\text{D}}(t)\rangle\langle \xi_{\text{D}}(t')\rangle=(1-\delta^2)e^{-2\widetilde{\nu}|t-t'|}$ when $t,t'\to \infty$~\cite{Bena2006,HL06,Ridolfi11,Taitelbaum2020,Taitelbaum2023}, where $\langle \cdot \rangle$ here denotes the ensemble average and  $1/(2\widetilde{\nu})$ is the finite DMN  correlation time.

Following Eq.~\eqref{eq:KD(t)}, the binary carrying capacity switches back and forth at rates $\widetilde{\nu}^\pm$ between a value $K=K^+$ ($\xi_{\text{D}}=1$) in  a mild environment, and $K=K^-< K^+$ ($\xi_{\text{D}}=-1$) when environmental conditions are harsh,
according to $K^+\xrightleftharpoons[\widetilde{\nu}^-]{\widetilde{\nu}^+}K^-$
As the DMN, when the time-switching $K(t)$  is  at stationarity, it takes the value $K=K^{\pm}$
with probability $(1\pm \delta)/2$. Its expected value and variance are time independent, and respectively read $\langle K \rangle_{2}=\left(\frac{1-\delta}{2}\right) K^- + \left(\frac{1+\delta}{2}\right) K^+$, and ${\rm var}(K)_{2}=(1-\delta^2)(K^+-K^-)^2/4$, while its auto-covariance is $\langle K(t)K(t')\rangle_{2}-\langle K(t)\rangle_{2}\langle K(t')\rangle_{2}=\left(\frac{K_{+}-K_{-}}{2}\right)^2\left(1-\delta^2\right)e^{-2\widetilde{\nu}|t-t'|}$ ($t,t'\to \infty$)~\cite{Bena2006,HL06,Ridolfi11,Taitelbaum2020,Taitelbaum2023}.

The time-switching $K(t)$ drives the overall population  size $N(t)$, and is hence  responsible for the coupling of demographic  and environmental fluctuations~\cite{Wienand2017,Wienand2018,west2020,Taitelbaum2020,Shibasaki2021,Taitelbaum2023,Hernandez2023,Asker2023,Hernandez2024,Asker2025,Hernandez2026}. 
Upon ignoring all fluctuations, the   mean-field dynamics of the population subject to a constant carrying capacity $K(t)=\bar{K}\gg 1$ satisfy the rate equations \eqref{eq:MF} for $N$
and the fraction $x=N_S/N$ of $S$ individuals; see Sec.~\ref{Sec:Static-MF}. When $\bar{K}$
is large but finite, as explained in Sec.~\ref{Sec:Static-Moran}, the population dynamics
can be aptly approximated by a Moran process \cite{Moran,Ewens,Blythe07,melbinger2010,cremer2011,Wienand2017,Wienand2018,west2020} by assuming a constant population size  $N=\bar{K}$. In this case the Moran probability and mean time for the fixation can be computed analytically
~\cite{antal2006fixation,traulsen2009stochastic,Wienand2017,Wienand2018,Taitelbaum2020,Hernandez2023,Asker2023,Hernandez2024,Asker2025,Hernandez2026}; see Sec.~\ref{Sec:Static-Moran} and Appendix~\ref{appendix:MFT-static}.

When the population size is sufficiently large for demographic fluctuations to be negligible, but is still subject to environmental variability (switching
of $K(t)$), the population size dynamics is well approximated by the two-state piecewise deterministic Markov process ($N$-PDMP)~\cite{davisPiecewiseDeterministicMarkovProcesses1984,hufton2016,Wienand2017,Wienand2018,west2020,Taitelbaum2020,Taitelbaum2023,Hernandez2023,Asker2023,Hernandez2024,Asker2025,Hernandez2026} that for binary switching 
is defined by 
\begin{equation}
\label{eq:PDMP-2state}
    \dot{N}=
    \begin{cases}
    N\left(1-\frac{N}{K^-}\right) & \text{if } \xi_{\text{D}}=-1, \\
    N\left(1-\frac{N}{K^+}\right)  & \text{if } \xi_{\text{D}}=1. 
    \end{cases}
\end{equation}
In the $N$-PDMP approximation, the population size  satisfies a deterministic logistic equation in each environmental state $\xi_{\text{D}}=\pm 1$, subject to the time-switching carrying capacity $K(t)$  given by Eq.~\eqref{eq:KD(t)}.
The case of the $N$-PDMP approximation for multi-state switching models with $(2n+1)$ environmental states is discussed in Sec.~\ref{Sec:PSD}. 
While it ignores the effect of demographic noise, the 
$N$-PDMP approximation  captures many properties of the (marginal) quasi-stationary
distribution $p_{\widetilde{\nu}^{\pm}}(N)$ of the population size in binary environments, including that $p_{\widetilde{\nu}^{\pm}}(N)$ is bimodal  under slow switching ($\widetilde{\nu}\ll 1$), while it is unimodal and centred around the effective carrying capacity
\begin{equation}
 \label{eq:curlyK2}
 {\cal K}_2=\left(\frac{1-\delta}{2K^-}+\frac{1+\delta}{2K^+}\right)^{-1}
\end{equation}
in the  fast environmental switching regime ($\widetilde{\nu}\gg 1$)~\cite{Wienand2017,Wienand2018,west2020,Taitelbaum2020,Taitelbaum2023,Hernandez2023,Asker2023,Hernandez2024,Asker2025,Hernandez2026}. 
The marginal probability density of the $N$-PDMP \eqref{eq:PDMP-2state} can be obtained analytically~\cite{Wienand2017,Wienand2018,west2020,Taitelbaum2020}:
\begin{equation}
\begin{aligned}
   p^{\text{PDMP}}_{{\nu}^{\pm}}(N)&= \frac{\mathcal{Z}}{N^2} \left(\frac{K^+ - N}{N}\right)^{\widetilde{\nu}^+-1} \left(\frac{N-K^-}{N}\right)^{\widetilde{\nu}^--1},
    \label{eq:NPDMP_marg}
\end{aligned}
\end{equation}
where $\mathcal{Z}$ is the normalisation constant and 
$N\in [K_-,K_+]$. In addition, to capturing many features of $p_{\widetilde{\nu}_{\pm}}$,  \eqref{eq:NPDMP_marg}  can be used to 
accurately approximate the average population size:
$\sum_{N=0}^{\infty} N p_{\widetilde{\nu}_{\pm}}(N)\approx
\int_{K_-}^{K_+} N p^{\text{PDMP}}_{\widetilde{\nu}_{\pm}}(N)~ dN$
~\cite{Wienand2017,Wienand2018,Taitelbaum2020}. 
Eq.~\eqref{eq:NPDMP_marg} of the binary PSD
has also been used to approximately compute the fixation probability and uMFT  in two-state switching  models~\cite{Wienand2017,Wienand2018,west2020,Taitelbaum2020,Taitelbaum2023,Asker2025}; see Sec.~\ref{Sec:Fixation}.

It is worth noting that DMN is commonly used to model evolutionary processes because of its simplicity
and ability to capture the conditions used in laboratory-controlled experiments.
Even if these are generally carried out in periodically changing environments~\cite{acar_stochastic_2008,Lambert2014,abdul2021}, it has 
been shown that letting $K$ switch periodically between $K^+$ and
$K^-$ with a frequency $\widetilde{\nu}$ leads to essentially the same dynamics as with Eq.~\eqref{eq:KD(t)}~\cite{Taitelbaum2020}.
Moreover, the relationship between DMN and other forms of environmental noise is well documented~\cite{HL06,Ridolfi11,Taitelbaum2023}.

\subsection{Multi-state environmental variability}
\label{appendix:EV}
In this work, environmental variability is encoded in  the continuous-time Markov
process $\xi(t)=i\in\{-n,\dots,0,\dots,n\}$  that is a $(2n+1)$-state coloured noise (non-zero correlation time).
Here, the environmental noise  $\xi(t)$ is initialised in its stationary distribution (see below) and is therefore  a $(2n+1)$-state generalisation of the classical (two-state) DMN noise used in numerous studies~\cite{Bena2006,Ridolfi11,HL06,acar_stochastic_2008,hufton2016,Wienand2017,Wienand2018,west2020,Shibasaki2021,Hernandez2023,Asker2023,Hernandez2024,Asker2025,Hernandez2026}.

The generator matrix  associated with the multi-state  switching of $\xi(t)$ according to 
Eq.~\eqref{eq:xi} is the $(2n+1)\times (2n+1)$ stochastic matrix ${\bm Q}$
whose entries $Q_{j,k}$ are the switching rates of \eqref{eq:xi}~\cite{allenIntroductionStochasticProcesses2010b}. For convenience, we choose the indices $j,k=0,\dots,2n$ so that, when $j\neq k$, $Q_{j,k}$ corresponds to the rate of the transition from state $j-n$ to state $k-n$, i.e. $j-n \stackrel{Q_{j,k}}{\longrightarrow} k-n$.
Thus, the  generator matrix  ${\bm Q}=(Q_{j,k})$ of  $\xi(t)$ defined by the transitions \eqref{eq:xi} is
\begin{align*}
\label{eq:Q}
Q_{j,j+1} &=
\begin{cases}
\nu^-_{2n+1}, & 1\le j\le n-1,\\
\omega^-_{2n+1}, & n+1\le j\le 2n-1,\\
\omega^-_{2n+1}, & j=n,\\
\nu^-_{2n+1}, & j=0,
\end{cases}
\\[1ex]
Q_{j,j-1} &=
\begin{cases}
\nu^+_{2n+1}, & 1\le j\le n,\\
\omega^+_{2n+1}, & n+1\le j\le 2n-1,\\
\omega^+_{2n+1}, & j=2n, 
\end{cases}
\end{align*}
with $Q_{jj}=-\sum_{k\neq j} Q_{jk}$. 
The stationary distribution  of $\xi(t)$ is the normalised left eigenvector ${\bm \pi}$ of the tridiagonal matrix ${\bm Q}$ associated with eigenvalue 0. 
When $\nu^-_{2n+1}/\nu^+_{2n+1}=\omega^-_{2n+1}/\omega^+_{2n+1}=(1+\delta)/(1-\delta)$,
we find that  
  ${\bm \pi}=(\pi_j)$, where $\pi_j\equiv  \pi_{i+n}$ and
\begin{equation}
\label{eq:pi-ap}
\begin{aligned}
\pi_{i+n}&=
\begin{cases}
\left(\frac{1-\rho}{1-\rho^{(2n+1)}}\right)\rho^{i+n}, \quad  &\text{ if } \rho\neq 1 \; (\delta\neq 0) \;\\
\frac{1}{2n+1}, \quad &\text{ if } \rho= 1, \; (\delta=0)\,
\end{cases}
\end{aligned}
\end{equation}
with $\rho\equiv (1+\delta)/(1-\delta)$, and  $i=-n,\dots,n$. 
For notational simplicity,  we denote the stationary distribution of $\xi$ by $\pi_i$; see Eq.~\eqref{eq:pi}. The stationary  average of $\xi$ is 
\[\langle \xi \rangle=\sum_{i=-n}^n i \pi_i
=\begin{cases}
\frac{\rho}{1-\rho} -n +\frac{(2n+1)\rho^{2n+1}}{\rho^{2n+1}-1}, \quad &\text{ if } \rho\neq  1, \\
0, \quad &\text{ if } \rho= 1, \,
\end{cases}
\]
with $\langle \xi \rangle>0$ if $\delta>0$ and $\langle \xi \rangle<0$ when $\delta<0$.
The stationary distribution  ${\bm \pi}$ can also be used to compute the stationary autocovariance  $A(t)\equiv {\rm lim}_{t'\to \infty}\langle  \xi(t')\xi(t'+t)\rangle - \langle \xi \rangle^2$, that decays as
$A(t)\sim e^{\lambda_1 t}$, where $\lambda_1$ is the eigenvalue of ${\bm Q}$ with largest non-zero real part. In general, when $\delta\neq 0$ and $\epsilon\neq 0$, the finite correlation time of $\xi(t)$ is $t_c\propto 1/\nu_{2n+1}$ with a proportionality factor that  depends non-trivially on $n,\delta$ and $\epsilon$. This is to be compared with the 
correlation time  $t_{\text{c}}=1/2\widetilde{\nu}$ of the dichotomous noise $\xi_{\text{D}}$
~\cite{Bena2006,HL06,Ridolfi11,Wienand2017,Wienand2018,Taitelbaum2020}; see Appendix~\ref{appendix:background}.

The multi-state carrying capacity, driven by $\xi$, can be written as
 \begin{equation}
  \label{eq:K}
 K(\xi(t))= \begin{cases}
         K^{-}+\left(\frac{K_0-K^{-}}{n}\right)(n+\xi(t)) & \text{if }  \xi(t)=i\leq 0,\\
        K_0+\left(\frac{K^{+}-K_{0}}{n}\right)\xi(t)&\text{if } \xi(t)=i> 0,
        \end{cases}
 \end{equation}
 with $K_i\equiv K(\xi(t)=i)$ given by Eq.~\eqref{eq:Ki}, and its  
 stationary probability distribution is therefore also $\bm{\pi}$, with $\lim_{\to \infty}\mathbb{P}\left(K(t)=K_i\right)=\pi_i$ given by Eq.~\eqref{eq:pi}. The average stationary carrying capacity  is therefore $\langle K \rangle_{2n+1}=\sum_{i=-n}^{n}  K_i \pi_i$ while its stationary variance  is ${\rm var}(K)_{2n+1}=
 \sum_{i=-n}^{n}  K_i^2 \pi_i-\langle K \rangle_{2n+1}^2$.

To meaningfully compare the properties of multi-state and binary switching models, it is useful to compute the mean times for the transitions from $K=K^{\mp}$ to $K=K^{\pm}$. In binary switching models, where the transitions $K^-\xrightleftharpoons[\widetilde{\nu}^+]{\widetilde{\nu}^-}K^+$
occur at rates $\widetilde{\nu}^{\pm}$, the mean waiting times for the reactions $K^{\mp} \to K^{\pm}$ thus are $1/\widetilde{\nu}^{\mp}$. In $(2n+1)$-switching models,
the transition $K^- \to K^+$ consists of $n$  moves, each of mean waiting time $1/\nu^-_{2n+1}$, followed by $n$ other transitions of average waiting time $1/\omega^-_{2n+1}$; see Fig.~\ref{fig:Fig1-cartoon}(a), yielding a mean waiting time
$n(1/\nu^-_{2n+1} +1/\omega^-_{2n+1})$.
Similarly, the mean waiting time for the  transition $K^+ \to K^-$
in multi-state switching models is $n(1/\nu^+_{2n+1} +1/\omega^+_{2n+1})$.
Therefore, by setting
$1/\widetilde{\nu}_{\pm}=n(1/\nu^{\pm}_{2n+1} +1/\omega^{\pm}_{2n+1})$, the mean times for the transition between the extreme rates of feast ($K=K^+$) and famine ($K=K^-$) is the same in binary and multi-state models. With $\nu_{2n+1}^{\pm}=\nu_{2n+1}(1\mp\delta)$ and $\omega_{2n+1}^{\pm}=(1+\epsilon)\nu_{2n+1}^{\pm}$,
we obtain the expression of Eq.~\eqref{eq:nutilde}.

For the $N$-PDMP-based approximation of Sec.~\ref{Sec:intermediate}, we need to compute the average number of switches ${\cal N}_{\text{sw}}$ occurring on the evolutionary timescale $t\sim 1/s$. This is 
obtained by multiplying $1/s$ with the total escape rate, given by $\sum_{i=-n}^{n}\pi_i \psi_i$, where $\psi_i=\nu^+_{2n+1}+\nu^-_{2n+1}$ when $i<0$, $\psi_i=\omega^+_{2n+1}+\omega^-_{2n+1}$ when $i>0$, and $\psi_{-n}=\nu^-_{2n+1}, \psi_{0}=\nu^-_{2n+1}+\omega^+_{2n+1},  \psi_{n}=\omega^+_{2n+1}$, and  $\pi_i$ is given by Eq.~\eqref{eq:pi} (see also Eq.~\eqref{eq:pi-ap}). The average number of switches during time $1/s$ is therefore
${\cal N}_{\text{sw}}=\frac{1}{s}\sum_{i=-n}^{n}\pi_i \psi_i=\nu_{2n+1}'/s$.
The rescaled switching rate used in the approximations of Eqs.~\eqref{eq:phiPDMP} and \eqref{eq:tau2n+1}
is therefore $\nu_{2n+1}'/s$, where
\begin{align*}
\hspace{-3mm}
\nu_{2n+1}' &= \nu_{2n+1}~\left(\frac{1-\rho}{1-\rho^{2n+1}}\right) \Bigg[ 1+\delta +\frac{2\rho}{1-\rho}(1-\rho^{n-1})(2+\epsilon) \nonumber\\ &\qquad +\rho^n \bigl[2+\epsilon(1+\delta)\bigr] +\rho^{2n}(1+\epsilon)(1-\delta) \Bigg],
\end{align*}
which is the expression of Eq.~\eqref{eq:nuprime}.  When $\delta=\epsilon=0$, we have $\rho\to 1$ and this expression simplifies to
$ \nu'
 =4n\nu_{2n+1}/(2n+1)$. Similarly to what was done in Refs.~\cite{Wienand2017,Wienand2018,west2020,Taitelbaum2020,Taitelbaum2023}, 
 we have also tried to use the approximation of Eq.~\eqref{eq:phiPDMP}
 with the simpler rescaled rate $\nu_{2n+1}/s$ instead of $\nu_{2n+1}'/s$ , and obtained results that, while in sound agreement with Gillespie simulations, were less accurate than those obtained from Eq.~\eqref{eq:phiPDMP} with the effective switching rate given by  Eq.~\eqref{eq:nuprime}.

\subsection{Master equation}
\label{appendix:ME}
The competition dynamics in the $(2n+1)$-state switching models is a continuous-time multivariate Markov process -- defined by the transition rates 
$T_{S/F}^{\pm}$ given by Eq.~\eqref{eq:Transition_rates} that satisfies the  master equation for the joint probability $p_{2n+1}(N_S, N_F,\xi,t)\equiv p_{2n+1}(\vec{N},\xi,t)$ of finding the population
in configuration  $\vec{N}=(N_S,N_F)$ and environmental state $\xi=i\in\{-n,\dots,n\}$.
The master equation for $(2n+1)$-state switching model reads
\begin{widetext}
 \begin{subequations}
  \label{eq:ME}
 \begin{align}
  \label{eq:ME1}
  \frac{\partial p_{2n+1}(\vec{N},\xi,t)}{\partial t}&=(\mathbb{E}^{-}_{S}-1)[T_{S}^{+}p_{2n+1}(\vec{N},\xi,t)] +
  (\mathbb{E}^{-}_{F}-1)[T_{F}^{+} p_{2n+1}(\vec{N},\xi,t)]\\
  &+
  (\mathbb{E}^{+}_{S}-1)[T_{S}^{-} p_{2n+1}(\vec{N},\xi,t)] +
  (\mathbb{E}^{+}_{F}-1)[T_{F}^{-} p_{2n+1}(\vec{N},\xi,t)]
  \nonumber\\ &+ (\nu^-_{2n+1} \mathbb{E}^{-}_{\xi}-\nu^+_{2n+1})p_{2n+1}(\vec{N},\xi,t)
  + (\nu^+_{2n+1} \mathbb{E}^{+}_{\xi}-\nu^-_{2n+1})p_{2n+1}(\vec{N},\xi,t), \text{ if $\xi\in\{1-n,\dots,-1$\}},\nonumber\\
  \label{eq:ME2}
  \frac{\partial p_{2n+1}(\vec{N},\xi=0,t)}{\partial t}&=(\mathbb{E}^{-}_{S}-1)[T_{S}^{+}p_{2n+1}(\vec{N},0,t)] +
  (\mathbb{E}^{-}_{F}-1)[T_{F}^{+} p_{2n+1}(\vec{N},0,t)]\\
  &+
  (\mathbb{E}^{+}_{S}-1)[T_{S}^{-} p_{2n+1}(\vec{N},0,t)] +
  (\mathbb{E}^{+}_{F}-1)[T_{F}^{-} p_{2n+1}(\vec{N},0,t)]
  \nonumber\\ & \nu^-_{2n+1} p_{2n+1}(\vec{N},-1,t) + \omega^+_{2n+1} p_{2n+1}(\vec{N},1,t) -(\nu^+_{2n+1} + \omega^-_{2n+1})p_{2n+1}(\vec{N},0,t),\nonumber\\
   \label{eq:ME3}
   \frac{\partial p_{2n+1}(\vec{N},\xi,t)}{\partial t}&=(\mathbb{E}^{-}_{S}-1)[T_{S}^{+}p_{2n+1}(\vec{N},\xi,t)] +
  (\mathbb{E}^{-}_{F}-1)[T_{F}^{+} p_{2n+1}(\vec{N},\xi,t)]\\
  &+
  (\mathbb{E}^{+}_{S}-1)[T_{S}^{-} p_{2n+1}(\vec{N},\xi,t)] +
  (\mathbb{E}^{+}_{F}-1)[T_{F}^{-} p_{2n+1}(\vec{N},\xi,t)]
  \nonumber\\ &+ (\omega^-_{2n+1} \mathbb{E}^{-}_{\xi}-\omega^+_{2n+1})p_{2n+1}(\vec{N},\xi,t)
  + (\omega^+_{2n+1} \mathbb{E}^{+}_{\xi}-\omega^-_{2n+1})p_{2n+1}(\vec{N},\xi,t)
  , \text{ if $\xi\in\{1,\dots,n-1$\}},\nonumber
  \end{align}
 \end{subequations}
\end{widetext}
 where $\mathbb{E}^{\pm}_{S/F}$ and $\mathbb{E}^{\pm}_{\xi}$ are shift operators such that
 $\mathbb{E}^{\pm}_{S}f(N_S,N_F,\xi,t)=
 f(N_S\pm 1,N_F,\xi,t)$, $\mathbb{E}^{\pm}_{F}f(N_S,N_F,\xi,t)=
 f(N_S,N_F\pm 1,\xi,t)$ and $\mathbb{E}^{\pm}_{\xi}f(N_S,N_F,\xi,t)=
 f(N_S,N_F,\xi\pm 1,t)$, with  $p_{2n+1}(\vec{N},\xi,t)=0$ whenever $N_S<0$ or $N_F<0$, or $\xi\notin \{-n,\dots,n\}$. The last lines of \eqref{eq:ME1}-\eqref{eq:ME3} encode  environmental switching respectively across harsh/mild states, and through the median state (see Sec.~\ref{Sec:Model-EV}). The transition rates are as in Eq.~\eqref{eq:Transition_rates}, and every time there is a switch $\xi =i \leftrightarrow \xi=i+1$, the carrying capacity appearing in $T_{S/F}^{-}$
 switches  $K_i \leftrightarrow K_{i+1}$ according to Eq.~\eqref{eq:Kswitch}.

 The class of individual-based multi-state switching models defined by Eqs.~\eqref{eq:Transition_rates}-\eqref{eq:Ki} satisfies 
the  master equation \eqref{eq:ME}, and 
has been simulated using the Gillespie algorithm~\cite{Gillespie76}; see  Appendix ~\ref{appendix:simulations}.

In Fig.~\ref{fig:Fig2-hist}, we show the histograms of the marginal quasi-stationary PSD $p_{\nu_{2n+1}}(N)$ for $t> 1/s$
obtained by marginalising the joint probability with respect to the environmental states and population composition, i.e.
$p_{\nu_{2n+1}}(N)=\sum_{i=-n}^{n} \sum_{\vec{N}}
\delta_{N_S+N_F=N}~ p_{2n+1}(\vec{N},\xi=i,t> 1/s)$, where  $\delta_{N_S+N_F=N}$ is the Kronecker delta function ensuring that the population size is $N$.

\section{Unconditional mean fixation time of in static and  fluctuating environments}
\label{appendix:MFT}
In this appendix, we discuss the properties of the mean time for either $S$ or $F$ to take over the entire population, i.e. the  unconditional mean
fixation time (uMFT). We first compute the uMFT  in a static environment, with a constant carrying capacity $K=\bar{K}$,
using an approximation based on the Moran process. We then use this result to study the uMFT in fluctuating
environments with a multi-state switching carrying capacity.
\subsection{uMFT in static environment: the Moran approximation}
\label{appendix:MFT-static}
The uMFT can  be computed exactly for a population of constant size $\bar{K}$ evolving according to the classical Moran process~\cite{Moran,Ewens}; see Sec.~\ref{Sec:Static-Moran}. When the initial fraction of $S$ individuals is $x_0=k/\bar{K}$, its expression, here denoted by $\tau_M$, defined as the expected value of the unconditional fixation times ${\rm min}\{t\geq 0: N_S(t)=\bar{K} \text{ or } N_S(t)=0\}$, reads~\cite{antal2006fixation,traulsen2009stochastic,Assaf2010,Taitelbaum2020,Asker2023,Asker2025} 
\begin{align}
 \label{eq:MFT_M}
\tau_{M}(x_0,\bar{K})&=  \phi_{M}\left(\frac{k}{\bar{K}},\bar{K}\right) \sum_{j=k}^{\bar{K}-1}\sum_{\ell=1}^{j} \frac{(1-s)^{\ell-j}}{\widetilde{T}_S^+\left(\frac{\ell}{\bar{K}},\bar{K}\right)} \\&- \left(1-\phi_{M}\left(\frac{k}{\bar{K}},\bar{K}\right)\right) \sum_{j=1}^{k-1}\sum_{\ell=1}^{j} \frac{(1-s)^{\ell-j}}{\widetilde{T}_S^+\left(\frac{\ell}{\bar{K}},\bar{K}\right)},\nonumber
 \end{align}
where $\phi_{M}\left(x_0=\frac{k}{\bar{K}},\bar{K}\right)$
is given by Eq.~\eqref{eq:phi_M} and $\widetilde{T}_S^+\left(\frac{\ell}{\bar{K}},\bar{K}\right)$ is from Eq.~\eqref{eq:MoRates} with $x=\ell/\bar{K}$. When
$0<s\ll 1$,
$\bar{K}s\gg 1$, with $x_0={\cal O}(1)$ (weak selection, large population, and initial condition well separated from the absorbing boundaries $0$ and $1$), we  have $\tau_{M}(x_0,\bar{K})\sim (\ln{\bar{K}})/s$: The uMFT has a weak dependence on
$\bar{K}$, with fixation occurring on a timescale $1/s$~\cite{Ewens,Blythe07,Wienand2017,Wienand2018,west2020,Asker2025}.

The Moran uMFT, given by Eq.~\eqref{eq:MFT_M}, is a good approximation of the uMFT in a static environment with a constant carrying capacity, $K=\bar{K}$, where the total population size fluctuates about  $\bar{K}$, i.e. $N\approx\bar{K}$. In this case, when $\bar{K}s\gg 1$ and  $0<s\ll 1$ (and $x_0={\cal O}(1)$), fixation occurs on a timescale $1/s$~\cite{melbinger2010,cremer2011,Wienand2017,Wienand2018,Taitelbaum2020,Hernandez2023,Asker2023,Hernandez2024,Asker2025,Hernandez2026}.

\begin{figure*}
\centering
\setlength{\unitlength}{1cm}
\begin{overpic}[width=0.32\textwidth]{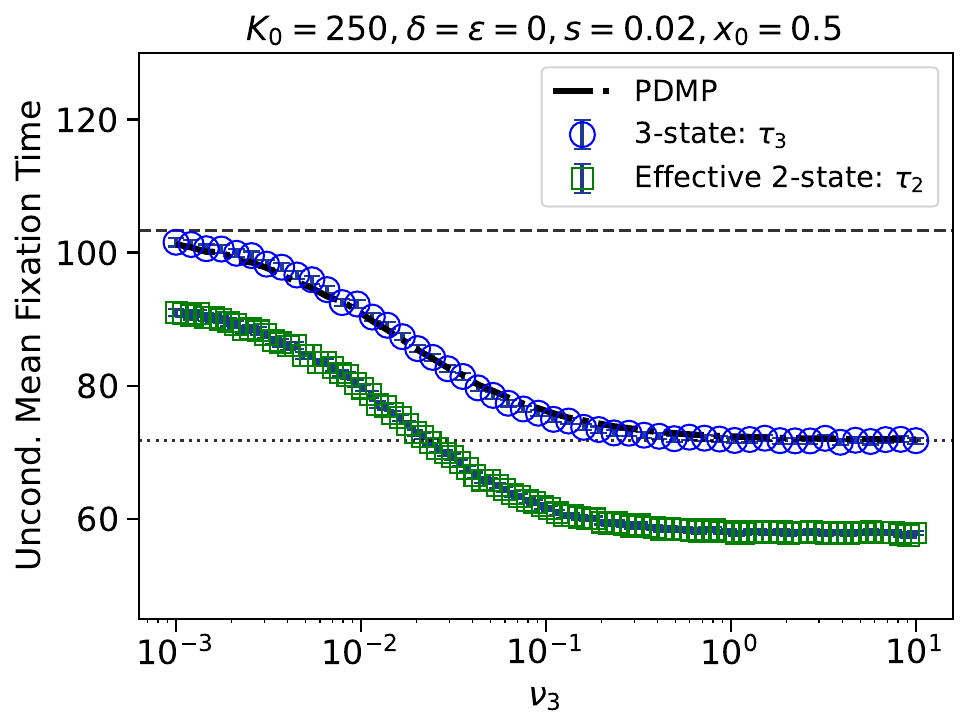
}
    \put(12,76){(a)}
\end{overpic}
\hfill
\begin{overpic}[width=0.32\textwidth]{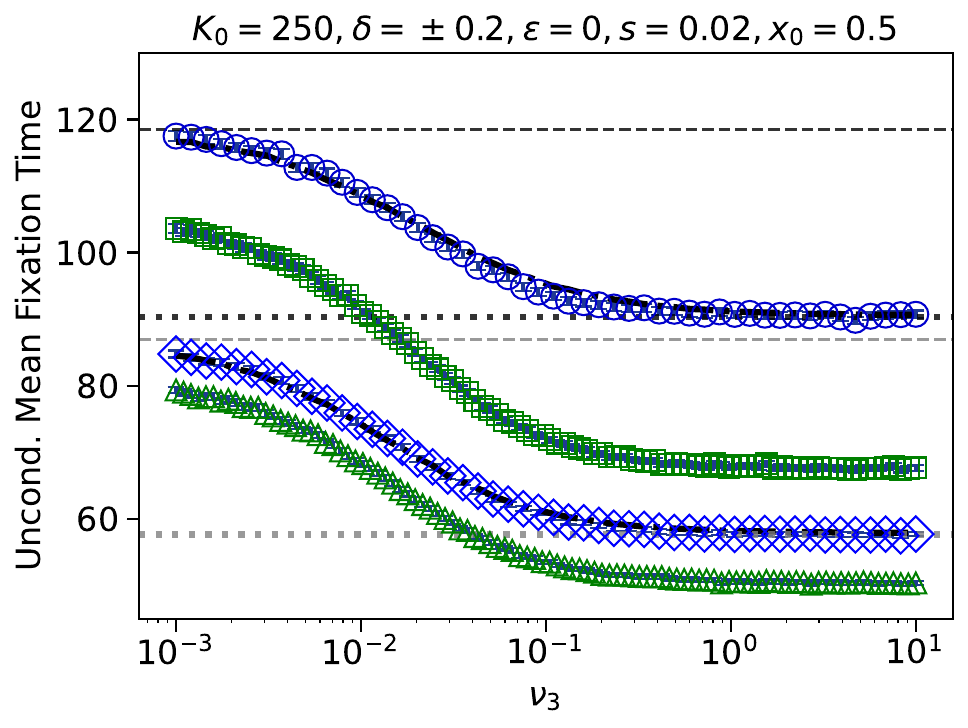
}
    \put(12,76){(b)}
\end{overpic}
\hfill
\begin{overpic}[width=0.32\textwidth]{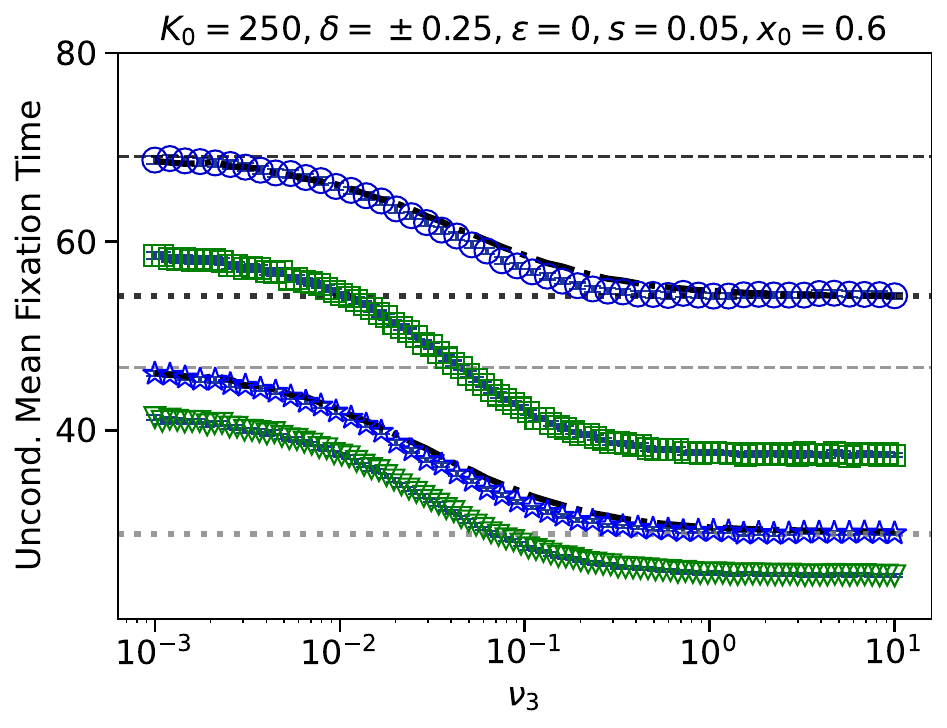
}
    \put(10,76){(c)}
\end{overpic}
\hfill
\vspace{1.5em}

\begin{overpic}[width=0.32\textwidth]{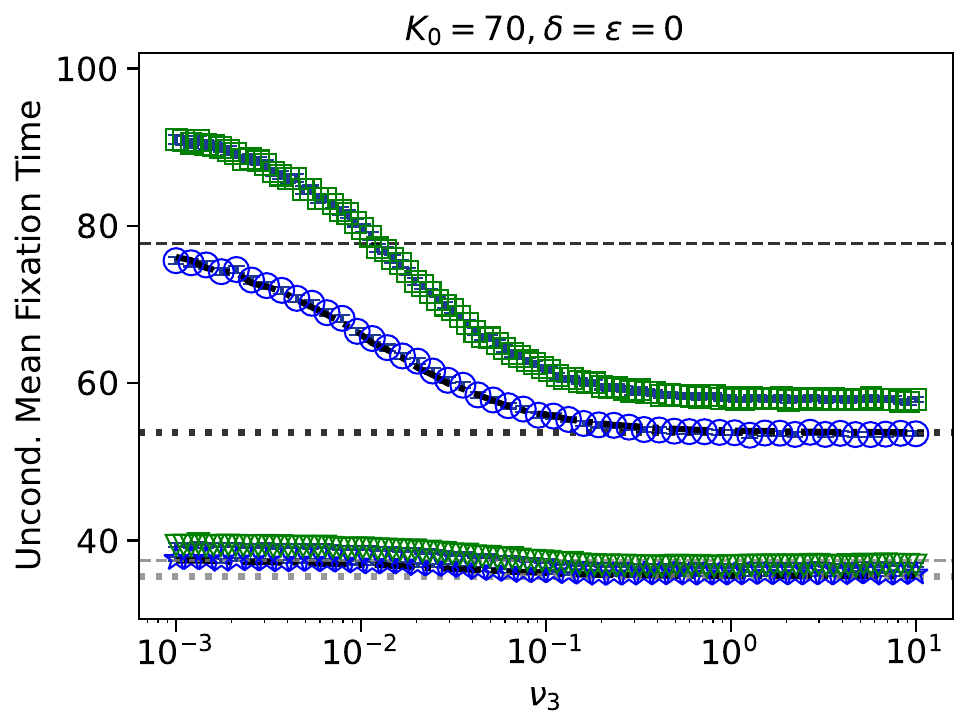
}
    \put(12,76){(d)}
\end{overpic}
\hfill
\begin{overpic}[width=0.32\textwidth]{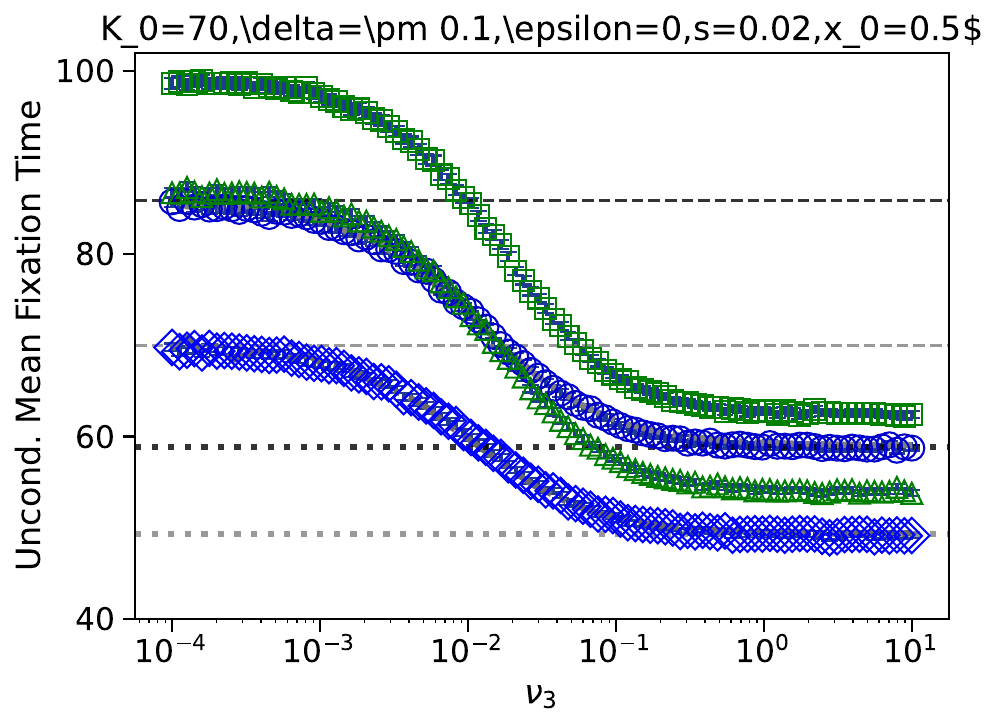
}
    \put(12,76){(e)}
\end{overpic}
\hfill
\begin{overpic}[width=0.32\textwidth]{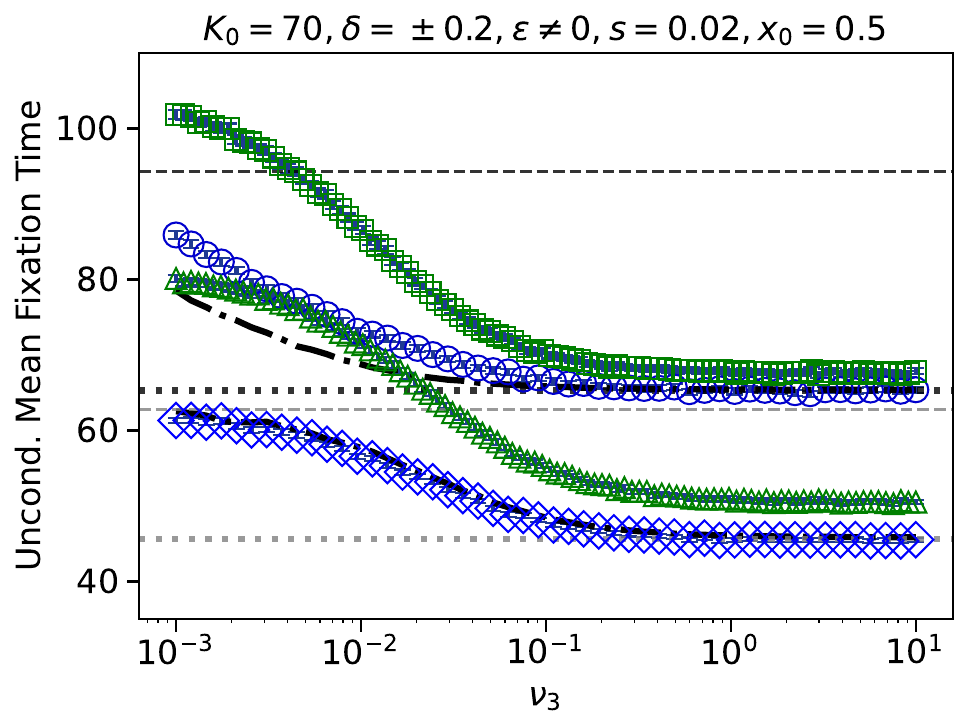
}
    \put(12,76){(f)}
\end{overpic}
\caption{Unconditional mean fixation time   in the three-state switching model as a function of $\nu_3$.
The centre state carrying capacity is  $K_0=250$ in (a)-(c), while  $K_0=70$ in (d)-(f). Here, $n=1$, and  
$(K^-,K^+)=(50,450)$ in all panels except (c) where $(K^-,K^+)=(25,475)$.
Symbols are from full stochastic simulations and black dashed-dotted curves (in all panels), often almost indistinguishable from markers, are from the $N$-PDMP-based approximation; see Eq.~\eqref{eq:phiPDMP}. 
In blue, is shown $\tau_{3}$ vs. $\nu_3$  for different parameter sets  $(\delta,\epsilon, s, x_0)$ that  are the same as in Fig.~\ref{fig:Fig5_n1} (see also the legends).
Green symbols  show simulation data  for  $\tau_2$, the uMFT of the effective two-state model; see text and Eq.~\eqref{eq:nutilde}. In (b,c,e,f), results for $\tau_2$ are shown as triangles when $\delta=-0.2$ and squares when  $\delta=0.2$. In (d), simulations data of $\tau_2$  are shown as squares for $(s, x_0)=(0.02,0.5)$ and downside triangles for $(s, x_0)=(0.1,0.7)$. For the $N$-PDMP-based approximation of  $\tau_2$, see Refs.~ \cite{Wienand2017,Wienand2018,Taitelbaum2020}.
The horizontal dashed and dotted lines are eyeguides showing
 $\tau_3^{0}$ (dashed) and $\tau_3^{\infty}$ (dotted).
Error bars are included but 
typically too small to see.}
\label{fig:Fig7_FigMFT_n1}
\end{figure*}
\subsection{uMFT under  environmental switching}
\label{appendix:MFT-switching}
Similarly to what is done in Sec.~\ref{Sec:Fixation} for the $S$ fixation probability, the Moran uMFT, given by Eq.~\eqref{eq:MFT_M},  can be used to obtain suitable approximations  of  the uMFT
under  $(2n+1)$-state environmental switching,  denoted by $\tau_{2n+1}$ and defined as the expected value of ${\rm min}\{t\geq 0: N_S(t)=N(t)>0 \text{ or } N_F(t)=N(t)>0 \}$. To study $\tau_{2n+1}$, it is again convenient to distinguish the regimes $\nu_{2n+1}\ll s$, $\nu_{2n+1}\gg s$, and $\nu_{2n+1}\sim s$.
Since the uMFT in static environments scales with $1/s$ for weak selection and  a sufficiently large population,
we expect a similar behaviour in fluctuating
environments for all  $n$, when $0<s\ll 1$, $s\langle N\rangle > 1$, and $x_0={\cal O}(1)$, as in the binary case~\cite{Wienand2017,Wienand2018,Taitelbaum2020,Asker2025}.

\begin{figure*}
\centering

\setlength{\unitlength}{1cm}

\begin{overpic}[width=0.32\textwidth]{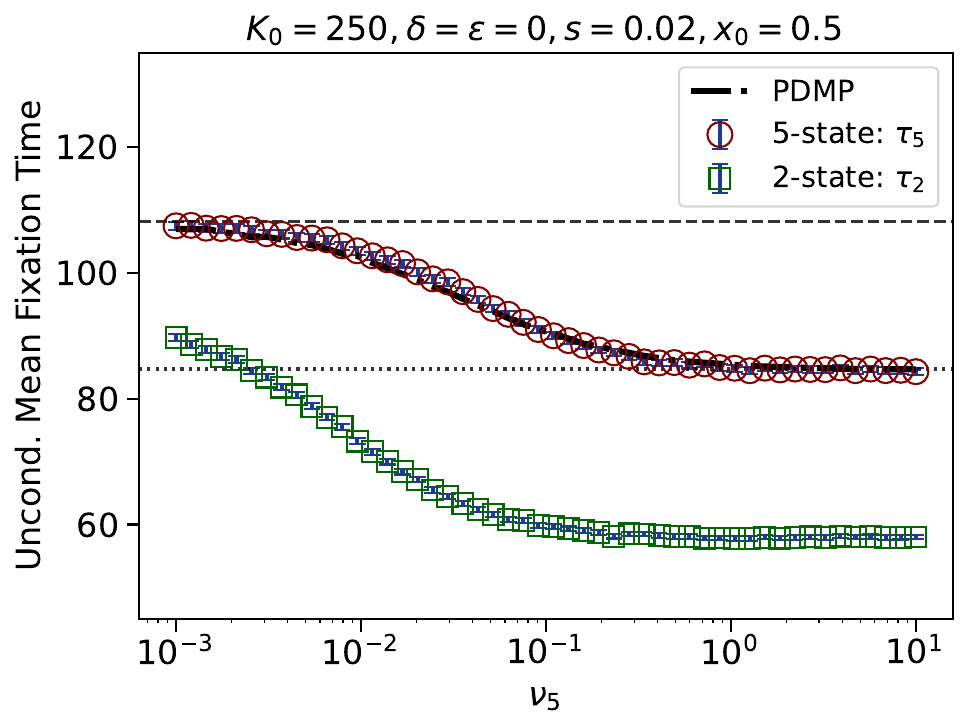
}
    \put(12,76){(a)}
\end{overpic}
\hfill
\begin{overpic}[width=0.32\textwidth]{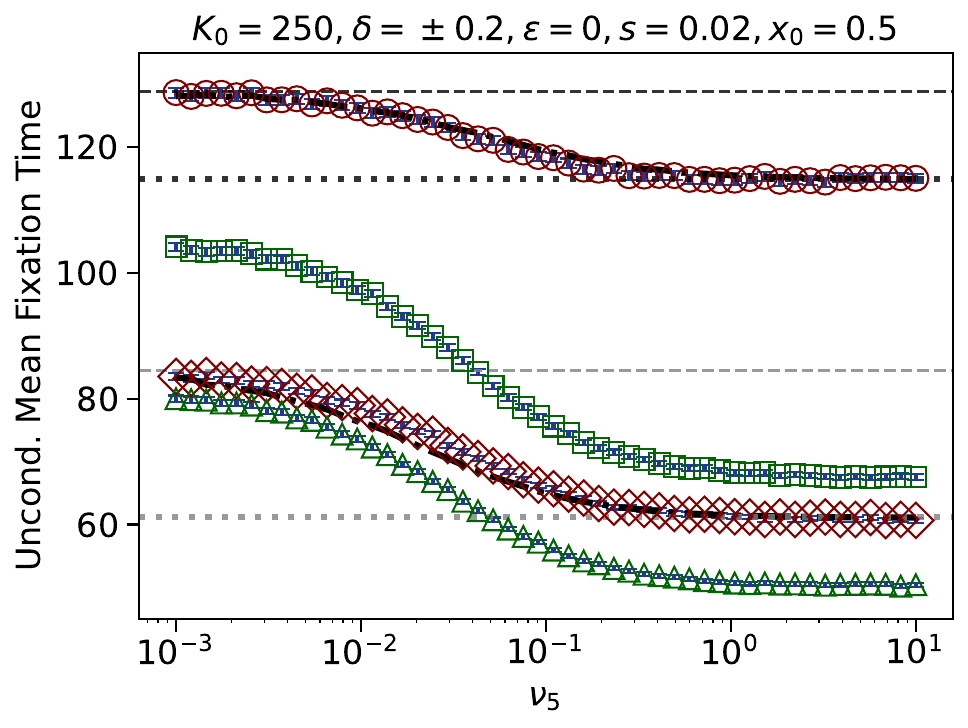
}
    \put(12,76){(b)}
\end{overpic}
\hfill
\begin{overpic}[width=0.32\textwidth]{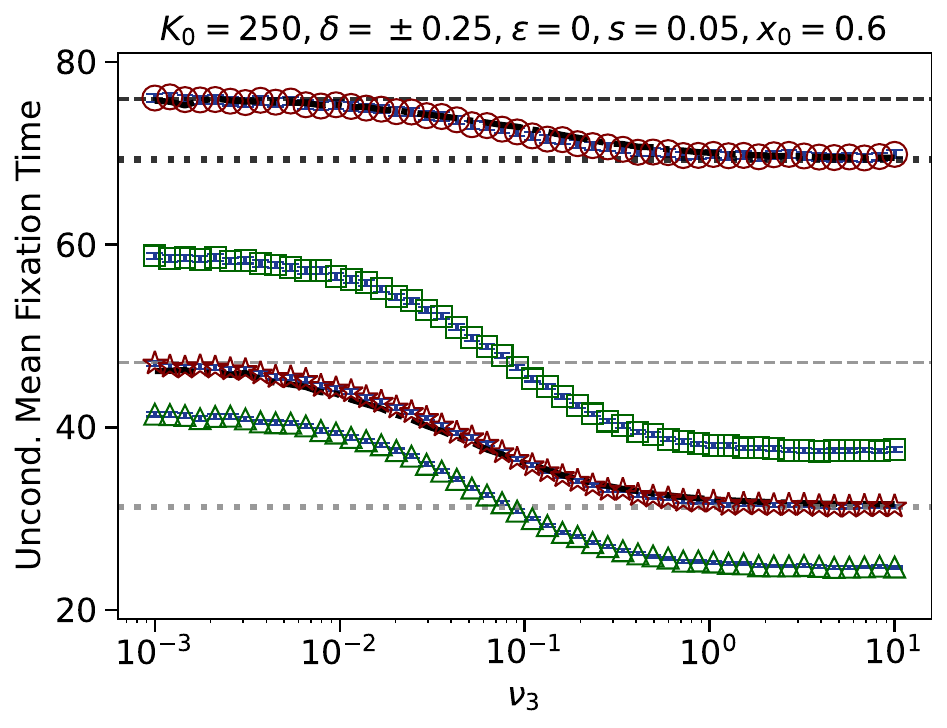
}
    \put(10,76){(c)}
\end{overpic}
\hfill
\vspace{1.5em}

\begin{overpic}[width=0.32\textwidth]{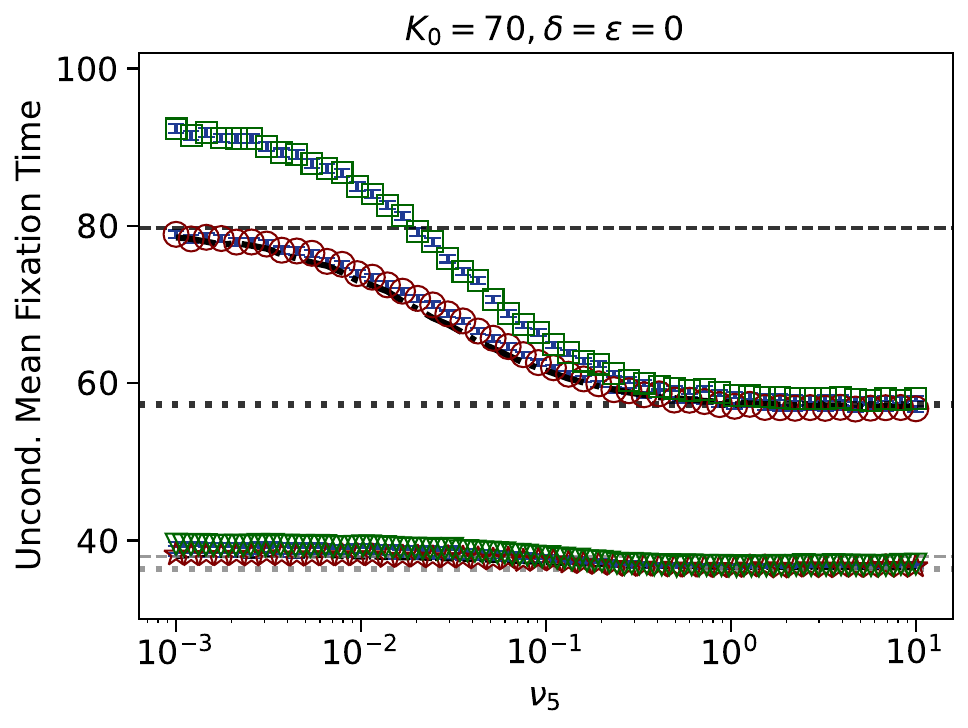
}
    \put(12,76){(d)}
\end{overpic}
\hfill
\begin{overpic}[width=0.32\textwidth]{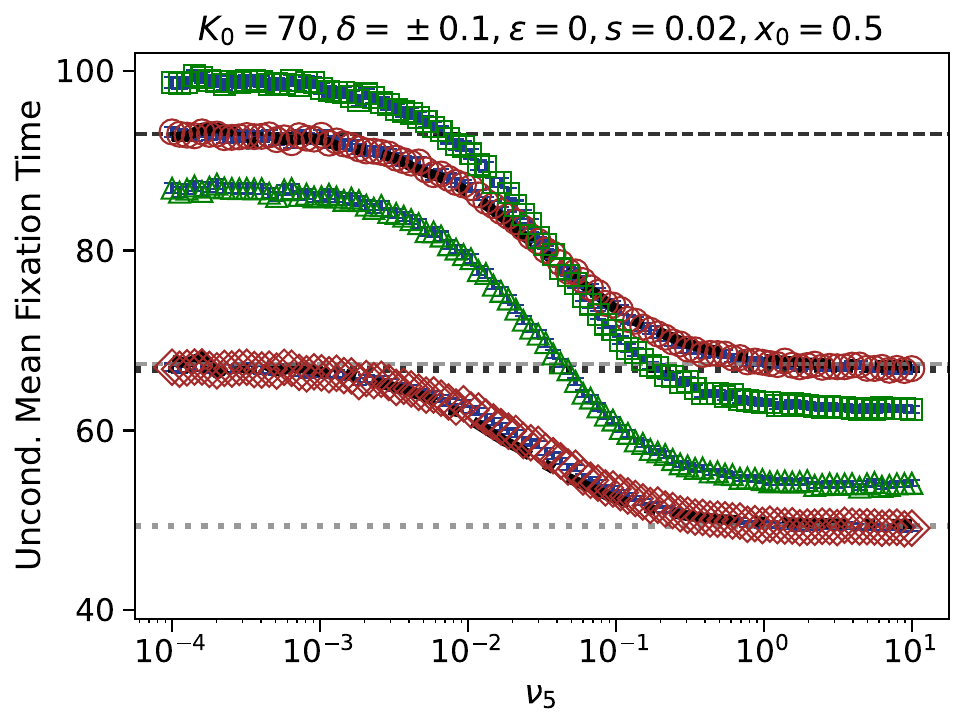
}
    \put(12,76){(e)}
\end{overpic}
\hfill
\begin{overpic}[width=0.32\textwidth]{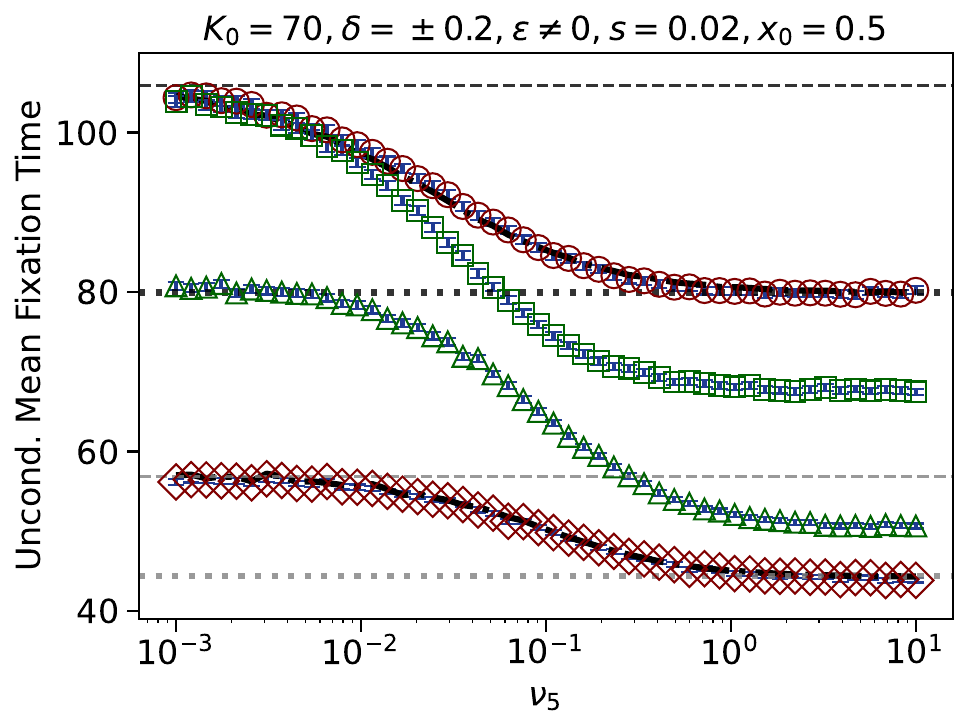
}
    \put(12,76){(f)}
\end{overpic}
\caption{Unconditional mean fixation time in the five-state switching model as a function of $\nu_5$.
The centre state carrying capacity is  $K_0=250$ in (a)-(c), while  $K_0=70$ in (d)-(f). Here, $n=1$, and  
$(K^-,K^+)=(50,450)$ in all panels except (c) where $(K^-,K^+)=(25,475)$.
Symbols are from full stochastic simulations and dashed-dotted curves (in all panels),  almost indistinguishable from markers, are from the $N$-PDMP-based approximation; see Eq.~\eqref{eq:phiPDMP}. 
In brown, is shown $\tau_{5}$ vs. $\nu_5$  for different parameter sets   $(\delta,\epsilon, s, x_0)$ that  are the same as in Fig.~\ref{fig:Fig6_n2} (see also the legends).
Green symbols  show simulation data  for  $\tau_2$, the uMFT of the effective two-state model; see text and Eq.~\eqref{eq:nutilde}. In (b,c,e,f), results for $\tau_2$ are shown as triangles when $\delta=-0.2$ and squares when  $\delta=0.2$. In (d), simulations data of $\tau_2$  are shown as squares for $(s, x_0)=(0.02,0.5)$ and downside triangles for $(s, x_0)=(0.1,0.7)$. For the $N$-PDMP-based approximation of  $\tau_2$, see Refs.~ \cite{Wienand2017,Wienand2018,Taitelbaum2020}.
The horizontal dashed and dotted lines are eyeguides showing 
$\tau_5^{0}$ (dashed) and $\tau_5^{\infty}$ (dotted); see Eq.\eqref{eq:tau2n+1}. 
Error bars are too small to see.
\label{fig:Fig8_FigMFT_n2}
}
\end{figure*}
In the slow switching regime ($\nu_{2n+1}\ll s$), 
it is likely that no
environmental switches occur prior to fixation. In this regime,
a Moran approximation for the uMFT is obtained from
Eq.~\eqref{eq:MFT_M} for a population of
size $N\approx K_i$ with a probability $\pi_i$, yielding
$\tau_{2n+1}^0=\sum_{i=-n}^{n} \tau_M(K_i)~\pi_i$ (compare with Eq.~\eqref{eq:phi0}).  For notational simplicity, we have dropped the dependence on $x_0$.

Under fast environmental switching ($\nu_{2n+1}\gg s$), the  carrying capacity self-averages and the population experiences the
effective carrying capacity ${\cal K}_{2n+1}$ given by Eq.~\eqref{eq:curlyK}. A Moran approximation of the uMFT is thus given by Eq.~\eqref{eq:MFT_M} for a population of
size $N\approx {\cal K}_{2n+1}$, yielding $\tau_{2n+1}^{\infty}=\tau_M({\cal K}_{2n+1})$ (compare with Eq.~\eqref{eq:phiinf}).

In the intermediate switching regime ($\nu_{2n+1}\sim s$), when $s\ll 1$,
we can use the $N$-PDMP-based approximation employed in Sec.~\ref{Sec:Fixation}; see Eq.~(\ref{eq:phiPDMP}). Exploiting the timescale separation between $N$ and $x$, and using the $N$-PDMP density (see Eq.~\eqref{eq:PDMP-nstate}) to approximate the
 stationary PSD,   
we can write $\tau_{2n+1}^{\text{PDMP}}\equiv \int_{K^-}^{K^+} \tau_M(N)~p^{\text{PDMP}}_{\nu'_{2n+1}/s}(N)~dN$,
%
using Eq.~\eqref{eq:nuprime}~\cite{Wienand2017,Wienand2018,west2020,Taitelbaum2020,Asker2025}.

The uMFT in the fluctuating multi-state switching models  with
$(2n+1)$ environmental states can therefore be approximated by
\begin{equation}
  \label{eq:tau2n+1}
\tau_{2n+1}(\nu)\approx
\begin{cases}
\sum_{i=-n}^{n} \tau_M(K_i)~\pi_i & (\nu\ll s),\\
\int_{K^-}^{K^+} \tau_M( N)~p^{\text{PDMP}}_{\nu_{2n+1}'/s}(N)~dN & (\nu\sim s), \\
\tau_M({\cal K}_{2n+1}) & (\nu\ll s),
\end{cases}
\end{equation}
using Eqs.~\eqref{eq:MFT_M}, \eqref{eq:pi}, \eqref{eq:Ki}, and \eqref{eq:curlyK}.
Simulation results of Figs.~\ref{fig:Fig7_FigMFT_n1} and \ref{fig:Fig8_FigMFT_n2} show that Eq.~\eqref{eq:tau2n+1}
is generally a good approximation as it faithfully captures the main features of the uMFT across all switching regimes. As expected, the uMFT is found to scale as $1/s$, with
$\tau_{2n+1}\approx \tau_{2n+1}^0\sim 1/s$ when $\nu_{2n+1}\ll s$ and  $\tau_{2n+1}\approx \tau_{2n+1}^{\infty}\sim 1/s$ when $\nu_{2n+1}\gg s$, as shown by the dotted and dashed lines. Moreover, $\tau_{2n+1}^{\text{PDMP}}$ correctly predicts that
the uMFT is a decreasing function of $\nu_{2n+1}$ 
(all other parameters kept fixed) in Figs.~\ref{fig:Fig7_FigMFT_n1} and \ref{fig:Fig8_FigMFT_n2}, and we find $\tau_{2n+1}\approx \tau_{2n+1}^{\text{PDMP}}\sim 1/s$ in all switching regimes. 
The uMFT at a given $\nu_{2n+1}$ increases with $n$ (all other parameters being kept fixed); compare, e.g., Figs.~\ref{fig:Fig7_FigMFT_n1}(a) and \ref{fig:Fig8_FigMFT_n2}(a).
Similarly, at a given $\nu_{2n+1}$, $\tau_{2n+1}$  increases with $\delta$; see, e.g.,  Fig.~\ref{fig:Fig7_FigMFT_n1}(b). 
Simulation results in Figs.~\ref{fig:Fig7_FigMFT_n1}, \ref{fig:Fig8_FigMFT_n2}
and Fig.\ref{fig:Fig10}(b) 
show no noticeable  dependence on the parameter $\epsilon$, when $-0.8\leq \epsilon\leq 5$.

\section{Simulation methods}
\label{appendix:simulations}
\begin{figure*}
\centering

\setlength{\unitlength}{1cm}

\begin{overpic}[width=0.45\textwidth]{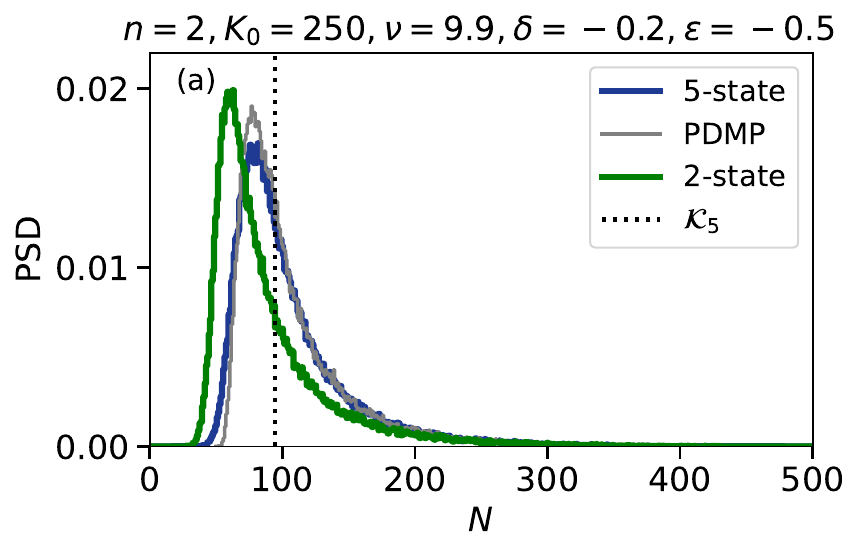}
\end{overpic}
\hfill
\begin{overpic}[width=0.45\textwidth]{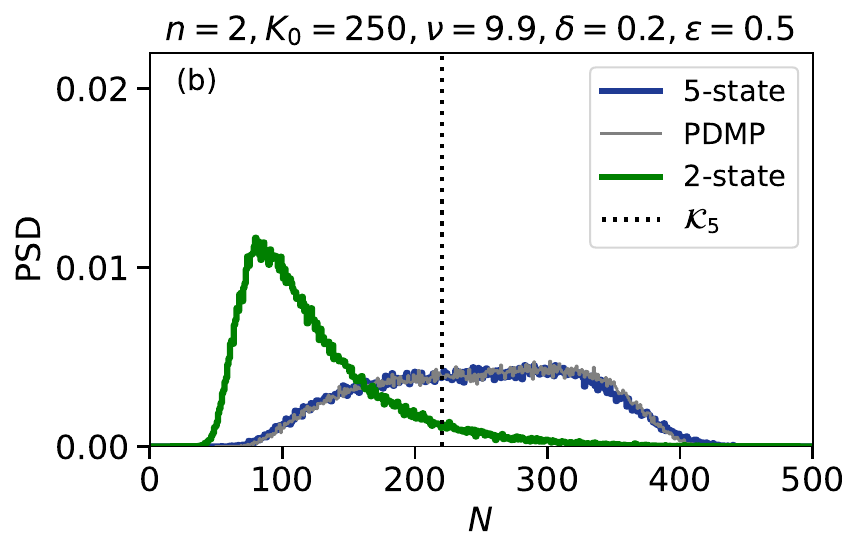}
\end{overpic}
\caption{Supplementary figure showing in dark blue the  PSD of the five-state switching model ($n=2$), $p_{\nu_5}(N)$ for $\nu_5=9.9$ (fast switching), with 
$(s,x_0,K^-,K_0,K^+)=(0.02,0.5,50,250,450)$. 
Other parameters are: $(\delta,\epsilon)=(0.2,0.5)$ in (a) and $(\delta,\epsilon)=(-0.2,-0.5)$ in  (b).
In grey is shown the $N$-PDMP approximation of the PSD (see Eq.~\eqref{eq:PDMP-nstate}), while the green histograms are from the effective two-state model (see text). Dotted lines are eyeguides indicating ${\cal K}_5\approx 94.70$ (a) and  ${\cal K}_5\approx220.18$ (b). 
These histograms have been obtained for $t> 2000$ (Appendix~\ref{appendix:simulations}) and have to be compared with those of Fig.~\ref{fig:Fig2-hist}(g),(h).  
}
\label{fig:Fig9}
\end{figure*}

\begin{figure*}
\centering

\setlength{\unitlength}{1cm}

\begin{overpic}[width=0.45\textwidth]{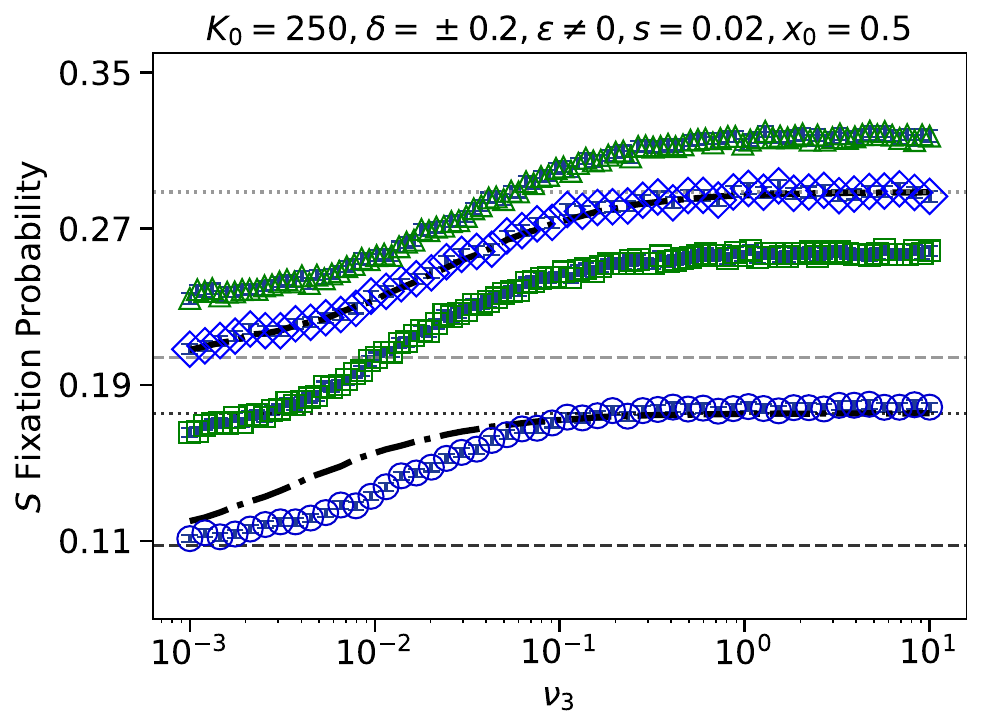}
    \put(1,73){(a)}
\end{overpic}
\hfill
\begin{overpic}[width=0.45\textwidth]{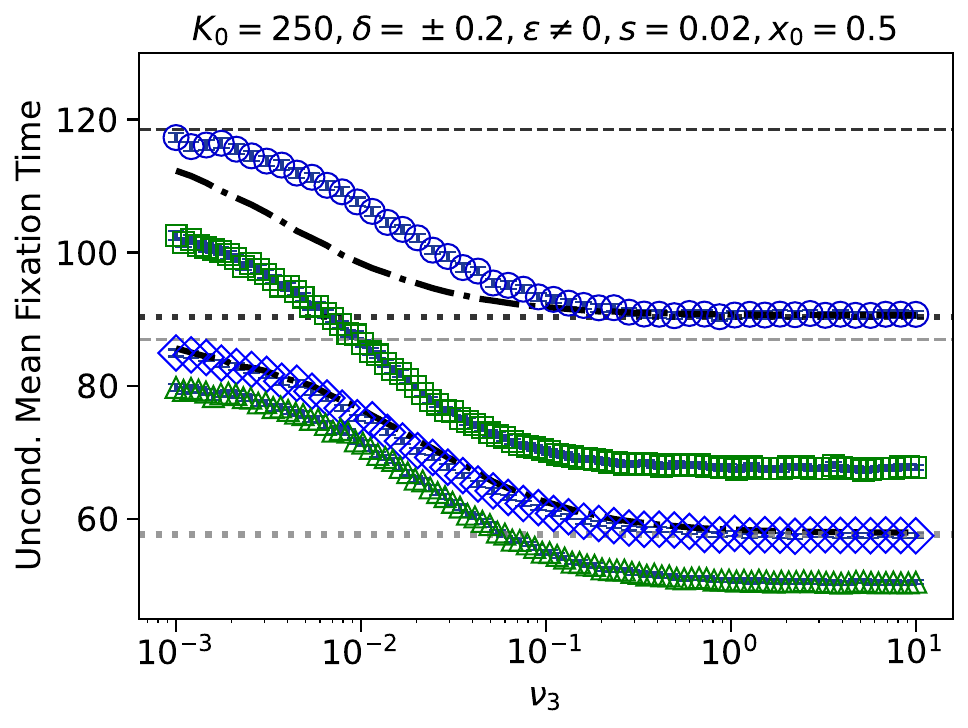}
    \put(1,73){(b)}
\end{overpic}
\caption{Supplementary figure showing in blue $\phi_{3}$ and $\tau_{3}$ vs $\nu_3$ under ternary switching  for $(\delta,\epsilon)=(0.2,5)$ (circles) and  $(\delta,\epsilon)=(-0.2,-0.5)$ (diamonds). Other parameters are
$(s,x_0,n, K^-,K_0,K^+)=(0.02,0.5,1,50,250,450)$.
The dashed-dotted  black curves show the $N$-PDMP approximations $\phi_{3}^{\text{PDMP}}$ in (a) and $\tau_{3}^{\text{PDMP}}$ in (b); see Eqs.~\eqref{eq:phiPDMP} and \eqref{eq:tau2n+1}. For $(\delta,\epsilon)=(-0.2,-0.5)$, $\phi_{3}^{\text{PDMP}}$ and $\tau_{3}^{\text{PDMP}}$ are almost indistinguishable from 
$\phi_{3}$ and $\tau_{3}$.
The green symbols show $\phi_{2}$ in (a) and $\tau_{2}$ in (b) for 
$(\delta,\epsilon)=(0.2,5)$ (squares) and  $(\delta,\epsilon)=(-0.2,-0.5)$ (triangles). 
}
\label{fig:Fig10}
\end{figure*}

To complement  our analytical  approaches, we have performed extensive stochastic simulations of the multi-state switching models using the Gillespie algorithm~\cite{Gillespie76}. These exactly mirrors the continuous-time  dynamics encoded in the master equation~\eqref{eq:ME}. The implementation of the simulations was efficiently optimised through just-in-time compilation (Numba)~\cite{codes-data-MM}. In Sections ~\ref{Sec:n1} and \ref{Sec:n2}, we have specifically focused on the switching models with three ($n=1$) and five ($n=2$) environmental states, whose PSD, average population size, and fixation properties were computed; see Figs.~\ref{fig:Fig2-hist} and \ref{fig:Fig3-meanN}, and Figs.~\ref{fig:Fig5_n1}-\ref{fig:Fig10}. The PSD, $S$ fixation probability and uMFT
 have been compared with their binary-state counterparts~\cite{Wienand2017,Wienand2018,Taitelbaum2020}.
The multi-state environmental noise $\xi$ and carrying capacity $K$; see Eqs.~\eqref{eq:Korder}-\eqref{eq:Ki} and 
\eqref{eq:K}, are always at stationarity, and therefore each simulation starts a time \(t=t_0=0\) with an initial value of $\xi(t_0)$ and carrying capacity $K(t_0)$ drawn from their  stationary distribution ${\bm \pi}$; see Eq.~\eqref{eq:pi}. In each run, the population has an initial  size  $N(t_0)=\langle K\rangle_{2n+1}$ that coincides with the average carrying capacity, and its composition is $N_S(t_0)=\text{round}\left(x_0 N(t_0)\right)$ and $N_F(t_0)= N(t_0)-N_S(t_0)$, where $x_0 N(t_0)$ is rounded to the nearest integer. and
Simulations were run in batches of $10^5$ realizations for each set of parameters $\{n,K^-,K^+,K_0,s,x_0,\nu_{2n+1},\delta,\epsilon\}$ by accounting for all the possible reactions that can take place at each time increment. For the class of individual-based models studied here, these are the four possible birth or death reactions with rates \(T^{\pm}_{S/F}(\vec{N}(t))\); see Eq.~\eqref{eq:Transition_rates}, and the switching of the multi-state carrying capacity $K(t)$ driven by the environmental switches of $\xi$.


In Figs.~\ref{fig:Fig5_n1}-\ref{fig:Fig8_FigMFT_n2} and \ref{fig:Fig10},
for each set $\{n,K^-,K^+,K_0,\delta,\epsilon\}$, simulation data 
for $\phi_3,\phi_5,\tau_3$ and $\tau_5$ are shown for $50$ values of $\nu_{3,5} \in [10^{-3}, 10]$ in all panels, excepts in panels (e) of these figures consisting of 100 values of $\nu_{3,5} \in [10^{-4}, 10]$. The panels (a)-(d),(f) of Figs.~\ref{fig:Fig5_n1} and \ref{fig:Fig7_FigMFT_n1} also contain simulation results for 
$\phi_2$ and $\tau_2$ obtained for $100$ values of $\nu_{3} \in [10^{-3}, 10]$ in panels (a)-(d),(f),  and for  $100$ values of $\nu_{3} \in [10^{-4}, 10]$ in Fig.~\ref{fig:Fig5_n1}(e) and \ref{fig:Fig7_FigMFT_n1}(e).  The panels (a)-(d),(f) of Figs.~\ref{fig:Fig6_n2} and \ref{fig:Fig8_FigMFT_n2} contain simulation data of 
$\phi_2$ and $\tau_2$ obtained for $50$ values of $\nu_{5} \in [10^{-3}, 10]$ in panels (a)-(d),(f),  and for  $100$ values of $\nu_{5} \in [10^{-4}, 10]$ in Fig.~\ref{fig:Fig6_n2}(e) and \ref{fig:Fig8_FigMFT_n2}(e). 
  Each data point in  Figs.~\ref{fig:Fig2-hist} and \ref{fig:Fig3-meanN}, and Figs.~\ref{fig:Fig5_n1}-\ref{fig:Fig10}  has been obtained by sampling ${\cal R}= 10^5$ realizations. 
The statistical errors on the simulation results on the fixation probability $\phi_{2n+1}$ and $\phi_{2}$
of Figs.~\ref{fig:Fig5_n1}, \ref{fig:Fig6_n2} and  \ref{fig:Fig10}
has been estimated using the classical Wald confidence interval method. Accordingly, the estimated error on  sample-averaged value $\hat{\phi}$ of $\phi$ is 
$z_{\alpha}~\sqrt{\frac{\hat{\phi}(1-\hat{\phi})}{{\cal R}}}$,
where we have chosen $z_{\alpha}=1.96$, which corresponds to a $95\%$ confidence interval. For  ${\cal R}= 10^5$, this yields  estimated errors of order $2\cdot 10^{-3}$, i.e around $1\%$ to $2\%$ of $\hat{\phi}$ in Figs.~\ref{fig:Fig5_n1}, \ref{fig:Fig6_n2} and  \ref{fig:Fig10}. These statistical errors are almost unnoticeable, suggesting that the statistical results are robust.

The histograms $p_{m}(N)$, with $m\in\{\nu_{2n+1},\widetilde{\nu}\}$ and $p^{\text{PDMP}}_{\nu_{2n+1}}(N)$
of Figs.~\ref{fig:Fig2-hist} and \ref{fig:Fig9} have been obtained by sampling a long trajectory of $N(t)$ and $N^{\text{PDMP}}(t)$ after discarding an initial transient, i.e. for $t>t_{\text{burn}}=2000$ to ensure that $N(t)$ and $N^{\text{PDMP}}(t)$ have settled in their long-time dynamics. For the full individual-based switching models, the long trajectory $N(t)$ is obtained from  Gillespie simulations. For the $N$-PDMP, a long trajectory $N^{\text{PDMP}}(t)$ is generated by implementing the environmental switching according to the Gillespie algorithm and, between each switch, by solving  the logistic equation $\dot{N}^{\text{PDMP}}=N^{\text{PDMP}}(1-N^{\text{PDMP}}/K_i)$  in the current environmental state $i$;  see Eq.~\eqref{eq:PDMP-nstate}. The histograms representing $p_{m}$ were thus obtained by taking ${\cal R}\gg 1$ samples each spaced by $\Delta t$: 
$p_{m}(N)\approx \frac{1}{{\cal R}} \sum_{j=1}^{{\cal R}} \delta_{N,N_j}$,
where $N_j = N(t_{\text{burn}} + j\Delta t)$, and $\delta_{N,N_j}=1$ when $N=N_j$ and 0 otherwise. Similarly, the stationary $N$-PDMP density $p_{\nu_{2n+1}}^{\text{PDMP}}(N)$ is approximated numerically from a long $N$-PDMP trajectory (after discarding an initial transient  $t_{\text{burn}}$) by collecting a large number ${\cal R}$ of samples, each spaced by $\Delta t$. These  define the empirical measure $\frac{1}{{\cal R}}\sum_{j=1}^{\cal R} \delta(dN -N_j)$, where $\delta(\cdot)$ is a Dirac delta function. Computationally, 
the sampled values are grouped into bins of unit width, 
and the resulting counts are normalised to construct the histogram approximations of the stationary $N$-PDMP density shown in Fig.~\ref{fig:Fig2-hist}.
For each sampled trajectory, the total simulated time is therefore 
$t_{\text{burn}} + {\cal R} \,\Delta t$. In Figs.~\ref{fig:Fig2-hist} and \ref{fig:Fig3-meanN},  
we have used ${\cal R}=10^5$ and $\Delta t=10$. 
Within this approach, the  average population size of Fig.~\ref{fig:Fig3-meanN} was
computed as the discrete time average from  a long trajectory of $N(t>t_{\text{burn}})$ and $N^{\text{PDMP}}(t>t_{\text{burn}})$ 
according to $\langle N \rangle_{2n+1} \approx \frac{1}{{\cal R}} \sum_{j=1}^{{\cal R}} N(t_{\text{burn}} + j\Delta t)$ and $\langle N \rangle_{2n+1}^{\text{PDMP}} \approx \frac{1}{{\cal R}} \sum_{j=1}^{{\cal R}} N^{\text{PDMP}}(t_{\text{burn}} + j\Delta t)$, with $t_{\text{burn}}=5000$. The  estimated 
error on $\langle N \rangle$ 
decreases with the number ${\cal R}_{\text{ind}}$ of independent samples, as $1/\sqrt{{\cal R}_{\text{ind}}}$. This implies that this approach is  well-suited to efficiently compute the average population size when the switching rate is not too low. In  Fig.~\ref{fig:Fig3-meanN}, for ${\cal R}= 10^5$ and $\Delta t= 10$, the estimated error scales as $10^{-3}/\sqrt{\nu_{2n+1}}$ and hence decreases by a factor 100 when $\nu_{2n+1}$ varies from $0.01$ to $100$. In the limit $\nu_{2n+1} \to 0$, the average population size is given  by
$\langle N\rangle_{2n+1} \approx \langle K \rangle_{2n+1}$; see Fig.~\ref{fig:Fig3-meanN} (dotted lines)

The number of independent samples ${\cal R}_{\text{ind}}$ is large when ${\cal R}\gg 1$
and the switching rate is not too low. Therefore,  within the approach outlined above, the $N$-PDMP-based approximation of the fixation probability and uMFT are efficiently computed 
by averaging over ${\cal R}$ samples of a $N-$PDMP trajectory $N^{\text{PDMP}}(t>t_{\text{burn}})$  obtained for the rescaled switching rate $\nu'/s$ (see Eq.~\eqref{eq:nuprime}) according to
$
\phi^{\text{PDMP}}_{2n+1} = \int_{K^-}^{K^+} \phi_M(N)~p^{\text{PDMP}}_{\nu'_{2n+1}/s}(N) dN \approx \frac{1}{{\cal R}} \sum_{j=1}^{{\cal R}} \phi_M(N_j^{\text{PDMP}})$ and 
$
\tau^{\text{PDMP}}_{2n+1} = \int_{K^-}^{K^+} \tau_M(N)~p^{\text{PDMP}}_{\nu'_{2n+1}/s}(N) dN \approx \frac{1}{{\cal R}} \sum_{j=1}^{{\cal R}} \tau_M(N_j^{\text{PDMP}})$, where 
$\phi_M(N_j^{\text{PDMP}})$ and $\tau_M(N_j^{\text{PDMP}})$ are the Moran fixation probability and uMFT (given by Eqs.~\eqref{eq:phi_M} and \eqref{eq:MFT_M}) evaluated at $N_j^{\text{PDMP}} = N^{\text{PDMP}}(t_{\text{burn}} + j\Delta t)$. In Figs.~\ref{fig:Fig5_n1}-\ref{fig:Fig8_FigMFT_n2} and Fig.~\ref{fig:Fig10},
this method was used for $\nu'/s\sim 0.1-1000$, with ${\cal R}= 10^5$, $\Delta t= 10$ and $t_{\text{burn}}=2000$, and proved to be fast and reliable: It provided results that are
 generally in very good agreement with those of Gillespie simulations (and analytical computations), but obtained much faster (by a factor of up to 100).
%


%
\section{Supplementary figures}
\label{appendix:add-figures}

To analyse a possible dependence of the PSD, $S$ fixation probability
and uMFT on the parameter $\epsilon$,
we have reproduced the histograms of Fig.~\ref{fig:Fig2-hist}
and computed $\phi_{2n+1}$ and $\tau_{2n+1}$ 
for different values of $\epsilon$, with $-0.8\leq \epsilon\leq 5$, and obtained results that are essentially the same as those in Fig.~\ref{fig:Fig2-hist} and Figs.~\ref{fig:Fig5_n1}-\ref{fig:Fig8_FigMFT_n2} where $\epsilon=0$. This is illustrated in Fig.~\ref{fig:Fig9}(a),(b) where we have used the same parameters  as in  Fig.~\ref{fig:Fig2-hist}(g),(h), but now set $\epsilon=\pm 0.5$ rather than $\epsilon=0$.  In Fig.~\ref{fig:Fig9}(c),(d) we also show the simulation data of  $\phi_{3}$ and $\tau_{3}$ for the same parameters  as in  Figs.~\ref{fig:Fig5_n1}(b) and \ref{fig:Fig7_FigMFT_n1}(b), but  now with $\epsilon=- 0.5$ and $\epsilon=5$ instead of $\epsilon=0$.
The histograms of Fig.~\ref{fig:Fig9}(a),(b) 
are identical to those of
Fig.~\ref{fig:Fig2-hist}(g),(h), and the results for $\phi_{3}$ and $\tau_{3}$ in
Fig.~\ref{fig:Fig9}(a),(b) are essentially the same as in Fig.~\ref{fig:Fig5_n1}(b) and Fig.~\ref{fig:Fig7_FigMFT_n1}(b). 
This suggests that the PSD, $\phi_{2n+1}$ and $\tau_{2n+1}$ have no noticeable  dependence on the parameter $\epsilon$ for $-0.8\leq \epsilon\leq 5$.


\bibliographystyle{unsrt}

\bibliography{bibliography}
\end{document}